\documentclass{svjour3}                     
\usepackage{fix-cm}
\smartqed  
\usepackage{graphicx}
\newif\ifSPACEHACK
\SPACEHACKfalse

\newif\ifDEBUG
\DEBUGtrue

\newif\ifANONYMOUS
\ANONYMOUSfalse

\usepackage{amsmath,amssymb,amsfonts}
\usepackage{algorithmic}
\usepackage{graphicx}
\usepackage{textcomp}
\usepackage{xcolor}
\usepackage{booktabs} 
\usepackage{xspace} 
\usepackage[normalem]{ulem}
\usepackage{makecell}
\usepackage{tcolorbox}
\usepackage{enumitem}
\usepackage{siunitx}
\usepackage[vskip=1em,font=itshape,leftmargin=2em,rightmargin=2em]{quoting}
\usepackage{fontawesome}
\usepackage{hyperref}
\usepackage{lscape}
\usepackage{afterpage}

\usepackage{makecell}
\usepackage{multirow}
\usepackage{booktabs}
\usepackage[T1]{fontenc}

\usepackage{natbib}
\setcitestyle{aysep={}} 

\usepackage[noadjust]{cite}

\usepackage{soul}

\ifDEBUG
    \newcommand{\JD}[1]{\textcolor{purple}{[JD:#1]}}
    \newcommand{\AG}[1]{\textcolor{olive}{[AG:#1]}}
    \newcommand{\WJ}[1]{\textcolor{teal}{[WJ:#1]}}
    \newcommand{\GKT}[1]{\textcolor{brown}{[GKT:#1]}}
    
    \newcommand{\NV}[1]{\textcolor{red}{[NV: #1]}}
    
    \newcommand{\TODO}[1]{\hl{#1}}
\else
    \newcommand{\JD}[1]{}
    \newcommand{\AG}[1]{}
    \newcommand{\WJ}[1]{}
    \newcommand{\GKT}[1]{}
    \newcommand{\NS}[1]{}
    \newcommand{\NV}[1]{}
    
    \newcommand{\TODO}[1]{#1}
\fi

\newcommand{\myparagraph}[1]{\vspace{0.20cm}\noindent\textbf{#1} \noindent{}}

\ifSPACEHACK
    \titlespacing*\section{0pt}{5pt plus 1pt minus 1pt}{3pt plus 1pt minus 1pt}
    \titlespacing*\subsection{0pt}{4pt plus 1.5pt minus 1.5pt}{4pt plus 1.5pt minus 1.5pt}
    \titlespacing*\subsubsection{0pt}{3pt plus 1pt minus 1pt}{3pt plus 1.5pt minus 1.5pt}
    \titlespacing*\paragraph{0pt}{1pt plus 1.5pt minus 1.5pt}{2pt plus 1.5pt minus 1.5pt}
\fi

\usepackage{balance} 
\makeatletter
\def\cl@chapter{}
\makeatother
\usepackage{cleveref}

\crefformat{section}{\S#2#1#3}
\crefname{figure}{Figure}{Figures}
\crefname{appendix}{Appendix}{Appendices}
\crefname{table}{Table}{Tables}
\crefname{algorithm}{Algorithm}{Algorithms}
\crefname{listing}{Listing}{Listings}
\crefname{theorem}{Theorem}{Theorems}
\crefname{thm}{Theorem}{Theorems}
\crefname{lemma}{Lemma}{Lemmata}
\crefname{equation}{Eqt.}{Eqts.}
\crefformat{Grammar}{Grammar #1}

\usepackage{enumitem}
\usepackage{booktabs}

\newcommand{\ie}{\textit{i.e.,} }
\newcommand{\eg}{\textit{e.g.,} }
\newcommand{\etal}{\textit{et al.}\xspace}
\newcommand{\etals}{\textit{et al.'s}\xspace}

\newcommand{\ParticipantQuote}[2]{
    \begin{quote}
    \emph{\textbf{#1}: ``#2''} 
    \end{quote}}

\newcommand{\new}[1]{\textcolor{black}{#1}}

\newcommand{\code}[1]{{\small\texttt{#1}}\xspace}

\newcommand\SampleNum{427\xspace}
\newcommand\BugNum{348\xspace}
\newcommand\IssueNum{334\xspace}
\newcommand\nonDefectIssueNum{93\xspace}
\newcommand\RepoNum{27\xspace}
\newcommand\SoloNum{19\xspace}
\newcommand\ZooNum{8\xspace}

\newcommand\LeaderNum{6\xspace}

\newcommand\ReproducedModelNum{7\xspace}

\newcommand\TotalModelsLOC{7,643\xspace}

\begin{document}

\title{Challenges and Practices of Deep Learning Model
Reengineering: A Case Study on Computer Vision
}

\authorrunning{Wenxin Jiang et al.}
\titlerunning{Challenge and Practices of DL Model Reengineering}

\ifANONYMOUS

\author{Anonymous Author.
}

\else
\author{Wenxin Jiang         \and
        Vishnu Banna    \and
        Naveen Vivek    \and
        Abhinav Goel    \and
        Nicholas Synovic    \and
        George K. Thiruvathukal    \and
        James C. Davis    \and
}
\fi


\institute{Wenxin Jiang \at
              Purdue University, West Lafayette, IN, USA \\
              \email{jiang784@purdue.edu}           
        \and
           Vishnu Banna \at
             Purdue University, West Lafayette, IN, USA \\
              \email{vbanna@purdue.edu}
        \and
           Naveen Vivek \at
              Purdue University, West Lafayette, IN, USA \\
              \email{vivek@purdue.edu}
        \and
           Abhinav Goel \at
              Purdue University, West Lafayette, IN, USA \\
              \email{goel39@purdue.edu}
        \and
           Nicholas Synovic \at
              Loyola University Chicago, Chicago, IL, USA \\
              \email{nsynovic@luc.edu}
        \and
           George K. Thiruvathukal \at
              Loyola University Chicago, Chicago, IL, USA \\
              \email{gkt@cs.luc.edu}
        \and
            James C. Davis \at
                Purdue University, West Lafayette, IN, USA \\
              \email{davisjam@purdue.edu}
}

\date{Received: date / Accepted: date}

\maketitle

\begin{abstract} \label{sec: abstract}
\emph{Context:}
Many engineering organizations are reimplementing and extending deep neural networks from the research community.
We describe this process as deep learning model reengineering.

\noindent
\emph{Problem statement:}
Deep learning model reengineering --- reusing, reproducing, adapting, and enhancing state-of-the-art deep learning approaches --- is challenging for reasons including 
under-documented reference models, 
changing requirements, 
and the cost of implementation and testing. 
In addition, individual engineers may lack expertise in software engineering, yet teams must apply knowledge of software engineering and deep learning to succeed. 

\noindent
\emph{Related works:}
Prior work has characterized the challenges of deep learning model development, but as yet we know little about the deep learning model reengineering process and its common challenges.
Prior work has examined on DL systems from a ``product'' view, examining defects from projects regardless of the engineers' purpose.
Our study is focused on reengineering activities from a ``process'' view, and focuses on engineers specifically engaged in the reengineering process.

\noindent
\emph{Methodology:}
Our goal is to understand the characteristics and challenges of deep learning model reengineering.
We conducted a case study of this phenomenon, focusing on the context of computer vision.
Our results draw from two data sources:
  defects reported in open-source reeengineering projects, and
  interviews conducted with open-source project contributors and the leaders of a reengineering team.
In the open-source data,
  we analyzed \BugNum defects from \RepoNum open-source deep learning projects.
Meanwhile, our reengineering team
  replicated \ReproducedModelNum deep learning models over two years; we interviewed 2 practitioners and \LeaderNum reengineering team leaders to understand their experiences.

\noindent
\emph{Results:}
Our results
  describe how deep learning-based computer vision techniques are reengineered,
  analyze the distribution of defects in this process,
  and
  discuss challenges and practices. 
We found that most defects (58\%) are reported by re-users,
and that reproducibility-related defects tend to be discovered during training (68\% of them are). 
Our analysis shows that most environment defects (88\%) are interface defects, and most of environment defects (46\%) are caused by API defects.
We also found that training defects have diverse symptomss and root causes.
We identified \new{four} main challenges in the DL reengineering process:
  model operationalization, 
  performance debugging,  
  portability of DL operations, 
  \new{and customized data pipeline}.
Integrating our quantitative and qualitative data, we propose a novel reengineering workflow.

\noindent
\emph{Future directions:}
Our findings inform several future directions, including:
  measuring additional unknown aspects of model reengineering;
  standardizing engineering practices to facilitate reengineering;
  and developing tools to support model reengineering and model reuse.

\end{abstract}

\section{Introduction} \label{sec:Intro}
Deep learning (DL) over neural networks 
achieves state-of-the-art performance on diverse tasks~\citep{Schmidhuber2015DLinNN}, including games~\citep{Dota2019, vinyals2019grandmaster}, language translation~\citep{Bahdanau2015, wu2016google}, and computer vision~\citep{Ren2017FasterRCNN, Redmon2018YOLO}.
After researchers demonstrate the potential of a DL approach in solving a problem, engineering organizations may incorporate it into their products. 
This software engineering task, of reusing, reproducing, and adapting state-of-the-art DL approaches, is challenging for reasons including
  mismatch between the needs of research and practice~\citep{Tatman2018TaxonomyofReproducibility4MLResearch,Huston2018AIfacesReproducibilityCrisis},
  variation in DL libraries~\citep{Pham2020AnalysisofVarianceinDLSWSystems} and other environmental aspects~\citep{Unceta2020EnvironmentalAdaptationDifferentialReplicationinML},
  and the high cost of model training and evaluation~\citep{Goyal2018ImageNetin1Hour}. 
An improved understanding of the DL engineering process will help engineering organizations benefit from the capabilities of deep neural networks.

\begin{figure}
    \centering
    \includegraphics[width=0.9\columnwidth]{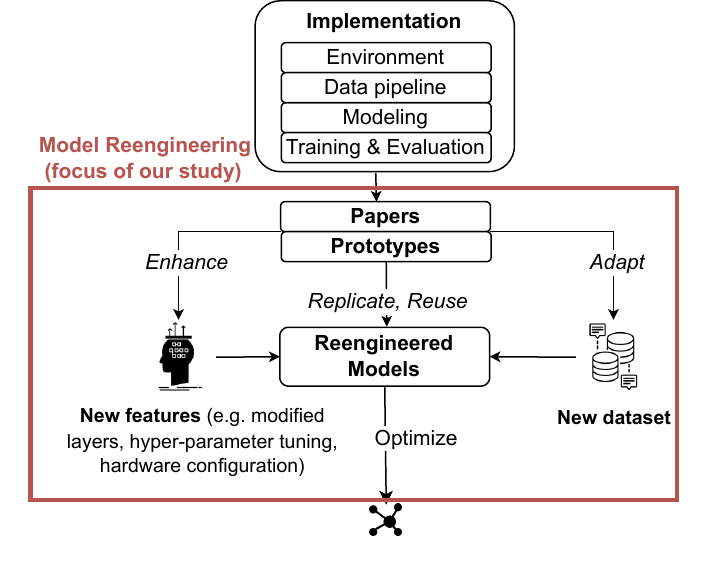}
    \caption{
        High-level overview of a DL model development and application life cycle.
        Prior work may have accidentally captured reengineering activities, but it did not describe reengineering activities as a distinct activity and was generally from ``product'' view.
        We specifically focus on model reengineering activities and ``process'' view(red box).
        We define DL \emph{reengineering} as reusing, replicating, adapting or enhancing existing DL models.}
    \label{fig:DL model lifecycle}
\end{figure}

As illustrated in \cref{fig:DL model lifecycle}, prior empirical studies have not fully explored the DL engineering process.
These works have focused on understanding the characteristics of defects during DL development. 
These works consider defect characteristics and fix patterns,
  both in general~\citep{Humbatova2020TaxonomyofRealFaultsinDLSystems, Sun2017RealBugsforML, Zhang2020ProgramFailuresofDLjobs}
  and by DL framework~\citep{Zhang2018TFBugs, Islam2019DLBugCharacteristics}.
In addition, these works focused primarily on 
the ``product'' view of DL systems, which provides an incomplete view of software engineering practice.
\new{Prior work conducted a comprehensive literature review of existing failure studies and revealed a gap in the understanding of the causal chain of defects~\citep{amusuo2022SoftwareFailureAnalysis,Anandayuvaraj2022ReflectingonRecurringIoTDevelopmentFailures}. This gap suggests the need for an approach that goes beyond simply analyzing the product itself. A ``beyond-the-product'' interpretation of failures is needed to gain a more detailed understanding of how defects arise.}
\new{On the other hand, a ``process'' view encapsulates the practices, activities, and procedures undertaken in the course of software development~\citep{pressman2005SEBOOK}. It illuminates the steps followed, strategies employed, and challenges encountered during the process of creating software. While the ``product'' view focuses on the resulting defects in a DL system, the ``process'' view would allow us to delve into the steps that led to those defects\new{~\citep{leveson2016EngineeringASaferWorld, leveson1995safeware}} and the practices that contributed to their resolution. This perspective also offers a deeper understanding on reengineering DL models, including how existing models are reused, adapted, or enhanced. It provides a framework to analyze not just the final product, but the entire journey of software development, offering more holistic insights.}

\begin{tcolorbox} [width=1.0\linewidth, colback=blue!07!white, top=1pt, bottom=1pt, left=2pt, right=2pt]
\new{\textbf{Definition}:
We define \textbf{DL model reengineering process} as: 
\textit{reusing, replicating, adapting, or enhancing an existing DL model}.}
\end{tcolorbox}

In this paper, we conducted a case study~\citep{SIGSOFT2020EmpiricalStandards4SEResearch} to examine the \textbf{Deep Learning reengineering process}: activities to
\textit{reuse, replicate, adapt, or enhance an existing DL model}.
We used a mixed-methods approach and drew from two complementary data sources~\citep{MixedMethodsResearch}.
First, to explore the characteristics, challenges, and practices of CV reengineering,
we analyzed \BugNum defects from \RepoNum open-source DL projects (\cref{Method: Bugs}).
Second, we describe the qualitative reengineering experiences of two open-source engineers and a DL reengineering team (\cref{Method: Case Study}). 
Combining these data, we report the challenges and practices of DL reengineering from a ``process'' view (\cref{sec:Result}). 
From our defect study, we observed that DL reengineering defects varied by DL stage (\cref{RQ1 results}):
 environmental configuration is dominated by API defects (\cref{RQ4 results});
 the data pipeline and modeling stages are dominated by errors in assignment and initialization (\cref{RQ2 results});
 and errors in the training stage take diverse forms (\cref{RQ4 results}).
The performance defects discovered in the training stage are the most difficult to repair (\cref{RQ3 results}).
From our interview study we identified similar challenges, notably in model implementation and in performance debugging (\cref{RQ5 results}).
These problems often arose from a lack of portability, \eg to different hardware, operating environment, or library versions.
Interview subjects described their testing and debugging techniques to address these challenges.
Synthesizing these data sources, we
  propose an idealized DL reengineering workflow~\cref{sec:Discussion}.
The difficulties we identify in DL reengineering suggest that researchers should investigate DL software testing, and that techniques to support the reuse of pre-trained models could mitigate many problems.

\ul{In summary, our main contributions are}:
	\begin{itemize}
        \item We conducted the first study that takes a ``process'' view of DL reengineering activities (\cref{sec:Method}).
       \item We analyze \BugNum defects from \RepoNum repositories in order to describe the characteristics of the defects in DL reengineering projects
	    (\cref{RQ1 results}--\cref{RQ4 results}).
	    \item We complement this quantitative failure study with qualitative data on reengineering practices and challenges (\cref{RQ5 results}). Our subjects included 2 open-source contributors \new{and 4 industry} practitioners. 
        \new{To have a more comprehensive perspective,}
        we also coordinated a two-year engineering effort by a student team to enrich this part of the study. To the best of our knowledge, this second approach is novel.
	    \item We illustrate the challenges and practices of DL model reengineering with a novel reengineering process workflow. We propose future directions for practical and empirical research based on our results and analysis (\cref{sec:Discussion}).
	\end{itemize}


\section{Background and Related Work} \label{sec:Background}

\subsection{Empirical Studies on Deep Learning Engineering Processes} \label{sec:Background-Empirical}

DL models are being adopted across many companies. With the demand for engineers with DL skills far exceeding supply~\citep{Nahar2022CollaborationChallengesinBuildingMLSystems}, companies are looking for practices that can boost the productivity of their engineers. 
Google~\citep{Breck2017MLtestscore},  Microsoft~\citep{Amershi2019SE4MLCaseStudy}, and SAP~\citep{Rahman2019MLSEinPractice} have provided insights on the current state of the DL development and indicate potential improvements.
Breck \etal indicated that it is hard to create reliable and production-level DL systems~\citep{Breck2017MLtestscore}.
Amershi \etal proposed the requirements of model customization and reuse, \ie adapting the model on different datasets and domains~\citep{Amershi2019SE4MLCaseStudy}.
Rahman \etal pointed out that knowledge transfer is one of the major collaboration challenges between industry and academia~\citep{Rahman2019MLSEinPractice}.
Our work identifies challenges and practices of knowledge transfer from academia to industry and supports creating reliable customized models.

In addition to views of the industry,
\new{academic researchers have also conducted empirical studies and supplied strategies to solve some DL engineering challenges}.
Zhang~\etal
illustrated the need for cross-framework
differential testing and the demand for facilitating debugging and profiling~\citep{Zhang2019CommonChallengesinDevelopingDLApplications}.
Serban~\etal discussed the engineering challenges and the support of development practices, and highlighted \new{some low adopted effective practices on testing, automating hyper-parameter optimization and model selection~\citep{Serban2020SEBPinMLAdoptionEffects}.}
Lorenzoni~\etal showed how DL developers could benefit
from a traditional software engineering
approach and proposed improvements in the ML development workflow~\citep{Lorenzoni2021MLModelDevelopmentfromSEPerspective}.
\new{These strategies are based on ``products'' rather than ``process''. We studied and proposed the problem-solving strategies from the process view.}

\cref{fig:DL model lifecycle} illustrates how our work differs from previous empirical studies. 
Although there have been many software engineering studies on DL, they 
collect data from a ``product'' view of DL systems~\citep{Islam2019DLBugCharacteristics, Shen2021DLCompilerBugs, Chen2022DLFrameworkBug, Lorenzoni2021MLModelDevelopmentfromSEPerspective}.
These works sample open-source defects or Stack Overflow questions and report on the features, functionalities, and quality of DL software.
Some work mentioned specific reengineering activities, such as model customization~\citep{Amershi2019SE4MLCaseStudy} and knowledge transfer of DL technologies~\citep{Rahman2019MLSEinPractice}.
However, there is no work \new{focusing on the ``process'' of these activities} yet.
We conduct the first empirical study of DL software from a ``process'' view which focuses on the activities and processes involved in creating the software product.
We specifically examine the DL reengineering activities and process by sampling the defects from open-source research prototypes and replications. 
We also collected qualitative data about the activities and process by interviewing open-source contributors and leaders of the student reengineering team.

\subsection{Reengineering in Machine Learning and Deep Learning} 
Historically, \emph{software reengineering} referred to replicating, understanding, or improving an existing implementation~\citep{SEReengineering}.
This process arises from needs including optimization, adaptation, and enhancement~\citep{Jarzabek1993SWReengineering4Reusability, Byrne1992ConceptualFoundation4SWReengineering, Tucker2010CaseStudyinSWREengineering}. 
Today, we observe that ML and DL 
engineers are reusing, replicating, adapting, and enhancing existing models to understand the algorithms and improve their implementation~\citep{Amershi2019SE4MLCaseStudy, Alahmari2020RepeatabilityofDLModels}.
The maintainers of major ML frameworks, including TensorFlow and Pytorch, store reengineered models within official GitHub repositories~\citep{TFMGGithub, TorchVisionGithub}.
Many engineering companies, including 
  Amazon~\citep{Amazon2018MLModelManagementChallenge},
  Google~\citep{Intro2TFMG},
  and Meta~\citep{FacebookAIReproducibility} 
are engaged in forms of \emph{DL reengineering}.
For example, Google sponsors the TensorFlow Model Garden which provides ``a centralized place to find code examples for state-of-the-art models and reusable modeling libraries''~\citep{Intro2TFMG}. Many research papers use PyTorch because it is easy to learn and has been rapidly adopted in research community, but many companies prefer TensorFlow versions because of it provides better visualization and robust deployment~\citep{Ryan2023PytorchvsTF}.
\new{Our work identified the differences between traditional software reengineering and DL reengineering process, in terms of the goals and its causal factors.}
\new{In this work we conceptualize reengineering a DL model and developing a new DL model as distinct engineering activities. These activities do overlap in some ways. Reengineering involves refining an existing DL model implementation for improved performance, such as modifying architecture or tuning hyperparameters. Conversely, developing a new DL model usually involves building a model from scratch, potentially drawing upon high-level concepts or algorithms from existing models but not directly modifying their code. A prime example of the blurred boundary is seen in \href{https://github.com/ultralytics/yolov5}{\code{ultralytics/yolov5}}, where significant enhancements to a replication of \href{https://github.com/ultralytics/yolov3}{YOLOv3} codebase led to models considered ``new'' \new{(the YOLOv5 model).}}
\new{We highlight that adapting and enhancing can sometimes be considered as developing a new DL model, specifically when a model from one domain is adapted to another domain. In this study, such adaptation is considered as part of the reengineering process.}
\new{However, despite the overlap between reengineering and developing new models, our failure study results reveal distinct patterns of issues and challenges faced in these two engineering activities, highlighting the importance of treating them as separate tasks with unique considerations.}

ML reengineering has received much attention in the research community~\citep{bhatia2023towardsMLpipelines, Huston2018AIfacesReproducibilityCrisis, Pineau2020, Gundersen2018ReproducibilityinAI, Gundersen2018onReproducibleAI}. 
Pineau \etal noted three needed characteristics for ML software: reproducibility, re-usability, and robustness. They also proposed two problems regarding reproducibility: an insufficient exploration of experimental variables, and the need for proper documentation~\citep{Pineau2020}. 
Gundersen \etal surveyed 400 AI publications, and indicated that documentation facilitates reproducibility.
They proposed a reproducibility checklist~\citep{Gundersen2018ReproducibilityinAI, Gundersen2018onReproducibleAI}.
Chen \etal highlights that the reproducibility of DL models is still a challenging task as of 2022~\citep{Chen2022TowardsTrainingReproducibleDLModels}.
Consequently, the characteristics of the reengineering process are significant for both practitioners~\citep{DeterminedAIReproducibility, FacebookAIReproducibility} and researchers~\citep{MLReproducibilityChallengeSpring2021, CMUMLReproducibility}.
\new{Previous research on machine learning research repositories on GitHub explored contributions from forks by analyzing the commits and PRs, but found few actually contributed back~\citep{bhatia2023towardsMLpipelines}. 
Our study takes a different data collection approach by examining defect reports from downstream users in the upstream repository, offering a new perspective on the reengineering process and feedback from downstream engineers.}
\new{
The DL reengineering process and its associated challenges and practices have been unexplored.
Prior work has used the concept of DL reengineering as a particular method of reusing DL models~\citep{Qi2023DNNReengineering}. Our research uniquely defines and investigates the process of DL model reengineering. This reengineering process and its associated challenges and practices have been unexplored, but we highlight its significance in the software engineering field.
}

The combination of the software, hardware, and neural network problem domains exacerbates the difficulty of deep learning reengineering.
The DL ecosystem is evolving, and practitioners have varying software environments and hardware configurations~\citep{Boehm2010ChangingNatureofSWEvolution}. 
This variation makes it hard to reproduce and adapt models~\citep{Goel2020LPDL}.
Additionally, neural networks are reportedly harder to debug than traditional software, \eg due to their lack of interpretability~\citep{Bibal2016InterpretabilityofMLModelsandRepresentations, DoshiVelez2017RigoriousInterpretableML}.


To facilitate the reengineering of DL systems, researchers advise the community to increase the level of portability and standardization in engineering processes and documentation~\citep{Gundersen2018ReproducibilityinAI, Pineau2020, mmdnn}. 
Microsoft indicated that existing DL frameworks focus on runtime performance and expressiveness and neglect composability and portability~\citep{mmdnn}.
The lack of standardization makes finding, testing, customizing, and evaluating models a tedious task~\citep{Gundersen2018onReproducibleAI}.
These tasks require engineers to ``glue'' libraries, reformat datasets, and debug unfamiliar code ---  a brittle, time-consuming, and error-prone approach~\citep{Sculley2014MLTechDebt}.

To support DL engineers, we conducted a case study on the defects, challenges, and practices of DL reengineering. 



\subsection{Deep Learning Defects and Fix Patterns}
Prior work focused on the general DL development process, studying the defects, characteristics, and symptoms. 
Islam~\etal demonstrated DL defect characteristics from 5 DL libraries in GitHub and 2716 Stack Overflow posts~\citep{Islam2019DLBugCharacteristics}. 
Humbatova~\etal analyzed data from Stack Overflow and GitHub to obtain a taxonomy of DL faults~\citep{Humbatova2020TaxonomyofRealFaultsinDLSystems}. 
By surveying engineers and researchers, Nikanjam~\etal specified eight design smells in DL programs~\citep{Nikanjam2021DesignSmellinDL}.

Furthermore, researchers conducted works on DL fix patterns~\citep{Sun2017RealBugsforML, Islam2020RepairingDNN:FixpatternsChallenges}. 
Sun~\etal analyzed 329 defects for their fix category, pattern, and duration~\citep{Sun2017RealBugsforML}.
Islam~\etal considered the distribution of DL fix patterns~\citep{Islam2020RepairingDNN:FixpatternsChallenges}.
Their findings revealed a distinction between DL defects and traditional ones.
They also identified challenges in the development process: fault localization, reuse of trained models, and coping with frequent changes in DL frameworks.
Some development defects studied from prior work are ``wrong tensor
shape''~\citep{Humbatova2020TaxonomyofRealFaultsinDLSystems} and ``\code{ValueError} when performing \code{matmul} with TensorFlow''~\citep{Islam2019DLBugCharacteristics}.
However, typical reengineering defects have not been well discovered, such as ``user's training accuracy was lower than what was claimed'' or ``training on a different hardware is very slow'' (\cref{table: IssueExamples}).

Prior work considered the defects in the DL model development process, without distinguishing which part of the development process they were conducting.
In this work, we focus on defects arising specifically during the DL model reengineering process (\cref{fig:DL model lifecycle}).
We use defect data from GitHub repositories.
We also collect interview data, providing an unusual perspective ---
prior studies used data from
  Stack Overflow~\citep{Islam2019DLBugCharacteristics, Zhang2018TFBugs, Humbatova2020TaxonomyofRealFaultsinDLSystems},
  open-source projects~\citep{Zhang2018TFBugs, Islam2019DLBugCharacteristics, Sun2017RealBugsforML, Humbatova2020TaxonomyofRealFaultsinDLSystems, Shen2021DLCompilerBugs, Garcia2020AVBugs},
  and surveys~\citep{Nikanjam2021DesignSmellinDL}.
\section{Research Questions} \label{sec:RQ}



To summarize the literature:
%
DL reengineering is common \new{as a process} in engineering practice, but is challenging for reasons including (under-) documentation, shifting software and hardware requirements, and unpredictable computational expense.
Prior work has studied the problems of DL engineering,
  \eg considering
    the characteristics of DL defects~\citep{Zhang2020NumericalBugsinNN}
    sometimes distinguished by DL framework~\citep{Zhang2018TFBugs}.
However, this prior work has not distinguished between the activities of DL development and DL reengineering, and has not examined DL reengineering specifically.



In this work, we define \textbf{DL model reengineering process} as: 
\textit{reusing, replicating, adapting, or enhancing an existing DL model}
We ask:


\begin{itemize}[leftmargin=50pt, rightmargin=10pt]


    
    \item[\textbf{RQ1}]  \textit{How do defects manifest in Deep Learning reengineering?}
    \item[\textbf{RQ2}] \textit{What types of Deep Learning reengineering defects are frequent?}
    \item[\textbf{RQ3}] \textit{What are the symptomss of Deep Learning reengineering defects?}
    \item[\textbf{RQ4}] \textit{What are the root causes of Deep Learning reengineering defects?}
    \item[\textbf{RQ5}] \textit{What are the challenges and practices in Deep Learning reengineering?}

\end{itemize}


    
    
    
    
    
    

    

\section{Model of Deep Learning Reengineering} \label{sec: ReengModel}
\new{We introduce the model of DL reengineering in this section. We include our preliminary analysis on open-source projects (\cref{sec:PreAnaOSProjects}), and discuss about the reengineering concepts (\cref{sec:ReengModel-ReengConcepts}). Finally, we also talked about the relationships of open-source reengineering projects \cref{sec:ReengModel-OSRelationships} and repository types (\cref{sec:ReengModel-RepoTypes}).}
\new{\subsection{Preliminary Analysis on Open-source Projects}}\label{sec:PreAnaOSProjects}
\new{In order to understand the characteristics of deep learning projects involving reengineering activities, we first conducted a preliminary study to obtain an overview of the common trends and challenges associated with reengineering in the context of DL applications~\citep{kitchenham2002preliminaryGuidlinesforEMSE, easterbrook2008selectingEMMethodsforSEResearch}.}


\new{For our preliminary analysis, we focused on popular GitHub projects implementing algorithms from the YOLO family~\citep{jiang2022YOLOreview}, which were included in \cref{table:repo info}. We chose the YOLO family because of its popularity and evolution over time, as it represents state-of-the-art in computer vision tasks within deep learning. The YOLO family's prevalence in open-source projects provides a rich dataset for analysis, and its multiple versions, each introducing new concepts and improvements, make it ideal for studying DL software evolution and reengineering.}

\new{By examining the most-commented issues in these projects, we were able to identify recurring concepts and connect them with concepts introduced by prior work, including reengineering concepts (\cref{sec:ReengModel-ReengConcepts}), open-source relationships (\cref{sec:ReengModel-OSRelationships}), and repository types (\cref{sec:ReengModel-RepoTypes}).
Putting these concepts together, we introduce our model of DL reengineering.}

\subsection{Reengineering Concepts} \label{sec:ReengModel-ReengConcepts}
\new{Prior work shows that studying defect manifestation, types, symptoms, and root causes separately can help understand the characteristics of defects~\citep{Islam2019DLBugCharacteristics, Wang2017ConcurrencyBugsinNodejs, Liu2014PerformanceBugs4SmartPhoneApps}.
To better characterize DL reengineering defects, we first introduce the relevant concepts specific to DL reengineering process --- defect \emph{reporters} (\cref{table: reporter})
  and
  defect \emph{manifestations} (\cref{table: manifestation}).
These concepts are based on our literature review and observations in preliminary analysis on open-source projects.
} 

\cref{table: reporter} defines four different types of \textit{Defect reporters} based on their reengineering activities. Prior work indicates that there are inconsistencies in the use of these terms~\citep{Gundersen2018ReproducibilityinAI}. 
Our definitions are mainly based on prior work\citep{Pineau2020, Li2020ModelAdaptation, GithubLabels, Amershi2019SE4MLCaseStudy, Alahmari2020RepeatabilityofDLModels} and our observations in preliminary analysis.

{
\begin{table}[h]
\small
\centering
\caption{
    Reporter types in the DL reengineering ecosystem. The reporter types are determined by whether they use the same code, dataset, and algorithm compared to the upstream project.
}
\vspace{-2mm}
\label{table: reporter}
\begin{tabular}{p{4cm} p{7cm}}
\toprule
\textbf{Reporter Category} & \textbf{Description} \\
\toprule 
\textbf{Re-user} & Uses the same code and dataset. This is the common-case behavior associated with pure re-use. \\
\\
\textbf{Adaptor} & Adapts code to other tasks (different dataset) and finds inconsistency compared to expectations.\\
\\
\textbf{Enhancer} & Adds new features (\eg layer modification, hyper-parameter tuning, multi-GPU training configuration).\\
\\
\textbf{Replicator} & Uses the same algorithm, data, and configuration, in distinct implementation (\eg TensorFlow vs. PyTorch).\\
\bottomrule
\end{tabular}
\end{table}
}

\cref{table: manifestation} defines three types of \textit{Defect manifestation}. We followed prior work studying the manifestation of defects~\citep{Liu2014PerformanceBugs4SmartPhoneApps}.
\new{The most obvious manifestation during reengineering process is whether the model crashes when it is executed. This is a common defect when running any DL model~\citep{Islam2019DLBugCharacteristics, Zhang2018TFBugs, guan2023RealWorldBugsinMLmodel}. 
\new{Prior work has also noted the difficulty in reproducing DL models, and we would like to understand how these difficulties impact the reengineering process~\citep{Sculley2015HiddenTechnicalDebtinMLSystems, Liu2020ReplicabilityandReproducibilityofDLinSE}.}
Additionally, our preliminary analysis also revealed a significant number of issues related to performance improvement or new feature addition to the original implementation. Often labelled as ``enhancement'' on GitHub, these issues typically emerge when features are added or when environments change. Since enhancements form an integral part of the reengineering process, we included them as a category of manifestation.
}

Manifestation provides an unusual view combining symptoms, root causes, and relevant contexts. We believe studying defect manifestations can give us a unique perspective of reengineering process.

{
\begin{table}[h]
\small
\centering
\caption{
    Defect manifestations and relevant definitions, determined by the runnability of code and the data used to train the model. 
}
\label{table: manifestation}
\begin{tabular}{p{4cm} p{7cm}}
\toprule
\textbf{Defect Category} & \textbf{Description} \\
\toprule
\textbf{Basic defects} & The code does not run (\eg it crashes, behaves very incorrectly, or runs out of memory).\\
\\
\textbf{Reproducibility defects} & Using the same data, the code runs without basic defects, but does not match the documented performance (\eg accuracy, latency).\\
\\
\textbf{Evolutionary defects} & The code and/or data has been changed to adapt to the user’s needs. It runs without basic defects, but does not match the specification/desired performance.\\

\bottomrule
\end{tabular}
\end{table}
}

\subsection{Open-source Relationships} \label{sec:ReengModel-OSRelationships}
\begin{figure}[h]
    \centering
    \includegraphics[width=0.8\linewidth]{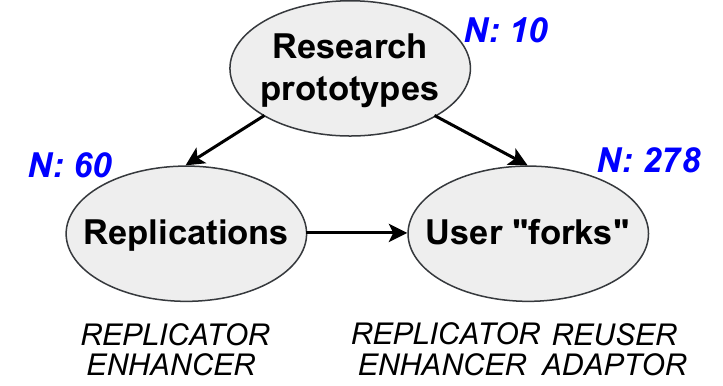}
    \caption{
    Relationship between CV reengineering repositories.
    The capitalized texts indicate what types of reporters commonly open issues in the above projects.
    Arrows indicate the dataflow of model reuse between upstream and downstream projects.
    \emph{N}: the number of defects resolved in each type of the projects we analyzed.}
    \label{fig:lifecycle}
\end{figure}

Based on the typical open-source software collaboration process~\citep{fitzgerald2006OSStransformation},
we modeled the expected relationship between CV reengineering repositories --- see~\cref{fig:lifecycle}. 
This figure depicts the relationship between three types of projects:
\begin{itemize}
    \item  \textit{Research Prototype}: an original implementation of a model.
    \item \textit{Replications}: a replication and possible evolution of the research prototype.
    \item \textit{User ``fork''}: a GitHub fork, clone, copy, etc. where users try to reuse or extend a model from the previous two types. 
\end{itemize}

 We expect the existence of two actions, ``\textit{REUSE} (fork)''  and ``\textit{REPORT (defect)}'', happening between different project types.
The down-stream projects reuse up-stream projects, and engineers open new issues when they encounter defects or identify necessary enhancements.
Typically the issues are reported in upstream projects, the maintainers explain the necessary fix patterns for reengineers,
and the defects are repaired in downstream ones.
Because our goal is to study reengineering defects, we do not study ``forks'', but instead examine the up-stream projects: research prototypes and replications. 





\subsection{Reengineering Repository Types} \label{sec:ReengModel-RepoTypes}
We identified two types of repositories during our preliminary analysis: \textit{zoo} and \textit{solo} repository.
A \emph{Solo} repository is either the research prototype from the authors or an independent replication.
\textit{Zoo repositories} all contain implementations of several models. 
A \textit{zoo repository} can be the combination of research prototypes and replications. 
For example, \code{TensorFlow Model Garden} contains different versions of YOLO~\citep{TFMGGithub}.
These \textit{zoo repositories} have been widely reused and contain many relevant issues (\cref{table:repo info}).

\section{Methodology} \label{sec:Method}

To answer our research questions, we used a case study approach~\citep{Perry2004CaseStudy4SE, Runeson2009Guidelines4CaseStudyinSE} that combined two perspectives~\citep{MixedMethodsResearch}.
For RQ1--RQ4, we analyzed DL reengineering defects in open-source reengineering projects.
However, this data source may be constraining because GitHub issues are known to be limited in the breadth of issue types and the depth of detail~\citep{Aranda2009SecretLifeofBugs}. 
To answer RQ5, we collected qualitative reengineering experiences from open-source contributors and the leaders of a DL reengineering team after they completed several projects.
These two perspectives are complementary:
  the open-source defects gave us broad insights into some kinds of reengineering defects,
  while
  the DL reengineering team gave us deep insights into the reengineering challenges and process.
\cref{fig:Methods-RQs} shows the relationship between our questions and methods.
To promote replicability and further analysis, all data is available in our artifact (\cref{sec:Reproducibility}).

{
\small
\renewcommand{\arraystretch}{0.9}
\begin{figure}[h]
    \centering
    \includegraphics[width=0.8\linewidth]{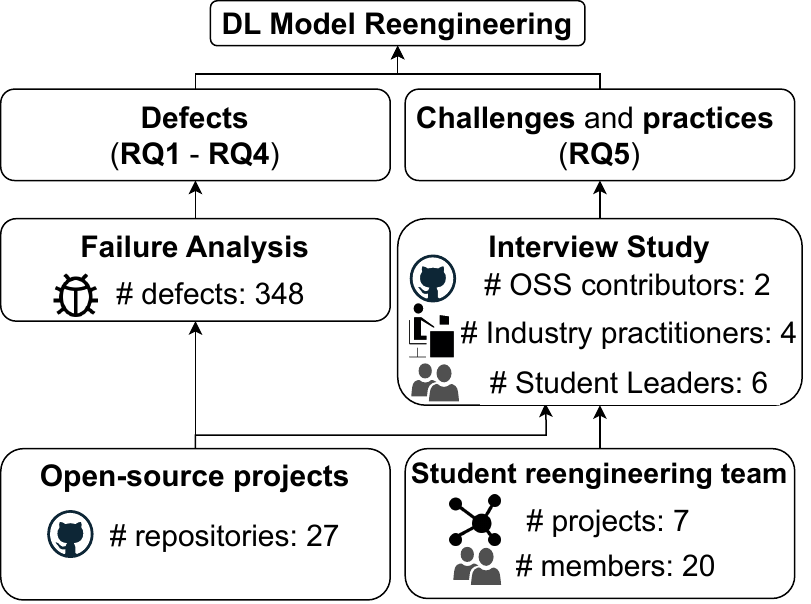}
    \caption{
    Relationship of research questions to methodology. The failure analysis is conducted on open-source \new{GitHub} projects, while the interview study is conducted on \new{2 contributors to the GitHub projects we studied, \new{4 industry practitioners recruited via social media platforms}, and 6 leaders from our reengineering team}.
    }
    \label{fig:Methods-RQs}
\end{figure}
}

In our case study, the phenomenon of interest was DL reengineering.
We situated this study in the context of Computer Vision (CV).
\new{CV plays a critical role in our society, with applications addressing a wide range of problems from medical to industrial robotics~\citep{Xu2021CVinConstructionACriticalReview}.
To scope our study, we focused on DL model reengineering in computer vision.
This case study approach gives us a comprehensive view of a particular DL application domain.
However, we perceive three reasons why our results may generalize to other DL application domains.
}
\new{\textbf{First},} as a major application domain of DL techniques~\citep{DL4CV, CVEngineer2021, Xu2021CVinConstructionACriticalReview}, CV serves as a microcosm of DL engineering processes~\citep{Amershi2019SE4MLCaseStudy, thiruvathukal2022lpcv}.\footnote{We acknowledge that there are approaches to computer vision that do not leverage deep learning~\citep{szeliski2022ComputerVision, forsyth2002ComputerVision}. In this study, we focused on engineers applying deep learning to problems in computer vision. }
\new{\textbf{Second}, techniques are shared between computer vision and other applications of deep learning. 
For example, the computer vision technique of Convolution Neural Networks (CNNs)~\citep{oShea2015introductiontoCNN} has been adapted and applied to identify features and patterns in Natural Language Processing and Audio recognition tasks~\citep{li2021CNNsurvey, alzubaidi2021DLreview}.
From other fields, neural network architectures such as transformers and large language models were initially designed for Natural Language Processing (NLP) and are now being applied to CV tasks~\citep{Cheng2022Mask2Former, Zou2023MetaSegmentEverything}.
}
\new{\textbf{Third}, the engineering process for deep learning models, encompassing stages such as problem understanding, data preparation, model development, evaluation, and deployment, is consistent across different domains, including both computer vision and natural language processing, thus further supporting the potential generalizability of our findings~\citep{Sculley2015HiddenTechnicalDebtinMLSystems, Amershi2019SE4MLCaseStudy}}
\new{Additionally, our focus aligns with prior \new{empirical software engineering studies of} deep learning, much of which has concentrated on the field of computer vision as a representative domain for deep learning studies.~\citep{ma2018deepgauge, ma2018mode, wei2022sebox4dl, Pham2020AnalysisofVarianceinDLSWSystems}.}
Our case study will thus provide findings for computer vision reengineering projects, which are important; and our results may generalize to other domains of deep learning.

{
\renewcommand{\arraystretch}{1}
\begin{figure}[h]
    \centering
    \includegraphics[width=0.9\linewidth]{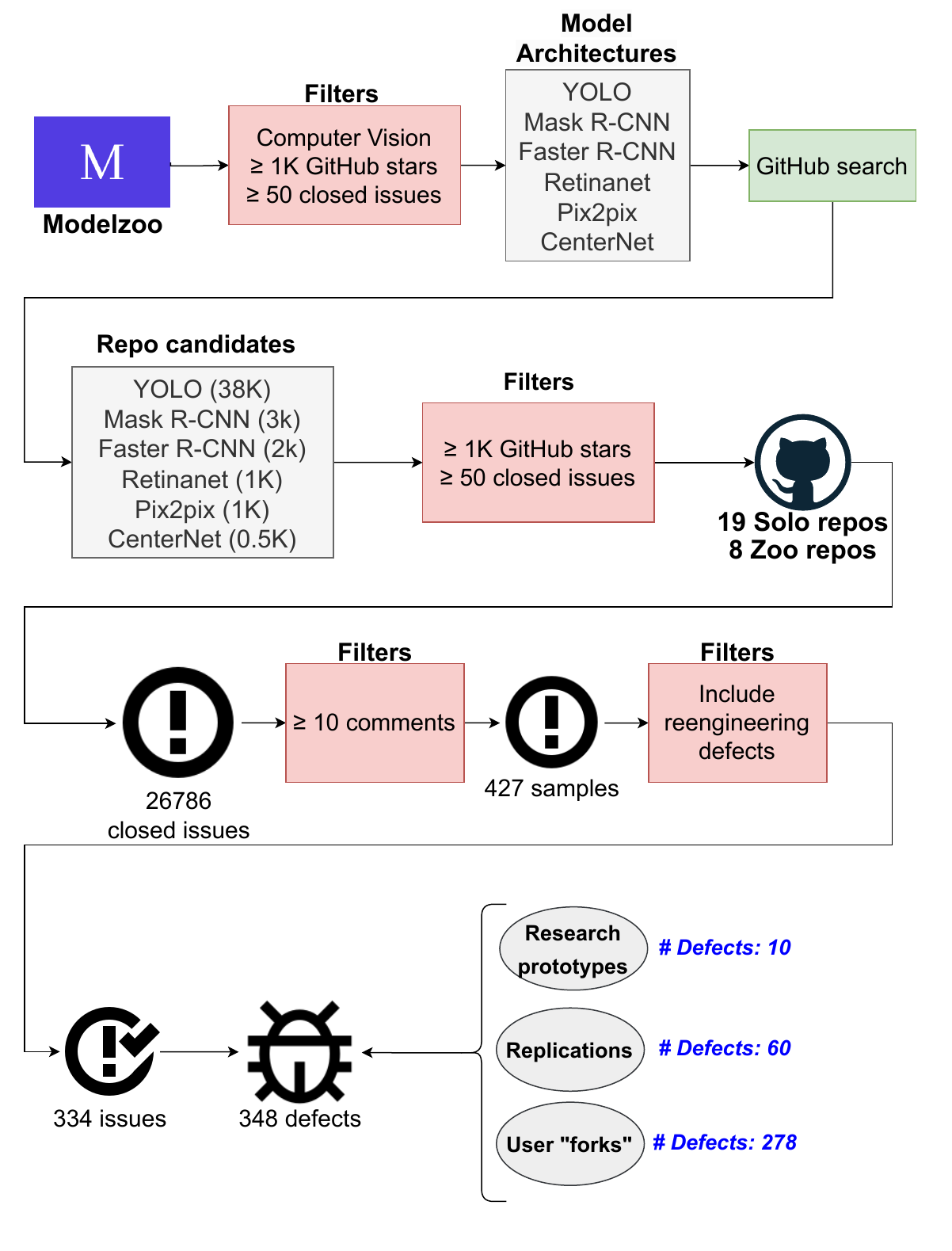}
    \caption{
    Overview of defect collection and distribution of the collected defects in three project types, as well as the number of projects and defects we got in each step. 
    The number of each model architecture after GitHub search (Most data were collected in 2021. Data associated with ``Repo candidates'' were collected in Feb, 2023)
    }
    \label{fig:BugStudyDataCollection}
\end{figure}
}

%

\subsection{RQ1-4: Failure Analysis on Computer Vision Reengineering} \label{Method: Bugs}

\cref{fig:BugStudyDataCollection} shows the overview of our data collection in the failure analysis (bug study).
We adapted previous failure analysis methods~\citep{Islam2019DLBugCharacteristics, Zhang2018TFBugs, Wang2020RegexBugs, Eghbali2020String-relatedBugs} to identify and analyze defects in DL reengineering. 
In total,
we examined \BugNum defects from \RepoNum repositories (\cref{table:repo info}).
The rest of this section discusses our selection and analysis method.

\subsubsection{Repository Selection} 

To find sufficient DL reengineering defects in this target --- research prototypes and replication projects --- we chose to look at CV models with substantial popularity and GitHub issues.
We proceeded as follows, along the top row of~\cref{fig:BugStudyDataCollection}:
\begin{enumerate}[leftmargin=1cm]
\item We started by selecting popular CV models from ModelZoo~\citep{ModelZooWeb}, a platform
that tracks state-of-the-art DL models. 
We looked at relevant GitHub repository and chose the CV models with over 1K GitHub stars and 50 closed issues.
\item We searched for implementations of these model architectures on GitHub.
\new{We utilized GitHub's code search feature, filtering by keywords associated with the model names.}
\item 
\new{For each architecture, we selected projects that implement the same model, each with a minimum of 1,000 GitHub stars and 50 closed issues. 
The projects were selected based on their popularity, as indicated by the number of stars~\citep{WhatsinaGithubStar}. 
However, to maintain the diversity of our data, we only chose up to the top five repositories for any given model. 
If there were more than five projects for a single model, we limited our selection to the five most popular ones. 
This approach helps to prevent bias in our results, such as over-representation of specific reproducibility issues or enhancements tied to a single model or architecture family.}
If there was only one implementation matching our criteria, we excluded that model in our analysis. 
\end{enumerate}

As shown in \cref{table:repo info} (last 4 rows), for two architectures this process did not always yield both prototypes and replications.
For \code{Pix2pix} we retrieved two prototypical implementations in different programming languages (\emph{viz.}\xspace Lua and Python). 
For \code{CenterNet} we retrieved two prototypes for different model architectures that share the same model name.
However, on inspection we found that these four repositories were actively reused, had many people engaged in the reengineering process, and had many reengineering defects.
Therefore, we included the issues from these repositories in our analysis.

As shown in \cref{fig:BugStudyDataCollection}, the same model architecture can be found in either \textit{zoo repositories} or \textit{solo repositories} during GitHub searching (\cref{sec:ReengModel-RepoTypes}).
Most of the repositories (19/27) we identified during the GitHub searching are \textit{solo repositories} which only implement a single model.
Both \textit{solo} and \textit{zoo repositories} have DL reengineering defects reported by down-stream replicators or users.
Therefore, we applied the same data collection methods to them and put the data together.
%

Overall, we examined \SoloNum \textit{Solo Repositories} and \ZooNum \textit{Zoo Repositories}.

\subsubsection{Issue Selection}
We applied two filters on issues in these repositories:
(1) Issue status is \textit{closed} with an associated \textit{fix}, to understand the defect type and root causes~\citep{Tan2014OSSBugCharacteristics}; 
(2) Issue has $\geq$ 10 comments, providing enough information for analysis.
We first used the two filters to filter the full set of issues in each repository, and then sampled the remainder. 

Our goal was to analyze at least 10\% of the issues for each reengineering project, but this was balanced against the wide range of qualifying issues for each repository. For example, \code{ultralytics/yolov5} has 279 qualifying issues while \code{yhenon/pytorch-retinanet} has only 4.
We first conducted a pilot study on 5 solo projects which includes 69 defects.\footnote{More details can be found in \cref{Method: Taxonomy}.} The pilot study indicated that choosing the 20 most-commented issues would cover roughly 10\% of the issues for these projects and give us plenty of data for analysis. 

\new{In order to achieve a sufficient confidence level from analyzing only 10\% of issues, we applied our filters uniformly across both solo and zoo repositories to prevent bias stemming from specific models or zoos. This approach allows us to focus on the broader process rather than being influenced by individual ``products''. The issues we analyzed were both representative of the most commonly encountered challenges in the field and diverse in terms of originating repositories, model types, and user engagement levels, including the most commented issues and those from different repositories and models. This strategy increases our confidence that the insights derived from this sample are reflective of the larger population of issues. 
}


For two zoo repositories (\code{tensorflow/models} and \code{pytorch/vision}), the most-commented issues are controversial API change requests, not reengineering defects, so we randomly sampled 10\% of the closed issues instead. For some smaller repositories, taking the top-20 qualifying issues consumed all available defects (\cref{table:repo info}).

For most of the selected repositories, we sorted the remaining issues by the number of comments and examined the 20 issues with the most comments.
This sample constituted $\geq$ 10\% of their issues.
For the zoo repositories from two major DL frameworks, \code{tensorflow/models} and \code{pytorch/vision}, the most-commented issues were API change requests, not defects.
Here, we randomly sampled 10\% of issues that met the first two filters.

\new{Another critical aspect of our data collection was to distinguish between reengineering and non-reengineering defects. 
\new{\cref{table: IssueExamples} provides examples of several defects of each kind.}
The criterion for this differentiation was based on whether the defect occurred during the process of reusing, replicating, adapting, or enhancing an existing deep learning model (\emph{reengineering defects}) or not (\emph{non-reengineering defects}). 
In this regard, reengineering defects were those that were directly related to these processes and hindered the correct function or performance of the model. For instance, a defect was considered as a reengineering defect if it pertains to a problem such as the model producing incorrect results when trained with a new dataset or the model failing to perform as expected after being adapted to a new use case.
\new{Lastly, there are also some issues classified as \emph{non-defects} that we excluded from our study.}
Examples include development questions or feature requests. 
These issues, although important in the broader software development process, are not directly related to the reengineering process and hence were not included in our study.
}
From these samples, we manually filtered out \nonDefectIssueNum non-defect issues, \eg development questions and feature requests.
\emph{78\% (\IssueNum/\SampleNum)} of sampled issues included a reengineering defect.


{
\small
\renewcommand{\arraystretch}{0.4}
\begin{table}[h]
\centering
\caption{
  Examples of reengineering defects included in our study, and non-reengineering defects excluded from the study.
  A reengineering defect refers to an error or flaw that occurs during the process of reusing, replicating, adapting, or enhancing an existing deep learning model.}
\label{table: IssueExamples}
\begin{tabular}{p{0.2\linewidth} p{0.07\linewidth} p{0.31\linewidth} p{0.33\linewidth}}
\toprule
\textbf{Issue ID} & \textbf{Type} & \textbf{\new{Description}} & \textbf{Fix}                  \\
\midrule
\href{https://github.com/tensorflow/models/issues/6043}{\emph{tensorflow/models \#6043}}   
& Reeng. 
& The losses occasionally contained a NaN value when using a customized dataset.        
& Replace the empty sequence in tensor with zeros. \\
\\
    \\
\href{https://github.com/ultralytics/yolov3/issues/310}{\emph{ultralytics/yolov3 \#310}}
& Reeng. 
& User training accuracy was lower than what was claimed by the replicator in the documentation. 
& Load the original checkpoint weights before training.
\\
    \\
\href{https://github.com/matterport/Mask_RCNN/issues/1938}{\emph{matterport/Mask \_RCNN\#1938}}
& Reeng. 
& Training on the RTX 2080 Ti GPU with CUDA library version 9 is slow.
& Fix the environment by following provided instructions and reduce the training epochs.
\\
  \\
\href{https://github.com/facebookresearch/detectron2/issues/3225}{\emph{facebookresearch/ detectron2\#3225}}
& Reeng. 
& Documentation does not seem to have been updated to reflect the new config files (\code{.py} rather than \code{.yaml})
& Add documentation for new training script to use existing configuration files.
\\
\midrule

\href{https://github.com/facebookresearch/Detectron/issues/370}{\emph{facebookresearch/ Detectron \#370}} & Non-reeng. &
  The \code{Caffe2} library does not have GPU support.
  & Use Docker installation.
  \\
\\

\href{https://github.com/tensorflow/models/issues/1838}{\emph{tensorflow/models \#1838}} & Non-reeng. &
  \code{pip install} does not work to install a specific version of TensorFlow.
  & Install the package from PyPI.
  \\

\midrule
\href{https://github.com/xingyizhou/CenterNet/issues/241}{\emph{xingyizhou/ CenterNet \#241}} & Non-reeng. &
  The user asked about the implementation details of a loss function.
  & N/A
  \\
\\
\href{https://github.com/facebookresearch/detectron2/issues/8}{\emph{facebookresearch/ detectron2 \#8}} & Non-defect &
  The user requested ONNX support of an existing model.
  & N/A
  \\
\\
  \bottomrule
\end{tabular}%
\end{table}
}

{
\footnotesize
\small 
\renewcommand{\arraystretch}{0.5}
\begin{landscape}
\begin{table}[t]
\caption{
  The studied repositories and defects.
  The \emph{Closed issues (qualified)} indicates the total closed issues and those matching two filters (closed, $\geq10$ comments). 
  \emph{Samples} are sampled from the qualified closed issues.
  After these selection criteria, we manually identified \IssueNum GitHub issues from the samples that described at least one reengineering defect.
  Some GitHub issues contained multiple reengineering defects.
  \textdagger Repository \code{rwightman/pytorch-image-models} was converted to \code{huggingface/pytorch-image-models} after the study completed.
  }
\label{table:repo info}
\begin{tabular}{
    l
    c
    S[table-number-alignment = right]
    S[table-number-alignment = left]
    |
    p{0.1\linewidth}
    S[table-number-alignment = left]
    p{0.08\linewidth}
}
\toprule
    \textbf{Repository} &
      \textbf{Type} &
      \textbf{Stars (K)} &
      \multicolumn{1}{l|}{\textbf{Forks (K)}} &
      \textbf{Closed issues (qualified)} &
      \textbf{Samples} &
      \textbf{Reeng. issues (defects)} \\
\midrule
\textbf{Zoo Repos}\\
\midrule
\href{https://github.com/tensorflow/models}{tensorflow/models} \emph{(Google)}                    & Zoo & 71.8  & 45.0    & 5560 (580) & 58 & 51 (51) \\
\href{https://github.com/facebookresearch/Detectron}{facebookresearch/Detectron} \emph{(Facebook)}           & Zoo & 24.8  & 5.4   & 613 (27) & 20 & 15 (15) \\
\href{https://github.com/facebookresearch/detectron2}{facebookresearch/detectron2} \emph{(Facebook)}         & Zoo & 18.7  & 5.0     & 2605 (90) & 20 & 12 (13) \\
\href{https://github.com/open-mmlab/mmdetection}{open-mmlab/mmdetection}               & Zoo & 17.1  & 6.1   & 4348 (155) & 20 & 16 (16) \\
\href{https://github.com/huggingface/pytorch-image-models}{rwightman/pytorch-image-models\textsuperscript{\textdagger}}       & Zoo & 14.3  & 2.3   & 414 (11)  & 11 & 5 (5) \\
    \href{https://github.com/pytorch/vision}{pytorch/vision} \emph{(Facebook)}                      & Zoo & 10.2  & 5.3   & 1504 (139) & 20 & 14 (14) \\
\href{https://github.com/NVIDIA/DeepLearningExamples}{NVIDIA/DeepLearningExamples} \emph{(NVIDIA)}          & Zoo & 6.7   & 2.0     & 436 (22)  & 20 & 18 (18) \\
\href{https://github.com/qubvel/segmentation_models}{qubvel/segmentation\_models}          & Zoo & 3.5   & 0.8 & 167 (12) & 12 & 8 (8)  \\

\midrule
\textbf{YOLO} \\
\midrule
\href{https://github.com/ultralytics/yolov5}{ultralytics/yolov5}                   & Solo (Proto.) & 18.2  & 6.3   & 3795 (279) & 20 & 19 (21)\\
\href{https://github.com/ultralytics/yolov3}{ultralytics/yolov3}                   & Solo (Repl.) & 8.0     & 3.0     & 1671 (204) & 20 & 17 (21) \\
\href{https://github.com/qqwweee/keras-yolo3}{qqwweee/keras-yolo3}                  & Solo (Repl.) & 6.9   & 3.4   & 226 (16)  & 16 & 13 (13) \\
\href{https://github.com/eriklindernoren/PyTorch-YOLOv3}{eriklindernoren/PyTorch-YOLOv3}       & Solo (Repl.) & 6.3   & 2.5   & 557 (26)  & 20 & 18 (19) \\
\href{https://github.com/YunYang1994/tensorflow-yolov3}{YunYang1994/tensorflow-yolov3}        & Solo (Repl.) & 3.5   & 1.4   & 116 (3) & 3 & 3 (4)  \\

\midrule
\textbf{Mask R-CNN} \\
\midrule
\href{https://github.com/matterport/Mask_RCNN}{matterport/Mask\_RCNN}                & Solo (Proto.) & 20.9 & 10.2  & 783 (62) & 20 & 15 (16) \\
\href{https://github.com/CharlesShang/FastMaskRCNN}{CharlesShang/FastMaskRCNN}            & Solo (Repl.) & 3.1 & 1.1  & 55 (2) & 2 & 1 (1) \\
\href{https://github.com/TuSimple/mx-maskrcnn}{TuSimple/mx-maskrcnn}                 & Solo (Repl.) & 1.8 & 0.5  & 83 (7) & 7 & 6 (6) \\

\midrule
\textbf{Faster R-CNN} \\
\midrule
\href{https://github.com/ShaoqingRen/faster_rcnn}{ShaoqingRen/faster\_rcnn}             & Solo (Proto.)& 2.5   & 1.2   & 53 (5)  & 5  & 4 (4) \\
\href{https://github.com/rbgirshick/py-faster-rcnn}{rbgirshick/py-faster-rcnn}            & Solo (Repl.) & 7.5   & 4.1   & 253 (33)  & 20 & 17 (17)\\
\href{https://github.com/jwyang/faster-rcnn.pytorch}{jwyang/faster-rcnn.pytorch}           & Solo (Repl.) & 6.6   & 2.2   & 363 (34) & 20  & 17 (18) \\
\href{https://github.com/endernewton/tf-faster-rcnn}{endernewton/tf-faster-rcnn}           & Solo (Repl.) & 3.6   & 1.6   & 65 (15)  & 15  & 12 (12) \\
\href{https://github.com/chenyuntc/simple-faster-rcnn-pytorch}{chenyuntc/simple-faster-rcnn-pytorch} & Solo (Repl.) & 3.4   & 1.0     & 267 (2) & 2  & 1 (2) \\

\midrule
\textbf{Retinanet} \\
\midrule
\href{https://github.com/fizyr/keras-retinanet}{fizyr/keras-retinanet}                & Solo (Proto.)& 4.2   & 2.0     & 1227 (90) & 20 & 16 (18) \\
\href{https://github.com/yhenon/pytorch-retinanet}{yhenon/pytorch-retinanet}             & Solo (Repl.) & 1.8   & 0.3     & 78 (4) & 4  & 2 (2) \\

\midrule
\textbf{pix2pix} \\
\midrule
\href{https://github.com/junyanz/pytorch-CycleGAN-and-pix2pix}{junyanz/pytorch-CycleGAN-and-pix2pix} & Solo (Proto.) & 16.2  & 4.9   & 847 (48) & 20 & 15 (15) \\
\href{https://github.com/phillipi/pix2pix}{phillipi/pix2pix}                     & Solo (Proto.) & 8.7   & 1.6   & 119 (13) & 13 & 5 (5) \\

\midrule
\textbf{CenterNet} \\
\midrule
\href{https://github.com/xingyizhou/CenterNet}{xingyizhou/CenterNet}                 & Solo (Proto.) & 6.0     & 1.7   & 542 (17) & 17 & 12 (12) \\
\href{https://github.com/Duankaiwen/CenterNet}{Duankaiwen/CenterNet}                 & Solo (Proto.) & 1.8   & 0.6     & 68 (2) & 2  & 2 (2) \\

\midrule
\textbf{Total} & \textbf{}             & \textbf{}         & \textbf{}    & \textbf{26786}            & \textbf{\SampleNum}              & \textbf{\IssueNum (\BugNum)} \\ 
\bottomrule
\end{tabular}
    \label{tab:symbols}
\end{table}
\end{landscape}

}

\subsubsection{Issue Analysis: Instrument} \label{Method: Taxonomy}

We noticed that previous studies have proposed comprehensive taxonomies for DL defects.
\new{To provide more insights from a process view}, we reorganized the existing taxonomies by distinguishing between four DL stages: environment, data pipeline, modeling, and training~\citep{Amershi2019SE4MLCaseStudy, MLOpsWorkflow}.
We reused a taxonomy of general programming defects (\eg interface defects, non-functional defects) from~\citep{Thung2012BugsinMLSystems}.
To better characterize the defects, we revise our instrument by adding categories from other works, as shown in Table \ref{table: DefectSymptoms}, \ref{table: DefectTypes}, and \ref{table: root causes}.

\new{To promote the robustness of our analysis, we employed a saturation method.}
 We first conducted a pilot study to achieve taxonomic saturation, where we iteratively categorized issues until no new categories emerged.
\new{As a starting point, we included all the defect categories from prior DL taxonomies, recognizing that these could all theoretically occur in the reengineering process~\citep{Islam2019DLBugCharacteristics, Zhang2018TFBugs, Humbatova2020TaxonomyofRealFaultsinDLSystems, Thung2012BugsinMLSystems, Seaman2008DefectCategorization}.
}
We made several changes to the instrument as we analyzed the first 3 repositories, but these tailed off in the 4th repository and we did not need to make any changes in the 5th repository. 
After this, we \new{concluded that our taxonomy had reached saturation, adequately covering the range of DL reengineering defects encountered, and} decided to finalize the instrument.

Our pilot study indicated that DL reengineering defects are a subset of DL defects (as anticipated by \cref{fig:DL model lifecycle}) and existing taxonomies can categorize DL reengineering defects.
\new{However, we did not learn the distribution of defects among different DL reengineering activities, which raised questions about the specific challenges faced in different phases of reengineering. Therefore, we recognized the need to continue with the study to gain a deeper understanding of these distributions and the factors contributing to the occurrence of defects in DL reengineering processes.}

\new{With a stable taxonomy from the pilot study, we then applied it to the larger dataset. During this phase, we monitored the categorization process and assessed whether new categories were needed. For example, if too many issues were being classified as ``Other'', we revisited and revised the taxonomy. Throughout the analysis, we found that the taxonomy from the pilot study was largely applicable, and only minor adjustments were necessary. This approach provides a practical balance between manageable analysis effort and reliable insight, and helps to ensure the validity and comprehensiveness of our findings.
Our final taxonomy is available in \cref{sec:Reproducibility}.
}

{
\begin{table*}
\centering
\small
\caption{
    Taxonomy for defect symptoms.
    The symptoms were adapted from \citet{Islam2019DLBugCharacteristics} by distinguishing two types of \textit{Bad Performance}: \textit{Accuracy/Speed Below Expectations}, referring to the symptoms defined by \citet{Zhang2018TFBugs}. The \textit{expectations} can be different from the documentation if the code or data change.
    We also added \textit{Numerical Errors} based on \citet{Wardat2021DeepLocalize}.
    \textbf{Bold}: Changed categories.}
\label{table: DefectSymptoms}
\begin{tabular}{p{4cm} p{7cm}}
\toprule
\textbf{Defect Symptoms Category} & \textbf{Description} \\
\toprule
\textbf{Speed Below Expectations} & The code runs but the training/inference time does not match the expectation.\\
\\
\textbf{Accuracy Below Expectations} & The code runs but the evaluation results do not match the expected accuracy.\\
\\
\textbf{Numerical Error} & The results are Inf, NaN or Zero which are caused by division (\ie division by zero returns not-a-number value), logarithm (\ie logarithm of zero returns -$\infty$ that could be transformed into not-a-number); Or the results appear random for each running; Or floating point overflow.\\
\\
Crash & The system stops unexpectedly.\\
\\
Data Corruption & The data is corrupted as it flows through the model and causes unexpected outputs.\\
\\
Hang & The system ceases to respond to inputs.\\
\\
Incorrect functionality & The system behaves in an unexpected way without any runtime or compile-time error/warning.\\
\\
Memory exhaustion & The system halts due to unavailability of the memory resources. This can be caused by, either the wrong model structure or not having enough computing resources to train a particular model.\\
\\
Other & Other symptoms that do not fall into one of the above categories.\\
\bottomrule
\end{tabular}
\end{table*}
}

{
\begin{table*}
\centering
\small
\caption{
    Taxonomy for defect types. 
    We reused a taxonomy of general programming defects from \citet{Seaman2008DefectCategorization}, adding the ``Configuration'' category from \citet{Thung2012BugsinMLSystems}.
    For convenience, this table presents the combined taxonomy.
}
\label{table: DefectTypes}
\begin{tabular}{p{4cm} p{7cm}}
\toprule
\textbf{Defect Type Category} & \textbf{Description} \\
\toprule

Algorithm/method & An error in the sequence or set of steps used to solve a particular problem or computation, including mistakes in computations, incorrect implementation of algorithms, or calls to an inappropriate function for the algorithm being implemented.\\
\\
Assignment$/$Initialization & A variable or data item that is assigned a value incorrectly or is not initialized properly or where the initialization scenario is mishandled (\eg incorrect publish or subscribe, incorrect opening of file).\\
\\
Checking & Inadequate checking for potential defect conditions, or an inappropriate response is specified for defect conditions.\\
\\
Data & Defects in specifying or manipulating data items, incorrectly defined data structure, pointer or memory allocation errors, or incorrect type conversions.(\ie Array, Linked List, Stack, Queue, Trees, Graphs)\\
\\
External Interface & Defects in the user interface (including usability problems) or the interfaces with other systems. (e.g. API defects)\\
\\
Internal Interface & Defects in the interfaces between system components, including mismatched calling sequences and incorrect opening, reading, writing, or closing of files and databases.\\
\\
Logic & Incorrect logical conditions, including incorrect blocks, incorrect boundary conditions being applied, or incorrect expression.\\
\\
Timing/optimization & Errors that will cause timing or performance problems.\\
\\
Non-functional Defects & Includes non-compliance with standards, failure to meet non-functional requirements such as portability and performance constraints, and lack of clarity of the design or code to the reader.\\
\\
Configuration~\citep{Thung2012BugsinMLSystems} & Defects in non-code (\eg configuration files) that affects functionality.\\
\\
Other & Other defects that do not fall into one of the above categories.\\
\bottomrule
\end{tabular}
\end{table*}
}

{
\begin{table*}
\centering
\small
\caption{
    Taxonomy for defect root causes. We adapted the taxonomy from \citep{Humbatova2020TaxonomyofRealFaultsinDLSystems} by reorganizing the categories into four DL stages. We distinguish the categories of data preprocessing and corrupt data (data flow bug) based on \citep{Islam2019DLBugCharacteristics}.
    \textbf{Bold}: changed/new categories.
}
\label{table: root causes}
\begin{tabular}{p{0.8cm} |p{4cm} p{6cm}}
\toprule
\textbf{DL Stage} & \textbf{Root Cause Category} & \textbf{Description} \\
\toprule
\multirow{11}{*}{{\rotatebox[origin=c]{90}{\normalsize \textbf{Data pipeline}}}}
&
\textbf{Data preprocessing} & If an input to the deep learning software is not properly formatted, cleaned, well before supplying it to the deep learning model.\\
\\
& \textbf{Corrupt data (data flow bug)} & Due to the type or shape mismatch of input data after it has been fed to the DL model.\\
\\
& Training data quality & Due to the complexity of the data and the need for manual effort to ensure a high quality of training data (\eg to label and clean the data, to remove the outliers).\\
\midrule
\multirow{10}{*}{{\rotatebox[origin=c]{90}{\normalsize \textbf{Modeling}}}}
&
Activation function & Incorrectly selecting the activation function of neurons.\\
\\
& Layer properties & Some layer's incorrect inner properties (\eg input/output shape, input sample size, number of neurons in it).\\
\\
& Missing/Redundant/Wrong layer & Adding, removing or changing the type of a specific layer was needed to remedy the low accuracy of a network.\\
\midrule
\multirow{20}{*}{{\rotatebox[origin=c]{90}{\normalsize \textbf{Training}}}}
&
Optimizer & The selection of an unsuitable optimization function for model training.\\
\\
& 
Loss function & Wrong selection and calculation of the loss function.\\
\\
& 
Evaluation & Problems caused by testing and validation (\eg bad choice of performance metrics)\\
\\
&
Hyper-parameters & Incorrectly tuning the hyperparameters (\eg learning rate, batch size, number of epochs) of a DL model.\\
\\
& 
\textbf{Training configuration} & Wrong training scripts.\\
\\
& 
Other training process & Other faults in the training process which do not fall into one of the above categories (\eg wrong management of memory resources, wrong post-processing of the output)\\
\midrule
\multirow{12}{*}{{\rotatebox[origin=c]{90}{\normalsize \textbf{Environment}}}}
&
API defect & Caused by APIs, this includes API mismatch, API misuse, API change, \textit{etc.}\\
\\
&
GPU Usage bug & Wrong usage of GPU devices while working with DL (\eg wrong reference to GPU device, failed parallelism, incorrect state sharing between subprocesses, faulty transfer of data to a GPU device).\\
\\
& 
\textbf{Wrong environment configuration} & Incorrect setting of other configurations (\eg wrong operating systems, internal interface defects).\\
\midrule
\multirow{3}{*}{{\rotatebox[origin=c]{90}{\normalsize \textbf{Other}}}}
&
\textbf{Insufficient/Incorrect documentation} & Engineers misunderstood the documentation or they cannot find correct or sufficient instructions.\\

\bottomrule
\end{tabular}
\end{table*}
}
  
\subsubsection{Issue Analysis: Process}
  
Our classification and labeling process build on prior studies of DL defects~\citep{Islam2019DLBugCharacteristics, Humbatova2020TaxonomyofRealFaultsinDLSystems}.
A graduate researcher first developed an instrument and led a pilot study on 69 defects from 5 repositories.
To calibrate the analysis instrument, a team of 6 undergraduate researchers worked in pairs, and independently analyzed the same set of defects.
One graduate student supervised the team, helped to address uncertainties, and refined the instrument based on the discussions.
In our taxonomy, we consider the categories with less than five counted defects as \textit{low frequency categories} to better present our results.

We analyzed the data from the pilot study twice, once using the original instrument and once using the improved instrument. After modifying the instrument to address ambiguities, we had two of the researchers re-annotated the data using the improved definitions and clarifications. Most of our annotators were undergraduate students who were unable to continue working on the project after the summer, so the rest half of our study was done by the most experienced rater, who sought a second opinion on a few uncertainties.

We used the Cohen’s Kappa measure for inter-rater agreement~\citep{cohen1960coefficient}. 
The kappa statistics can represent the inter-rater reliability~\citep{mchugh2012interrater}.
 Using the original instrument, the Cohen’s Kappa Value was 0.46 (``moderate'').
 The Kappa value for the modified instrument was 0.79 (``substantial''). 
 We also measured 30\% of the rest study and the Cohen’s Kappa value was 0.70 (``substantial'').\footnote{\new{Due to a clerical error, a small fraction of issues ($\sim$30) were analyzed by only one researcher. We noticed this while reviewing our data. That researcher was our most experienced analyst and had high certainty on those defects, so we felt no second review was warranted.}}

The data and scripts used to calculate the Cohen's Kappa value are available (\cref{sec:Reproducibility}).

\subsection{RQ5: Interview Study on Computer Vision Reengineering} \label{Method: Case Study}
\new{To enrich our understanding of DL reengineering challenges and practices, we triangulated the failure analysis with a qualitative data source: interviews with engineers.
Our population of interest was engineers who are involved in reengineering activities in the context of computer vision.
We followed the standard recruiting methods of contacting open-source contributors and advertising on social media platforms.
We also complemented that data with interviews of a CV reengineering team composed of Purdue University undergraduate students.}


\subsubsection{External Subjects: Open-source Contributors and Social Media Recruits} \label{Method: Interviews}


We recruited participants from open-source model contributors
\new{who had contributed to the projects we studied in \cref{Method: Bugs}.
Our recruitment process started by sending emails to the primary contributors of each repository in \cref{table:repo info}.}
\new{Subsequently, we expand our search on popular platforms like Hacker News\footnote{https://news.ycombinator.com/} and Reddit\footnote{https://www.reddit.com/}.
\new{Out of the 25 open-source contributors we contacted, we received responses from 2 of them, giving us a response rate of 8\%.
We also received 7 responses from engineers via social media platforms.}
}

\subsubsection{Internal Subjects: Purdue's Computer Vision Reengineering Team} \label{Method: CVReengineeringTeam}

Motivated by the method of measuring software experiences in the software engineering laboratory described by~\citet{Valett1989SWMeasurementExperiencesinSELab},
our lab organized a CV reengineering team, focused on the replication of CV research prototypes.
The goal of this team is to provide high-quality implementations of state-of-the-art DL models. Most team members are third- and fourth-year undergraduate students. 
Their work is supervised and approved by Google engineers over weekly sync-ups.

The team’s industry sponsor (Google) uses our replication of \ReproducedModelNum models in production in their ML applications and publishes them open-source in one of the most popular zoo repository we studied in \cref{Method: Bugs}, TensorFlow Model Garden.
Each of the team’s projects had 1-2 student leaders. The team leaders were fourth-year undergraduates, sometimes working for pay and sometimes for their senior thesis. Team leaders typically worked for 15-20 hours per week over their 2 years on the team.
All team leaders contributed to at least two reengineering projects. All team members received team-specific training via our team’s 6-week onboarding course on DL reengineering.


The completed models required \TotalModelsLOC lines of code, measured using the cloc tool~\citep{ClocTool}. 
The estimated cost of the reengineering team's work was $\sim$\$105K (\$15K/model): \$40K on wages and \$65K on computation (\$5K for VM and storage, \$60K for hardware accelerator rental [e.g., GPU, TPU]).

\subsubsection{Data Collection}

To design our interview, we followed the guideline of \textit{framework analysis} which is flexible during the analysis process~\citep{Srivastava2008FrameworkAnalysis}.
A typical framework analysis include five steps: data familiarization, framework identification, indexing, charting, and mapping~\citep{ritchie2002qualitative}.
Our interview follows a three-step process modeled on the \textit{framework analysis} methodology:

\myparagraph{Data Familiarization and Framework Identification}
Based on the reengineering challenges and practices we identified from our literature review~\cref{sec:Background} and the results of open-source failure analysis~\cref{Method: Bugs}.
We created a \textit{thematic framework}, including the challenges and practices of bug identification and testing/debugging, and created a draft reengineering workflow as one theme of the reengineering practices.

\myparagraph{Interview Design:}
We designed a semi-structured interview protocol with questions that follow our identified themes of DL reengineering. 
We conducted two pilot interviews and then revised our framework and interview protocol by clarifying our questions.
The final interview protocol includes four main parts: demographic questions, reengineering process workflow, reengineering challenges, and reengineering practices.
\cref{table: InterviewProtocol} indicates the three main themes of our interview protocol.
%
These interviews were technical and focused, more closely resembling a structured interview than a semi-structured one.
During the interview, we provide relevant themes and show the draft of the reengineering workflow in a slide deck.\footnote{The final version of the reengineering workflow is shown in~\cref{fig:process workflow}.} 

\myparagraph{Filtering Interviews for Relevance:}
\new{We applied three inclusion criteria for the interview participants recruited from the social media platforms: (1) the participant should have industry experience, (2) hold at least a bachelor's degree, and (3) have experience in CV and/or DL reengineering. Based on these criteria, four of the seven subjects from the social media platform group met the requirements. Additionally, the subjects recruited from open-source contributors and our CV reengineering team already had relevant experience.}

\new{Overall, our qualitative data include 6 external interviews (2 open-source contributors, 4 industry practitioners) and \LeaderNum internal interviews (leaders of CV reengineering team after the team had been operating for 18 months).
\cref{tab: demographic} summarizes the demographic data of the 6 external participants.}

{
\begin{table}[h]
\centering
\caption{
  \new{
  Participant demographics.
  P1 and P2 are open-source contributors, P3-P6 are practitioners recruited from social media platforms. Most participants
    self-reported
    here} 
    intermediate or expert skills in deep learning (DL) and software engineering (SE).
  All participants have experience in reengineering Computer Vision (CV) models.
  Some of them have also applied deep learning to Natural Language Processing (NLP), Reinforcement Learning (RL), and Audio.
  } 
\label{tab: demographic}
\begin{tabular}{
    cccl
}
\toprule
    \textbf{ID} &
    \textbf{SE skill} &
    \textbf{DL skill} &
    \textbf{Domain}
    \\
\midrule
P1                    
    & Expert
    & Intermediate
    & CV
\\

P2           
    & Expert
    & Expert
    & CV, NLP
\\
\midrule

P3         
    & Beginner
    & Intermediate
    & Audio, CV, NLP
\\
P4               
    & Expert
    & Expert
    & CV, RL, NLP
\\
P5       
    & Expert
    & Intermediate
    & CV, NLP, RL
\\
P6                      
    & Expert
    & Intermediate
    & CV, RL
\\
\bottomrule
\end{tabular}
\end{table}
}

\subsubsection{Data Analysis}
The interview recordings were transcribed by a third-party service\footnote{\url{https://www.rev.com/}}.
One of the transcripts was in Chinese and so a researcher who is a native speaker listened to the recording.
Themes were extracted and mapped by one researcher, the same person who conducted the interviews. 
We compared those parts of each transcript side-by-side, and noted challenges or practices that were discussed by multiple leaders and extracted illustrative quotes. 
\new{After completing the interviews, we implemented a process of member checking with 6 team leaders to validate our findings~\citep{birt2016MemberChecking}. They agreed with our analysis and interpretations.}
By analyzing the interview transcripts,
we were able to understand the larger picture and summarize common challenges.

{
\begin{table*}
\centering
\small
\caption{
    Interview protocol addressing RQ5.
    We answer RQ5 by combining results from four kinds of questions: demographic questions, process workflow, challenges, and effective practices.
    Details of demographic questions can be found in our artifact (\cref{sec:Reproducibility}).
    The interview is semi-structured. The questions in italics are questions that all participants were asked. The other questions are examples of follow-up questions.}
\label{table: InterviewProtocol}
\begin{tabular}{p{2cm} |p{9cm}}
\toprule
\textbf{Themes} & \textbf{Questions}\\
\toprule
\multirow{11}{*}{\normalsize \textbf{Process}}
&
\textbf{Q1:} \textit{Can you talk me through the process that your team follows to re-engineer a machine learning model from research paper/existing implementation/another engineer’s project?}\\
\\
& \textbf{Q2:} \textit{Please take a look at our \textit{draft workflow}. Can you tell me if you think this is an accurate process workflow?}\\
\\
& \textbf{Q3:} \textit{Would you like to add any back-edges in this diagram?}\\
\\
& \textbf{Q4:} How does your team update new iterations of your model if it doesn’t work for the first time?\\
\midrule

\multirow{15}{*}{\normalsize  \textbf{Challenges}}
&
\textbf{Q5:} \textit{Which parts do you think are challenging when re-engineering a model}\\
\\
& \textbf{Q6:} Can you tell me about an error you found in TRAINING/MODELING/DATA PIPELINE?\\
\\
& \textbf{Q7:} Can you describe any challenges you met when implementing TRAINING/MODELING/DATA PIPELINE?\\
\\
& \textbf{Q8:} How do you address these challenges?\\
\\
& \textbf{Q9:} Have you met any challenges when integrating all components?\\
\\
& \textbf{Q10:} Can you think about 1-2 changes to the reengineering process that would make this process easier for you?\\
\midrule

\multirow{16}{*}{\normalsize  \textbf{Practices}}
&
\textbf{Q11:} \textit{How does your team work together to make the process more effective?}\\
\\
& \textbf{Q12:} How do you decide an existing implementation is trustworthy?\\
\\
& \textbf{Q13:} What do you find is helpful/problematic in a DL research paper?\\
\\
& \textbf{Q14:} What do you find is helpful/problematic in the documentation of DL models?\\
\\
& \textbf{Q15:} Are there existing tools (or other technologies) you found valuable/problematic for re-engineering?\\
\\
& \textbf{Q16:} How do you determine the acceptable trade-off between the performance of the model (accuracy/speed) and the cost of your team (time/money, etc.)?\\
\bottomrule
\end{tabular}
\end{table*}
}

\section{Results and Analysis} \label{sec:Result}

\subsection{\textbf{RQ1: How do defects manifest in Deep Learning reengineering?}} \label{RQ1 results}

\begin{figure}
    \centering
    \includegraphics[width=0.85\linewidth]{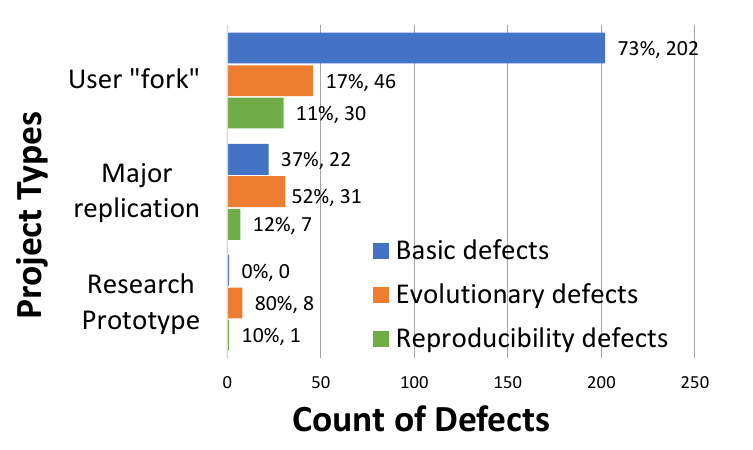}
    \caption{
        Manifestation in different project types.
        Most defects (80\%, 278/348) are identified in user ``forks'', among which 73\% (202/278) are basic defects.
        In this and subsequent figures, the indicated percentages are calculated relative to the corresponding type.
    }
    \label{fig:Manifestation - Project types}
\end{figure}

\begin{figure}[h]
    \centering
    \includegraphics[width=0.85\linewidth]{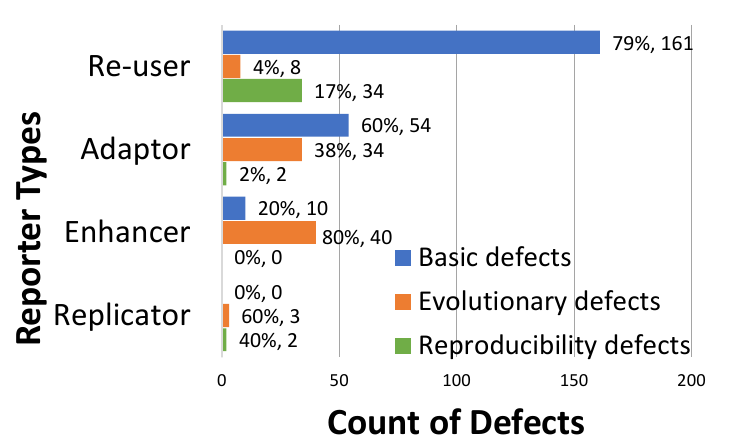}
    \caption{
        Manifestation vs. Reporter types.
        58\% (203/348) of the defects are reported by re-users.
        Less than 1\% are reported by replicators.
        Almost all reproducibility defects (89\%, 34/38) are reported by re-users.
    }
    \label{fig:Manifestation - Reporter types}
\end{figure}

\begin{figure}
    \centering
     \includegraphics[width=0.85\linewidth]{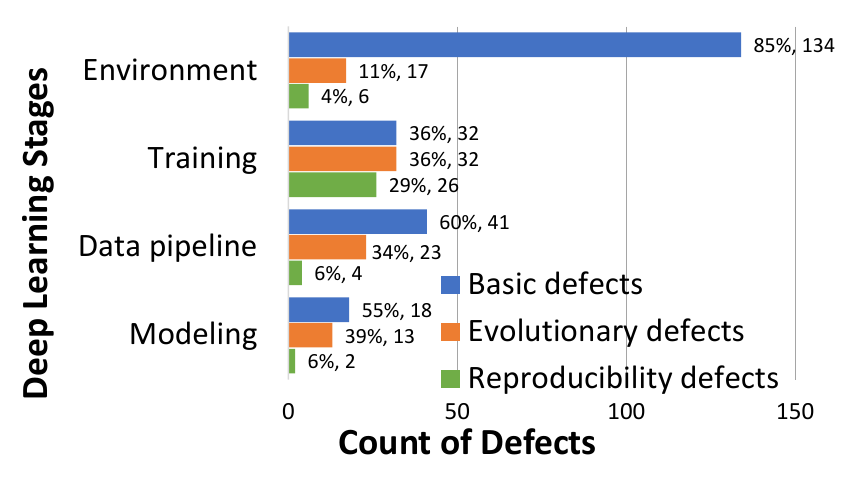}
     \caption{
         Manifestation by DL stage.
         Most \emph{reproducibility} (68\%, 26/38) and \emph{evolutionary} (38\%, 32/85) defects occur in Training stage. Data pipeline also has many \textit{evolutionary} (34\%) defects.}
     \label{fig:Manifestation - DL Stages}
\end{figure}

\begin{tcolorbox} [width=\linewidth, colback=yellow!30!white, top=1pt, bottom=1pt, left=2pt, right=2pt]
\textbf{Finding 1}: 
\textit{Project types:} User ``forks'' has the most basic (73\%) and reproducibility (79\%) defects (\cref{fig:Manifestation - Project types}).
\textit{Reporter types:} Most defects (58\%) are reported by re-users. Defects reported by replicators are rarely identified ($<$1\%). Almost all reproducibility defects are reported by re-users (\cref{fig:Manifestation - Reporter types}).
\textit{DL stages:} 91\% defects are reported during environment, training, and data pipeline. 
68\% reproducibility defects occur in the training stage (\cref{fig:Manifestation - DL Stages}).
\end{tcolorbox}

To understand the characteristics of DL model reengineering, we analyzed the distribution of defect manifestations in terms of reporter types and DL stages.

\myparagraph{Project types:}
Most defects are basic defects (73\%) located in user ``forks''.
Most reproducibility defects (79\%, 30/38) are identified in user ``forks''.
%
%
%
%
%
%
%
%
Most of the basic defects in user ``forks'' are due to the misunderstanding of the implementation, miscommunication, or insufficient documentation of the model(s) they are using.
In the discussions, we saw the owners of the repositories often tell the re-users to read the documentation, while the re-users would remark that the documentation is confusing.
This observation supports the suggestions from Gundersen~\etal on documentation improvement~\citep{Gundersen2018ReproducibilityinAI} and indicates there is still a need for detailed documentation and tutorials, especially for the reengineering process~\citep{FacebookAIReproducibility}.

\myparagraph{Reporter types:}
\cref{fig:Manifestation - Reporter types} indicates that almost all reproducibility defects are reported by re-users.
This finding somewhat follows from our reporter type definitions --- an adaptor uses a different dataset, and an enhancer adds new features to the model, so neither is likely to report a reproducibility defects.
However, the absence of reproducibility defects from replicators was more surprising.
Perhaps replicators are more experienced, and more likely to fix the problem themselves than to open an issue.
Alternatively, perhaps there are simply far more re-users in this community.


\myparagraph{DL stages:}
As indicated in \cref{fig:Manifestation - DL Stages}, 91\% (315/\BugNum) of defects are reported during environment, training, and data pipeline, 
26\% (90/\BugNum) of defects are reported in the training stage and 20\% (68/\BugNum) in the data pipeline stage.
However, only 9\% (33/\BugNum) of defects are reported in the modeling stage.

The majority (68\%) of reproducibility defects we found are located in the training stage. 
Differently, 60\% of defects in the data pipeline and 55\% in the modeling stages are basic defects, which also means that they are easier to identify. This kind of manifestation can be easily found from either the error messages or visualization of the input data.
For reproducibility efforts, it is less likely for re-users and replicators to encounter reproducibility defects because they are using the same data and model architectures between one another.


The data indicates that training stage is the most challenging, and data pipeline stage is the second in the reengineering process.
Most of the training defects do not result in crashes, but lead to the mismatches between the reimplementation and specification/documented performances.
The possible reason is that replicators can refer to the existing replications/prototypes, reuse the data pipeline, or use the same architecture (\eg backbones) when dealing with similar tasks.

\subsection{\textbf{RQ2: Frequent types of Deep Learning reengineering defects?}} \label{RQ2 results} 
\begin{figure}
    \centering
    \includegraphics[width=0.9\linewidth]{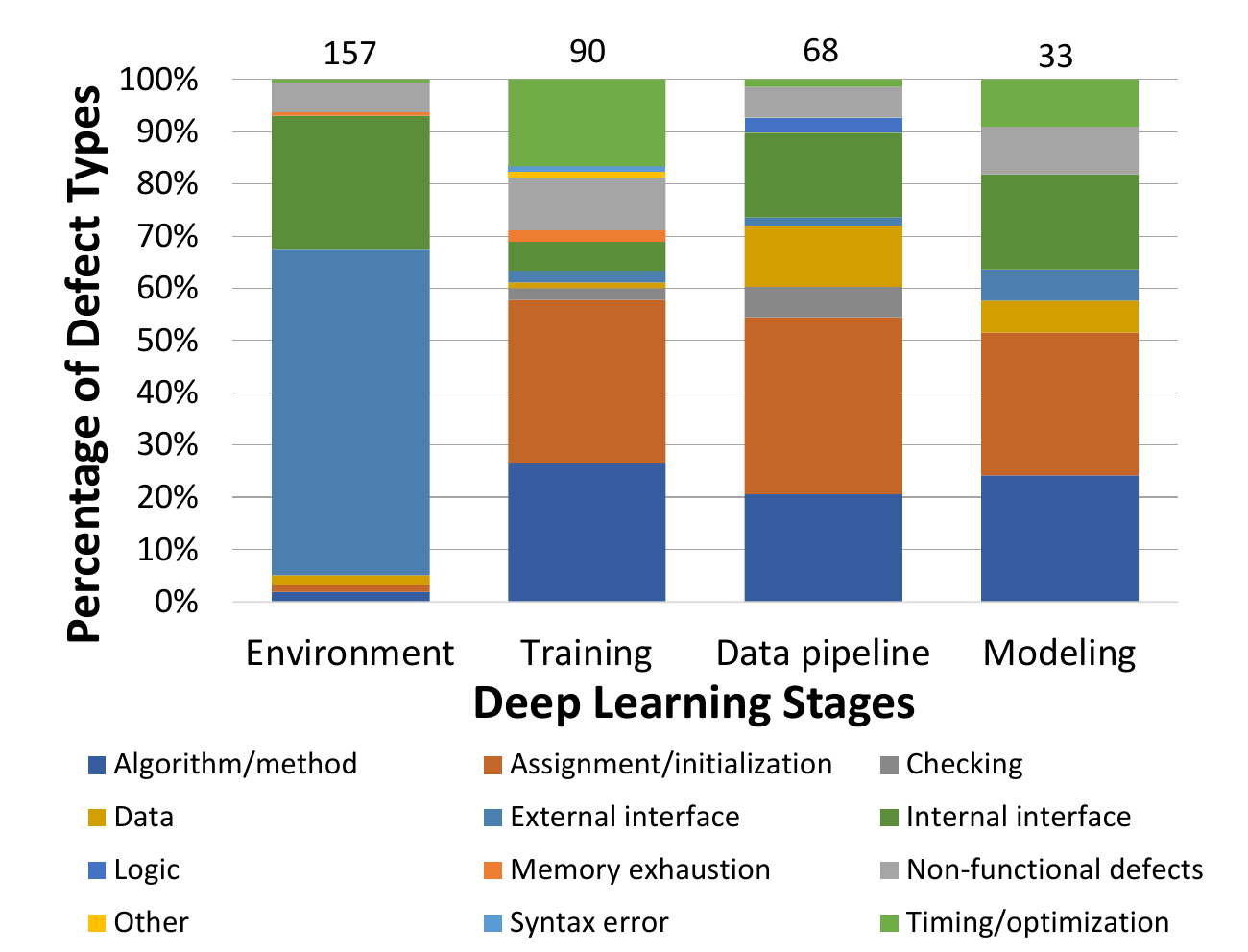}
    \caption{
        Defect types by DL stage.
        Most Environment defects are \emph{interface} defects.
        The Data Pipeline and Modeling stages have similar distributions oriented towards \emph{assignment/initialization} defects. 
        Training defects are diverse.
    }
    \label{fig:Bug types - general (percentage)}
\end{figure}

\begin{tcolorbox} [width=\linewidth, colback=yellow!30!white, top=1pt, bottom=1pt, left=2pt, right=2pt]
\textbf{Finding 2}:  Most Environment defects are \emph{interface} defects (88\%).
        The Data Pipeline and Modeling stages have similar distributions oriented towards \emph{assignment/initialization} defects. 
        Training defects are diverse. (\cref{fig:Bug types - general (percentage)})
\end{tcolorbox}

Here we consider the defect types by DL stage (\cref{fig:Bug types - general (percentage)}).
88\% of the defects in the environment configuration are interface defects. These defects can be divided into external interface defects (62\%, 98/157), \ie the user interface or the interfaces with other systems; and internal interface defects (25\%, 40/157), \ie the interface between different system components, including files and datasets~\citep{Seaman2008DefectCategorization}. 
For the external environment, engineers have to setup the DL APIs and hardware correctly before running the model.
However, 
there are often inconsistencies between different DL frameworks and hardware, leading to defects.
For the internal environment, engineers need to set up the modules and configure the internal paths,
but the documentation or tutorials of the model are sometimes incomprehensible and result in defects. 

During training there are relatively more algorithm/method and timing/ optimization defects, compared to other stages.
Engineers appear prone to make mistakes in algorithms and methods when adapting the model to their own datasets, or to fail to optimize the model in the training stage.


We observe that assignment/initialization defects account for 34\% (23/68) in data pipeline stage and 27\% (9/33) in the modeling stage. 
Internal interface defects also account for 16\% (11/68) of the defects in the data pipeline.
To fix assignment/initialization defects and internal interface problems, engineers only need to change the values or set up the modules and paths correctly.
These are relatively simple defects and simple tool support could help.




\subsection{\textbf{RQ3: What are the symptoms of Deep Learning reengineering defects?}} \label{RQ3 results}

\begin{figure}
    \centering
    \includegraphics[width=0.95\linewidth]{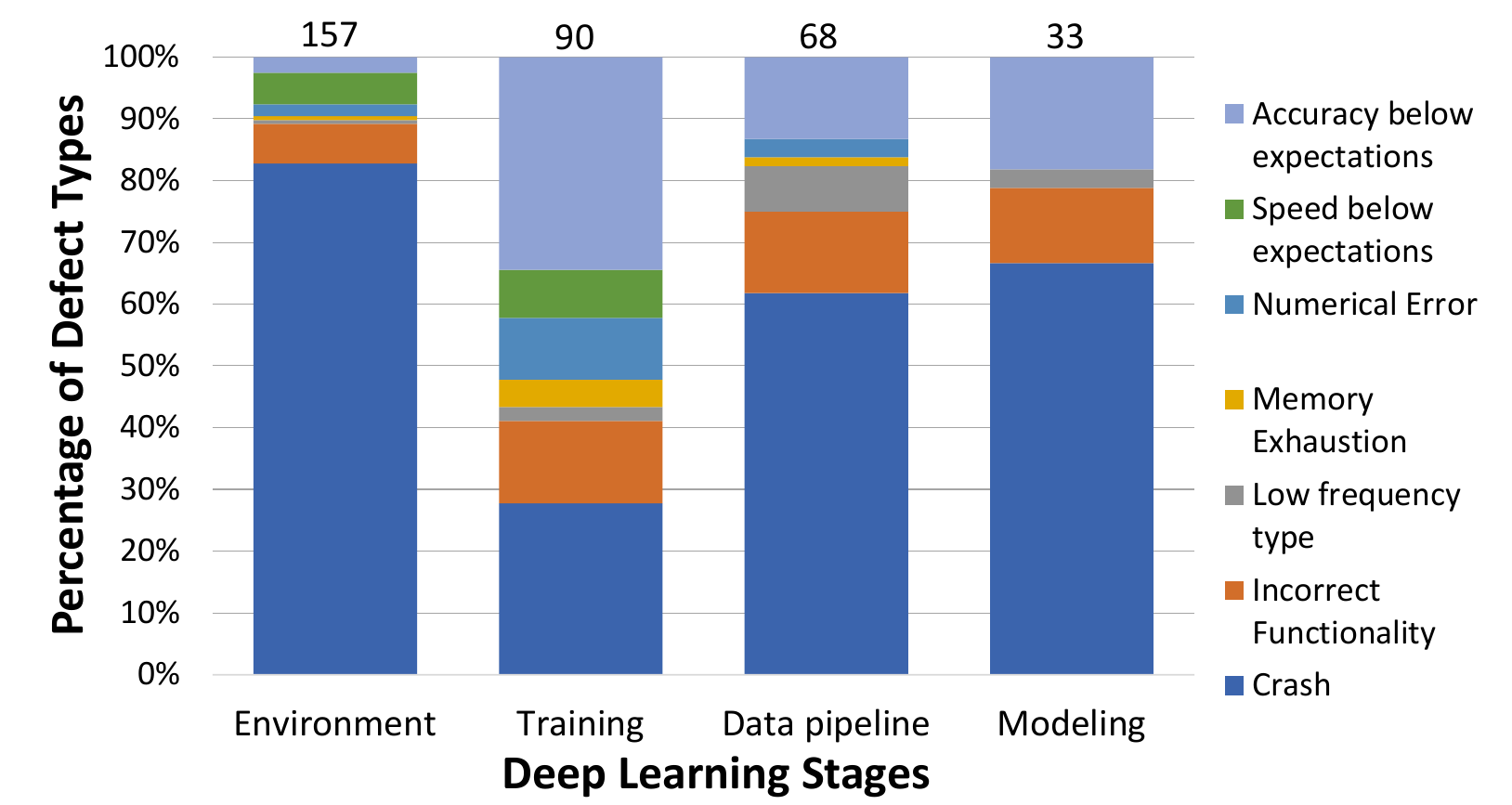}
    \caption{
        Defect symptomss~\citep{Islam2019DLBugCharacteristics, Zhang2018TFBugs}
        by DL stage.
        Across most stages, the most frequent symptoms is \emph{crash} (62-82\%).
        In Training, \emph{Accuracy below expectations} accounts for the largest proportion (34\%) of defects.
    }
    \label{fig: Symptomss}
\end{figure}

\begin{tcolorbox} [width=\linewidth, colback=yellow!30!white, top=1pt, bottom=1pt, left=2pt, right=2pt]
\textbf{Finding 3}:  
Across most stages except training, the most frequent symptom is \emph{crash} (62-83\%).
Training is the most challenging stage where \emph{Accuracy below expectations} accounts for the largest proportion (34\%) of defects. (\cref{fig: Symptomss})

\end{tcolorbox}

\cref{fig: Symptomss} shows the distribution of defect symptomss in different DL stages.
Our data shows that \textit{Crash} is a common. 83\% (130/157) defects result in crashes in environment, data pipeline, and modeling.
Most crashes happen due to incorrect environment configuration, \eg internal interface, APIs, and hardware.

In contrast, 72\% (65/90) of defects in the training stage do not result in crashes (34\% lead to \textit{Accuracy below expectations}).
The training defects are more likely to result in lower accuracy and incorrect functionality which are harder to identify. 
Locating the defect could be more time-consuming because the fixers have to train the model for many iterations and compare the accuracy or other metrics to see whether the training works properly.
Based on this, we believe that training is the most challenging stage.

Similar to the distribution shown in \cref{fig:Manifestation - DL Stages}, \cref{fig: Symptomss} indicates that reproducibility and evolutionary defects (\eg lower accuracy and incorrect functionality) can be located in any of the four stages.
For evolutionary defects, since the code or data has been changed based on the research prototypes or replications, the defects are more likely to be identified in the changed parts 
which can be easily found.
Nevertheless, for reproducibility defects, especially for the reimplementation built from scratch, it is hard to debug.



\subsection{\textbf{RQ4: What are root causes of Deep Learning reengineering defects?}} \label{RQ4 results}

\begin{tcolorbox} [width=\linewidth, colback=yellow!30!white, top=1pt, bottom=1pt, left=2pt, right=2pt]
\textbf{Finding 4}:  
Most Environment defects are caused by API defects (46\%, 73/157).
In the Data Pipeline, \emph{data preprocessing} defects
predominate (38\%, 26/68). 
Most Modeling defects are due to \emph{model/weight operations} (67\%, 22/33). 
Training defects have diverse causes. (\cref{fig:Root causes})

\end{tcolorbox}

\cref{fig:Root causes} shows the distribution of root causes in different DL stages. 
To answer RQ4, we analyzed the distributions of defects and present our findings by DL stage.

{
\small
\begin{figure}
    \centering
    \includegraphics[width=0.9\linewidth]{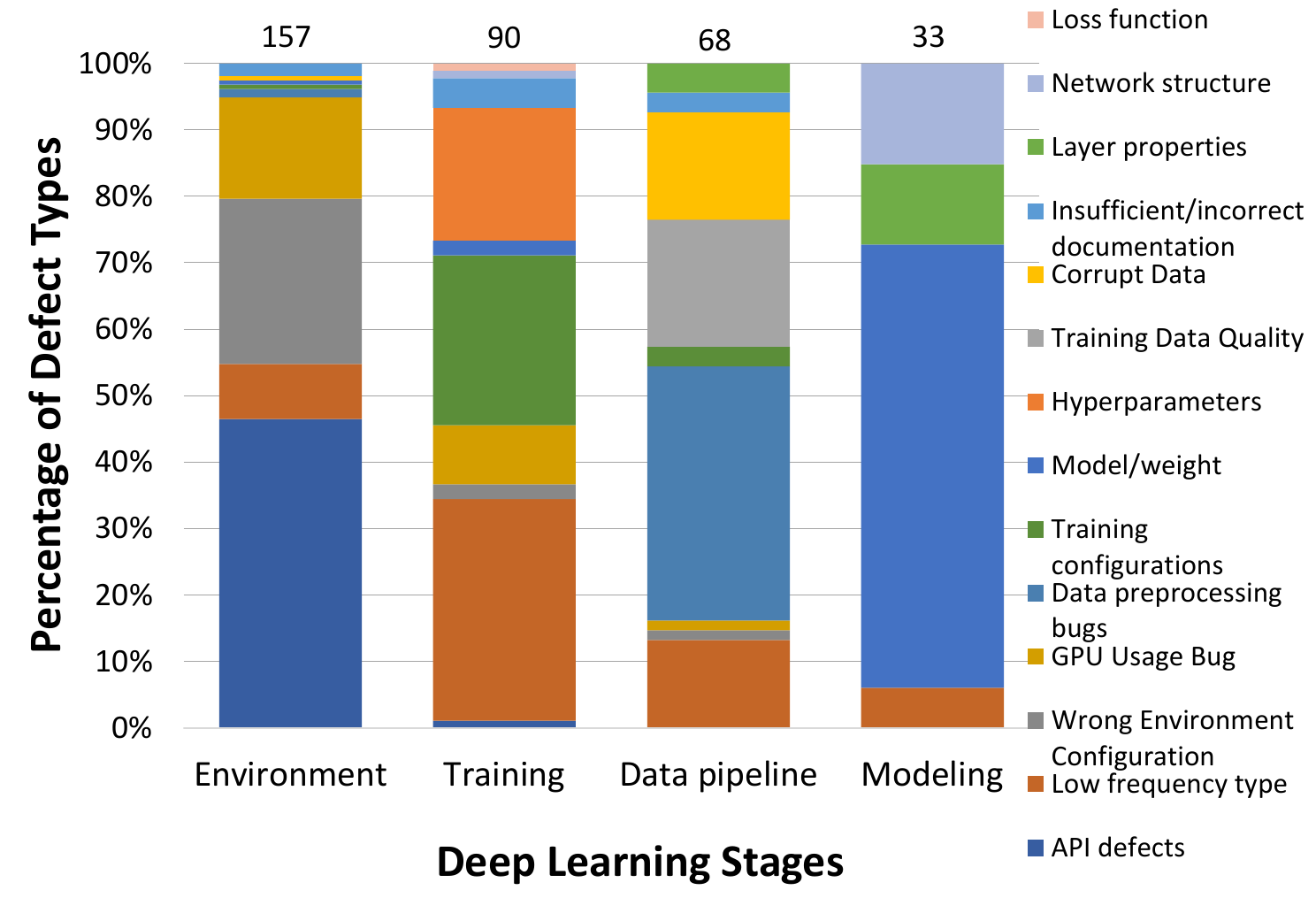}
    \caption{Root causes
    by DL stage.
    In the Data Pipeline, \emph{data preprocessing}, \emph{corrupt data} and \emph{data quality} predominate. 
    Most Modeling defects are due to \emph{model/weight operations} and \emph{network structure}. 
    Training defects have diverse causes. 
    }
    \label{fig:Root causes}
\end{figure}
}

\myparagraph{Environment}
\cref{fig:Root causes} shows that most of the environment defects are caused by API defects (46\%, 73/157), Wrong environment configuration (25\%, 39/157), and GPU usage defects (15\%, 24/157). Many reusers reported defects due to API changes and mismatches. Insufficient documentation can easily lead to misunderstandings and confusion.
The portability of models is another problem, especially for the GPU configuration.

\myparagraph{Data pipeline}
\cref{fig:Root causes} indicates three main root causes in the data pipeline: data preprocessing (38\%, 26/68), training data quality (19\%, 13/68), and corrupt data (16\%, 11/68).
Engineers are likely to have troubles in data processing and data quality. These defect types are especially frequent for adaptors who are using their own datasets. Datasets vary in format and quality compared to the benchmark CV datasets (\eg ImageNet~\citep{ImageNet}, COCO~\citep{COCO}). Therefore, before adaptors feed the data into the model, they have to first address the dataset format, shape, annotations, and labels. Moreover, customized datasets are less likely to have enough data to ensure a comparable level of accuracy, so it is necessary to use data augmentation in their data pipeline. However, as we observed, some engineers did not realize the significance of data augmentation and data quality in their reengineering tasks which lead to the lower accuracy.

\myparagraph{Modeling}
The main root causes in the modeling stage are \emph{Model$/$weight} (67\%, 22/33), layer properties (12\%, 4/33), and network structure (9\%, 5/33).
This cause represents the incorrect initialization, loading, or saving of models and weights.
We observed that some reengineering work moves from one DL framework to another. Though some tools exist for the interoperability of models between DL frameworks~\citep{mmdnn}, we did not see them in use, and engineers are still having troubles in the modeling. 

\myparagraph{Training}
There are multiple defect types contributing to training defects.
The top two are training configurations and hyper-parameters.
Training configurations include different training algorithms (\eg non-maximum suppression, anchor processing) and some specific parameters which are used to configure the training, but different from the hyper parameters. Putting together with hyper parameter tuning, these two causes are the pitfalls in the training stage. When engineers have different environment configurations or adapt the model to their own datasets, it is necessary to modify the training algorithms and tune the hyper parameters in a proper way so that the training results match their expectations.

\subsection{\textbf{RQ5: Challenges and practices in Deep Learning reengineering?}} \label{RQ5 results}
\begin{tcolorbox} [width=\linewidth, colback=yellow!30!white, top=1pt, bottom=1pt, left=2pt, right=2pt]

\textbf{Finding 5}:  
\textit{Challenges}:
The \new{four} main challenges in the DL reengineering process are model operationalization, performance debugging, portability of DL operations, \new{and customized data pipeline}.
\textit{Practices}: 
(Interface): Interface can help unify and automate model testing, especially testing of component integration.
(Testing): 
Validation was a common emphasis. Team employed complementary testing techniques (\ie unit, differential, and visual testing).
(Debugging): 
Comparing evaluation metrics after just 25\% of training is a shortcut.

\end{tcolorbox}

Our failure analysis shed some light on CV reengineering challenges, but little on the guidelines for ML reengineering requested in prior works~\citep{Rahman2019MLSEinPractice, Devanbu2020SE4DLVision}.
In this section, we describe those aspects based on the experiences of our CV reengineering team.
First, we describe three main reengineering challenges we found in our reengineering team.
Then, in~\cref{sec:ReengineeringPractices} we describe three practices we identified from interview study to mitigate those challenges, including interface, testing, and debugging.



\subsubsection{Reengineering Challenges} \label{sec:reengWorkflowAndChallenges}

We interviewed 2 open-source practitioners and \LeaderNum leaders of the team (\cref{Method: Case Study}).
\new{Our interview study identified 8 reengineering challenges within the process \new{(\cref{table:Challenges})}, and we specifically analyzed four of them which were mentioned by at least three participants.}


{
\small
\renewcommand{\arraystretch}{0.8}\
\begin{landscape}
\begin{table*}[ht]
\centering
\caption{
  \new{Challenges during deep learning reengineering process.
  The third column shows how many participants (of 12) mentioned the challenge.
  Relevant reengineering activities identified in interview data are also incorporated.
  }
  }
\label{table:Challenges}
\begin{tabular}{p{0.2\linewidth}p{0.3\linewidth}cp{0.26\linewidth}}
\toprule
\textbf{Challenge} & \textbf{Description} &\textbf{\# Participants} &\textbf{Reeng. Activities} \\
\midrule

\textbf{1.} Model Operationalization
& Unclear implementations, configurations, or scripts of the original model.
& 8
& Reuser, Replicator, Adaptor, Enhancer

\\
\\
\textbf{2.} Performance Debugging          
& Debugging and achieving expected end-to-end performance metrics.
& 8
& Reuser, Replicator, Adaptor, Enhancer
\\    
\\
\textbf{3.} Portability of DL Operations
& Discrepancies in function naming or signatures across frameworks, inconsistencies in behaviors of the `same' operations, limitations on specific hardware.
& 6
& Reuser, Replicator, Enhancer
\\
\\

\textbf{4.} Customized data pipeline
& Data processing, quality assurance, and pipeline implementation.
& 4
& Replicator, Adaptor, Enhancer
\\
\\
\textbf{5. }Team collaboration
& Inconsistent implementations among multiple team members.
& 2
& Replicator, Enhancer
\\
\\
\textbf{6. }Code Readability
& Obfuscated code.
& 1
& Enhancer
\\
\\
\textbf{7. }Evaluation Metrics
& Identifying the best metric to evaluate the model.
& 1
& Adaptor
\\
\\
\textbf{8. }Data availability.
& Unavailable training data.
& 1
& Adaptor
\\
\bottomrule
\end{tabular}
\end{table*}
\end{landscape}
}





\myparagraph{Challenge 1: Model Operationalization}

\ParticipantQuote{Practitioner 1}{The only way to validate that your model is correct is to...train it, but the pitfall is that many people only write
the code for the model without any training code...That,
in my opinion, is the biggest pitfall.}
\ParticipantQuote{\new{Practitioner 9}}{\new{I would say that understanding the previous work is often the most difficult part
...translating that information and transmitting that information in an efficient way can be very hard.}}
\ParticipantQuote{Leader 2}{So one challenge...digging through their code and figuring out what is being used and what isn't being used.}
\ParticipantQuote{Leader 3}{And in the paper, they...give a pseudo code for just iterating through all..., which they mentioned in the paper that is really inefficient. And they do mention that they have an efficient version, but they never explain how they do it [in the paper].}

\ParticipantQuote{Leader 5}{A lot of the work is figuring out what they did from just the paper and the implementation, which is not exactly clear what they're doing.}

There may be multiple implementations or configurations of the target model.
The interview participants found it hard to
distinguish between them, 
to identify the reported algorithm(s), 
and to evaluate their trustworthiness.
First, some research prototypes are not officially published. To correctly replicate the model, it is hard to identify which implementation they could refer to.
Moreover, the model may be different from the one in the paper. Leader 3 reported that the research prototype used a different but more efficient algorithm without any explanation.
Even for the original prototypes, the leaders reported that the prototype's model or training configuration may differ from the documentation (\eg the research paper).
These aspects made it hard for the reengineering team to understand which concepts to implement and which configuration to follow.

\myparagraph{Challenge 2: Performance Debugging}

\ParticipantQuote{Practitioner 1}{If your data format is complicated then the code around it would also be complicated and it would eventually lead to some
hard to detect bugs.}

\ParticipantQuote{\new{Practitioner 3}}{\new{[Performance bugs] are silent bugs. You will never know what happened until it goes to deployment.}}

\ParticipantQuote{Leader 1}{You can take advantage of...differential testing but then someone shows out a probabilistic process to this...it becomes really difficult because the testing is dependent upon a random process, a random generation of information.}
\ParticipantQuote{Leader 4}{Sometimes it gets difficult when the existing implementation... [doesn't] have the implementation for the baseline. So we have to figure out by ourselves. Paper also doesn't have that much information about the baseline...if you're not able to even get the baseline, then going to the next part is hard...}

Another main challenge we observed is matching the performance metrics of existing implementations, especially the original research prototype. 
These metrics include inference behavior (\eg accuracy, memory consumption) and training behavior (\eg time to train).
Multiple interview participants stated that performance debugging is difficult.
Even after replicating the model, performance still varies due to hardware, hyper-parameters, and algorithmic stochasticity.
Compared to Amershi \etals general ML workflow~\citep{Amershi2019SE4MLCaseStudy}, during reengineering we find that engineers focus less on neural network design, and more on model analysis, operation conversions, and testing.

\myparagraph{Challenge 3: Portability of DL Operations}

\ParticipantQuote{Practitioner 2}{The biggest problem that we encounter now is that some methods will depend on certain versions of the software...When you modify the version, it will not be reproducible...For example, the early TensorFlow and later TensorFlow have different back propagation when calculation gradient. Later updated versions will have different results even when running with the same code...Running our current code with CUDA 11.2 and PyTorch 11.9, the program can be executable, but the training will have issues....The inconsistency between PyTorch and Numpy versions could make the results the same every time.
}

\ParticipantQuote{Leader 3}{There was basically [a] one-to-one [mapping of] functions from DL\_PLATFORM\_1 [to our implementation in] DL\_PLATFORM\_2. But halfway through...we realized we needed to make it TPU friendly ...had to figure out a way to redesign...to work on TPU since the MODEL\_COMPO NENTS in DL\_PLATFORM\_1  were all dynamically shaped.}
\ParticipantQuote{Leader 2}{They don't talk about TPUs at all...If you're in a GPU environment, it's a lot easier also, but in order to convert it to TPU, we had to put some design strategies into place and test different possible prototypes.}
\ParticipantQuote{Leader 6}{The most challenging part for us is the data pipeline...a lot of the data manipulation...in the original C implementation is...hard [in] Python.}

Though engineers can refer to existing implementations, the conversion of operations between different DL frameworks is still challenging~\citep{mmdnn}.
The interview participants described four kinds of problems.
(1) Different DL frameworks may use different names or signatures for the same behavior. 
(2) The ``same'' behavior may vary between DL frameworks, \eg the implementation of a ``standard'' optimizer.\footnote{The reengineering team identified and disclosed three examples of this behavior to the owners of the DL framework they used. The documentation of a major DL framework was improved based on our reports.} 
(3) Some APIs are not supported on certain hardware (\eg TPU), 
and the behavior must be composed from building blocks.
(4) Out of memory defects could happen in some model architectures when using certain hardware (\eg TPU).
We opened the issue in a major DL framework and the maintainers are investigating.\footnote{See \url{https://github.com/tensorflow/models/issues/10528}.}

\myparagraph{\new{Challenge 4: Customized Data Pipeline}}

\ParticipantQuote{\new{Practitioner 1}}
{\new{If your data format is complicated then the code around it would also be complicated and it would, eventually, lead to some hard to detect bugs.}}

\ParticipantQuote{\new{Practitioner 4}}
{\new{Sometimes some data can mess the whole model prediction. We need to get them out of our project...It probably consumed 30\% of our entire project time.}}

\ParticipantQuote{\new{Practitioner 6}}
{\new{[Data pipeline] is the most challenging part because it’s not like
there’s a formula of things that you can do...When it comes to data pipeline, it’s
more of an art than science}}

\ParticipantQuote{\new{Leader 1}}
{\new{The data pipeline is the hardest to verify and check because testing it takes so long. You need to train the model in order to test the data pipeline.}}


\new{Customizing the data pipeline is a challenging stage in the DL reengineering process, especially when engineers want to adapt the original model to a new dataset. However, our qualitative data also suggest that this can occur during the replication and enhancement stages, where a customized data pipeline may be needed. For instance, replicating a model in a different framework requires corresponding data preprocessing. Similarly, enhancing a model to support different data formats also necessitates adjustments to the data pipeline.
The interview participants described challenges in data processing, data quality, and the implementation of the data pipeline. 
First, the data processing is challenging because there is no standardized solution. 
The practitioners, especially model adaptors, needs to process the data properly to match the original model input. 
Second, the data quality can affect the model performance a lot. For example, Practitioner 4 mentioned that some data can mess the whole model prediction and it is time consuming to address these issues.
Third, the data pipeline can be very different when using different programming languages and DL frameworks. Implementation and testing can be time-consuming.
}

\subsubsection{Reengineering Practices} \label{sec:ReengineeringPractices}
We also summarized three reengineering practices based on our interview analysis:

\myparagraph{Practice 1: Starting with the Interface}

\ParticipantQuote{Practitioner 2}{We will define the API for the [our own] interface first. In such case, if something goes wrong during the component integration, we will easily localize the defect.}

{\ParticipantQuote{\new{Practitioner 3}}{\new{We look for implementations which are very quick to start with, which have a very good API to start.}}

\ParticipantQuote{Leader 4}{If you're using existing implementations, you should take time to familiarize with the existing things [code base] that we already have.}
\ParticipantQuote{Leader 6}{The DL\_FRAMEWORK interface, for the MODEL\_ZOO specifically, I think that's very important.}

Interface can help unify and automate model testing. Typically, an existing code base has a unified structure which includes the basic trainer and data loader of a model. The interface is essential if the reengineering team is required to use existing code base or model zoos, such as Banna \etal describe for the TensorFlow Model Garden~\citep{banna2021experience}.
\textit{Practitioner 2} indicated that implementing interface APIs first can help a lot on bug localization during component integration. Their team also combine agile method~\citep{cohen2004introduction2AgileMethod} with the use of interface APIs which makes the team collaboration more effective. 
Our reengineering team shares similar experience by implementing interface API first. \textit{Leader 4} indicated that team members have to get familiar with the code base and relevant interface APIs first before the real model implementation. 

\myparagraph{Practice 2: Testing}

\ParticipantQuote{Leader 4}{Unit testing was eas[ier]....because if you're just trying to check what you're writing, it is easier than always trying to differentiate your test and match it with some other model.}

\ParticipantQuote{Leader 1}{You can load the...weights from the original paper. [It] might be a little bit difficult to do but you could do that [and] test the entire model by...comparing the outputs of each layer.'}

Based on the interviews and notes from weekly team meetings, we found that complementary testing approaches were helpful.
We describe the team's three test methods: unit testing, differential testing, and visual testing. 



\emph{Unit testing} can ensure that each component works as intended.
For the modeling stage, a \ul{pass-through test} for tensor shapes should be done first, to check whether the output shape of each layer matches the input shape of the next. 
They also do a \ul{gradient test} which calculates the model's gradient to ensure the output is differentiable.
%
%

\emph{Differential testing} compares two supposedly-identical systems by checking whether they are input-output compatible~\citep{Mckeeman1998DifferentialTesting}.
This is most applicable in the Modeling stage: 
the original weights/checkpoint can be loaded into the network, and the behavior of original and reengineered model can be compared.
This technique isolates assessment of the model architecture from the stochastic training component~\citep{Pham2020AnalysisofVarianceinDLSWSystems}. 

Differential testing is also applicable in Environment and Training stages to ensure the consistency of each function.
For example, when testing the loss function in the training stage, they generate random tensors, feed them into both implementations, and compare the outputs.
The outputs should match, up to a rounding error. 

\ParticipantQuote{Leader 4}{Both...implementations, even though we are doing the same model, the way they organized are very different....Differential testing can get difficult.}

\ParticipantQuote{Leader 5}{In the data pipeline, one of the big challenges is that the data pipeline is not deterministic, it's random. So it's hard to make test cases for it nor to see if it matches the digital limitation, because you can't do differential testing because it's random.}

However, differential testing does not apply to all DL components.
For example, the data pipeline has complex probabilistic outputs, 
where they found it simpler to do \emph{visual testing}.
For data pipeline, each preprocessing operation can be tested by passing a sample image through the original pipeline and the reengineering pipeline.
They visually inspect the output of two pipelines to identify noticeable differences.
This approach is applicable in CV, but may not generalize to other DL applications. 


Unit, differential, and visual testing are complementary. 
At coarser code granularities, automated differential tests are effective; at a fine-enough level of granularity the original and replicated component may not be comparable, and unit tests are necessary. 
The characteristics of a data pipeline are difficult to measure automatically, and visual testing helps here. 



\myparagraph{Practice 3: Performance Debugging}

\ParticipantQuote{Leader 1}{I think the biggest thing is logging the accuracy after every single change...then you know what changes are hurting your accuracy and which changes are helping, which is immensely helpful.}
\ParticipantQuote{Practitioner 1}{If your data format is complicated, then the code around it would also be complicated and it would, eventually, lead to some hard to detect bugs.}

The most common way to detect reproducibility and evolutionary defects is to compare model evaluation metrics.
This approach can be costly due to the resources required for training in a code-train-debug-fix cycle. 
However, as we noted from the team meetings,
70\% of the final results may be achieved within 10-25\% of training. 
They thus check whether evaluation metrics and trends are comparable after 25\% of training, shortening the debugging cycle.

Beyond this approach, they agree with prior works that record-keeping help model training and management~\citep{schelter2017automatically, vartak2016modeldb}.

\section{Discussion and Implications} \label{sec:Discussion}

\new{In this section, we first triangulate our findings into a reengineering workflow (\cref{sec:Workflow}). Then we compare our findings to prior works (\cref{sec:Discussion-CompareToPriorWork}) and propose future directions (\cref{sec:Implications}).}

\subsection{Triangulating our findings into a reengineering workflow} \label{sec:Workflow}


The two prongs of our case study identified similar challenges in \emph{model implementation and testing}, especially \emph{performance debugging}, which we show below in \cref{fig:process workflow}.
Our failure analysis found that most reengineering defects are caused by DL APIs and hardware configuration (see \cref{fig:Manifestation - DL Stages} and \cref{fig:Bug types - general (percentage)}), and we identified the similar challenge of \emph{portability} in our CV reengineering team.
Moreover, in the failure analysis we suggested that the training stage would be the most challenging because of the diversity of failure modes (cf. \cref{fig:Bug types - general (percentage)}, \cref{fig: Symptomss}, and \cref{fig:Root causes}). 
This stage was also highlighted by our CV reengineering team as the \emph{performance debugging} challenge.
In addition, as we noticed from our failure analysis, engineers applied multiple testing strategies, among which the most common ones are unit testing, differential testing, and visual testing. 
To support the reengineering works, we provide detailed strategies and insights based on our reengineering experience.


However, our two data sources were not fully in agreement.
The interviews with reengineering team leaders identified the \emph{model operationalization} challenge, while the failure analysis data provided no direct evidence of this challenge.
We believe that when engineers encounter model operationalization defects, they may not open an issue about it, or may be unable to identify the defect until testing.
Thus the \emph{operationalization} challenge may manifest as \emph{performance debugging} defects. 

To illustrate DL Reengineering challenges and practices, we developed a reengineering workflow based on our study on open-source projects and reengineering team.
One leader of our reengineering team described the reengineering workflow their sub-team followed.
The other leaders revised it until agreement.
Then we revised it based on our sub-team approaches and observations of open-source bug reports.

The resulting workflow (see \cref{fig:process workflow} below) has three main stages: 
  (1) Model selection and analysis;
  (2) Implementation and testing;
  and
  (3) Review and release.
In the first stage, the team identifies a candidate model for the desired task (\eg low-power object recognition), and determines its suitability on the target DL framework and hardware.
Existing implementations are examined as points of comparison.
In the second stage, the components of the system are implemented, integrated, and evaluated, typically with different personnel working on each component.
At the end of this stage, the model performance matches the paper to a tolerance of 1-3\%.
In the third stage, the model is tailored for different platforms, \eg servers or mobile devices, and appropriate checkpoint weights are published.
We included our checklists for these tasks in our artifact.
Note that this workflow is fairly linear, without the extensive iteration suggested by Amershi \etal for ML model development~\citep{Amershi2019SE4MLCaseStudy}.
As this workflow is focused on reengineering, the system requirements and many design elements were well understood; less iteration is necessary.
{
\begin{landscape}
\begin{figure*}[t]
    \centering
    \includegraphics[width=\linewidth]{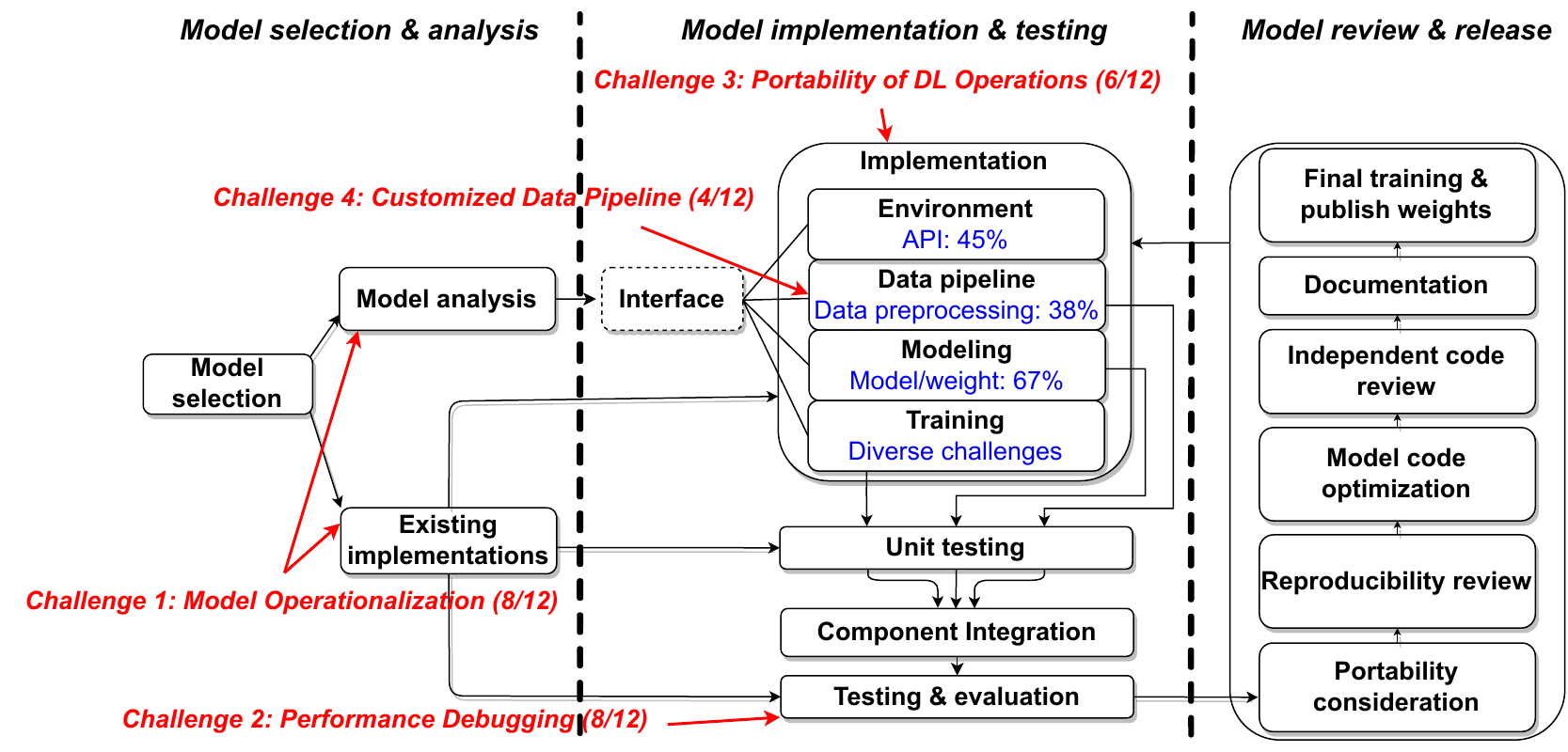}
    \caption{
    Reengineering workflow, divided into three main stages. 
    Forward edges denote the reengineering workflow.
    Dotted line denotes optional (interface).
    Back edges denote points where errors are often identified. 
    Red arrows indicate the main location of each challenge, \new{with the numbers in parentheses representing the count of participants who mentioned these challenges.}
    Blue text denotes the main root causes of failures in each DL stage (see \cref{fig:Root causes}).
    }
    \label{fig:process workflow}
\end{figure*}
\end{landscape}
}

\subsection{Comparison to Prior Works} \label{sec:Discussion-CompareToPriorWork}

\new{We compare our work to three kinds of studies:
  those on DL model development,
  those on traditional software reengineering, and
  those on pre-trained model reuse.
In~\cref{tab:comparison} we summarize our analysis of the similarities and difficulties between our findings and these domains.}

\begin{table}[h]
    \centering
    \caption{
        \new{Differences between our findings on DL reengineering and prior work. Within prior work, we consider DL development, traditional software reengineering, and pre-trained model reuse.}
    }
    \begin{tabular}{
        p{0.15\linewidth}p{0.35\linewidth}p{0.35\linewidth}
    }
        \textbf{Aspect} & \textbf{Prior findings} & \textbf{Our findings}  \\
        \toprule
        Goals & Traditional software reengineering focus on improving system quality, maintainability, and performance~\citep{rosenberg1996softwareReengineering, majthoub2018softwarereengineering}. & DL reengineering aims to facilitate further software reuse and support the ongoing evolution of the software. It is also a significant role of RTP pipeline. (\cref{sec: ReengModel}) \\
        \midrule
        Model Implementation & Prior work highlighted the challenges in selecting model hyperparameters and debugging training process~\citep{Zhang2019CommonChallengesinDevelopingDLApplications, Zhang2020ProgramFailuresofDLjobs, Zhang2018TFBugs}. & Performance debugging is also challging in DL reengineering process. Additional Challenges exist in model operationalization, portability of DL operations, and customized data pipeline.(\cref{RQ5 results}) \\
        \midrule
        Defect Distribution  & Previous studies found API and basic defects common, with fewer hyper-parameter and data quality defects\citep{Islam2016BugRelicationinCodeClones, Humbatova2020TaxonomyofRealFaultsinDLSystems}. & DL reengineering has more defects related to hyper-parameter tuning and training data quality. (\cref{RQ1 results}-\cref{RQ4 results})\\
        \midrule
        Testing Practice  & Pre-trained model reuse faces similar testing challenges but specific situations encountered are different~\citep{Braiek2020onTestingMLPrograms, Zhang2019MLTesting, Jiang2022PTMReuse}.
        & In DL reengineering, engineers can leverage existing implementations, so the application of effective testing methods, such as differential testing and metamorphic testing, could yield significant benefits. (\cref{RQ5 results}, \cref{sec:Workflow})\\
        \bottomrule
    \end{tabular}
    \label{tab:comparison}
\end{table}

\subsubsection{\new{Deep Learning Model Development}}
Our general findings on the CV reengineering process match some results from prior works.
For example, like prior bug studies on DL engineering~\citep{Islam2019DLBugCharacteristics,  Humbatova2020TaxonomyofRealFaultsinDLSystems}, we observed a large percentage of API defects (21\%, 74/\BugNum) within the CV reengineering process. In line with the results from Zhang \etal~\citep{Zhang2018TFBugs}, we also found basic defects are the most common manifestation.

However, we observed notably different proportions of defects, \eg by stage and by cause.
We found a higher proportion of hyper-parameter tuning defects (5\%, 18/348) in the DL reengineering process compared to the results of Islam \etal, who reported a proportion of less than 1\%~\citep{Islam2019DLBugCharacteristics}.
Additionally, the results presented by Humbatova \etal shows that 95\% of survey participants had trouble with training data~\citep{Humbatova2020TaxonomyofRealFaultsinDLSystems}. However, in our result, training data quality only accounts for 19\% (13/68) of defects in data pipeline! This difference may arise from context: 58\% (203/348) of the defects we analyzed were reported by re-users using benchmark datasets. 




Qualitatively, our in-depth study of a CV reengineering team identified more detailed challenges in DL reengineering.
For model configuration, we observed challenges in
    distinguishing models,
    identifying reported algorithm(s),
    and evaluating of trustworthiness.
These were not mentioned in prior works~\citep{Islam2020RepairingDNN:FixpatternsChallenges}.
We also refined portability challenges into three different types: different names/signatures for the same behavior, inconsistent and undocumented behaviors, and different behavior on certain hardware. The latter two were not identified before~\citep{Zhang2019CommonChallengesinDevelopingDLApplications}.
Finally, we noted the importance --- and difficulty --- of performance debugging in DL reengineering.


}

\new{Given these substantial differences, we make some conjecture as to causes.}
\new{DL reengineering process involves unique problem-solving strategies distinct from those used in normal DL development processes. 
These include the need to carefully review and modify hyperparameters from existing models and the data format to fit new datasets, which introduce additional complexity and the potential for data quality defects. 
Furthermore, identifying and rectifying defects even when using the same dataset can require advanced analytical skills and understanding of the initial model's behaviors. 
The reusability and trustworthiness of the starting models also require careful assessment, requiring advanced knowledge of machine learning principles and critical evaluation of existing implementations. 
In contrast, the typical development process often focuses on designing and training a model from scratch, where control over variables and initial conditions is more straightforward~\citep{Amershi2019SE4MLCaseStudy, Rahman2019MLSEinPractice}. 
\new{We conclude that problem-solving approaches in both the development and reengineering processes present distinct challenges: adapting existing components is a core engineering task; inventing new models leans more towards the realm of science.}
}

\subsubsection{\new{Traditional Software Reengineering}}
We also compare the DL reengineering to other reengineering works in the software engineering literature~\citep{singh2019OSSreengineering, bhavsar2020MLreengineering}.
We propose that the goal and focus are two main differences:
First, the goal of many reengineering projects is to improve software functionality, performance, or implementation~\citep{rosenberg1996softwareReengineering, majthoub2018softwarereengineering}, while the main goal we saw in DL reengineering was to support further software reuse and customization.
Second, other reengineering studies focus on the maintenance/process aspect, while in our study we saw that the DL reengineering activities focus on the implementation/evolution aspect~\citep{bennett2000softwareMaintenanceandEvolution}. 
One possible causal factor here is that the DL reengineering activities we observed were building on research products, while general software reengineering is revisiting the output from engineering teams.

\new{The Research-to-Practice (RTP) pipeline has seen increasing adoption in the industry through the rapid invention and deployment of DL techniques~\citep{wang2020synergybetweenDLandSE, bubeck2023sparksofAGI, zhou2022CVinManufacturing, dhanya2022DL4SmartAgriculturalApps}.}
\new{Our investigation of the reengineering process reveals its role in the RTP pipeline~\citep{grima2017RTPModel}. As Davis \etal summarize, the reuse paradigm of DL models encompasses conceptual, adaptation, and deployment reuse~\citep{davis2023PTMReuseChallengesandDirections}.
Our work draws attention to the significance of conceptual and adaptation reuse in the model reengineering process. In particular, our study underscores the ways in which practitioners transition knowledge from research prototypes into practical engineering implementations, such as replicating models in different frameworks or adapting models to custom datasets.}


\subsubsection{\new{Pre-trained Deep Learning Model Reuse}}
\new{Recently, researchers have begun studying the reuse challenges and practices of pre-trained DL model reuse~\citep{Jiang2022PTMReuse, Jiang2022PTMSupplyChain, Kathikar2023VulnerabilitiesofHFModels, davis2023PTMReuseChallengesandDirections}. 
Due to the engineering cost of employing deep learning models in practice, DL engineers tend to reuse open-source pre-trained models (PTMs) for downstream tasks~\citep{davis2023PTMReuseChallengesandDirections, Jiang2022PTMReuse, jiang2023ptmtorrent}.
Like DL reengineering, pre-trained model reuse also encompasses several activities such as conceptual, adaptation, and deployment reuse~\citep{davis2023PTMReuseChallengesandDirections}. 
Both share common hurdles including lack of information about original prototypes, inconsistencies, and trust issues~\citep{Montes2022DiscrepanciesAmongPTNN, Jiang2022PTMReuse}.
However, we note two difference between DL reengineering and pre-trained model reuse.
    First, DL reengineering can benefit significantly from effective differential and metamorphic testing techniques (\cref{sec:ReengineeringPractices}), as compared to pre-trained model reuse. The goal of reengineering a model is to achieve the same performance when reusing or replicating models, while that of reusing pre-trained models is mainly focused on downstream performances~\citep{Jiang2022PTMReuse}.
    \new{Second, prior work has explored effective methods for identifying suitable pre-trained models that can be adapted to specific downstream tasks~\citep{You2021LogME, You2021RankingandTuningPTMs, Lu2019AutoDNNSelection4EdgeInference}. The reengineering process also requires this model selection step (\cref{fig:process workflow}), and adaptation is a key component of the DL reengineering process. However, selecting a good research prototype is harder in reengineering process because model operationalization and portability of DL operations are the two most challenging parts. (\cref{sec:reengWorkflowAndChallenges}).
    }}

\new{
Another emerging trend in reengineering models is the use of so-called ``Foundation Models''~\citep{zhou2023comprehensive, yuan2021florence}. These models have unique attributes in model adaptation. For instance, large language models can be adapted using methods such as prompting or zero/few-shot learning~\citep{touvron2023llama, touvron2023llama2, Brown2020gpt, liu2022Ptuning, wei2021finetuned, abdullah2022chatgpt}. 
This adaptability distinguishes them from the more extended adaptation processes found in the DL reengineering techniques discussed in this work. The key advantage is that they leverage the knowledge and generalization capabilities embedded in their pretrained weights. This facilitates quicker and often more efficient adaptation to new tasks without the need for prolonged retraining or a customized data pipeline~\citep{wang2023cocktailsgd, shu2022testtimePromptTuningforZeroShotGeneralizationinVLModels}.
We therefore expect that the focus of reengineering with foundation models will predominantly be on adaptation rather than reuse, replication, or enhancement. This assertion is based on the primary objective of employing foundation models for downstream tasks~\citep{zhou2023comprehensive}. The portability of these foundation models is still an active area of research~\citep{pan2023understandingLLMinCodeTranslation, yuan2023powerofFM, wu2023visualChatGPT}. Such unique characteristics might necessitate different reengineering practices tailored to foundation models.
Regrettably, the reengineering phenomena of foundation models were not captured in our study because our data was collected in 2021, which was before the rise of foundation models such as LLMs.
}

\subsection{Implications} \label{sec:Implications}
Our empirical data motivates many directions for further research:




\subsubsection{Empirical Studies}
Our case study identified several gaps in our empirical understanding of DL reengineering. 
\emph{First}, to fully understand reengineering problems, it would be useful to investigate more deeply how experts solve specific problems.
Our data indicated the kinds of problems being solved during CV reengineering, but we did not explore the problem solution strategies nor the design rationale.
For example, reengineering defects include indications of expert strategies for choosing loss functions\footnote{See \url{https://github.com/ultralytics/yolov3/issues/2}.} and tuning hyper-parameters\footnote{See \url{https://github.com/junyanz/pytorch-CycleGAN-and-pix2pix/issues/150}.}. 
\emph{Second}, our findings motivate further interviews and surveys for DL engineers, \eg to understand the reengineering process workflow (\cref{fig:process workflow}) and to evaluate the efficiency benefits of our proposed strategies (among others~\citep{Serban2020SEBPinMLAdoptionEffects}).
\emph{Third}, the difficulties we measured in DL model reengineering imply that reengineering is hard. Monte \etal showed that collections of pre-trained models (\eg ONNX model zoo~\citep{ONNX} and Keras applications~\citep{KerasApplication}) may not be entirely consistent with the models they claim to replicate~\citep{Montes2022DiscrepanciesAmongPTNN}.
Prior work also indicated that discrepancies do exist in the DL model supply chain~\citep{2022JiangEmpirical, Jiang2022PTMReuse}. These defects could result in challenges of \textit{model operationalization} and \textit{portability} (\cref{sec:reengWorkflowAndChallenges}) and therefore impact the downstream DL pipelines~\citep{Xin2021ProductionMLPipelines, Gopalakrishna2022IoTPractices, Nikitin2022AutomatedEvolutionaryApproach4theDesignofCompositeMLPipelines}. 
There is no comprehensive study on how these discrepancies can affect downstream implementations and what are effective ways to identify them.
\new{It is important to mention that our study focuses on CV models and does not cover large language models which can be applied to artificial general intelligence systems. Consequently, we advocate for future research to explore the reengineering process of large language models to provide a more complete understanding of DL reengineering. 
In order to enhance the reengineering process and facilitate knowledge transfer, we foresee future work focusing on methods to improve the transferability and auditability of different deep learning systems.}


Given the substantial prior research on DL failures, both quantitative and qualitative, we were surprised by the differences between our results and prior work that we described in~\cref{sec:Discussion-CompareToPriorWork}.
Although we can only conjecture, one possible explanation of these differences is that prior work takes a product view in sampling data (cf. \cref{sec:Background-Empirical}), while our work's sampling approach takes a process view guided by the engineering activity.
The sampling methods between our work and previous studies vary in what they emphasize. 
Prior work used keywords to search for issues, pull requests, and Stack Overflow questions~\citep{Islam2019DLBugCharacteristics, Zhang2018TFBugs, Humbatova2020TaxonomyofRealFaultsinDLSystems, Sun2017RealBugsforML, Shen2021DLCompilerBugs} which provide them a product view of the deep learning failures.
In contrast, during the data collection, we identified the engineering activities and process \emph{first}, then categorized and analyzed the defects.
If the results of a failure analysis differ based on whether data is sampled by a product or a process perspective, the implications for software failure studies are profound --- most prior studies take a product view~\citep{amusuo2022SoftwareFailureAnalysis}.
This may bear further reflection by the empirical software engineering research community.


\subsubsection{DL Software Testing}
Our results shows the DL-stage-related characteristics of reengineering defects (\cref{RQ1 results}--\cref{RQ4 results}) and difficulties of debugging (\cref{RQ5 results}).
We recommend future directions on debugging tools for each DL stage, especially Data Pipeline and Training. 

Our analysis shows that most of the defects in data pipeline are due to the incorrect data pre-processing and low data quality.
There have been many studies on data management~\citep{Kumar2017DataManagementinML}, validation for DL datasets~\citep{Breck2019DataValidation4ML}, and practical tools for data format converting~\citep{Willemink2020MedicalImagingData4ML}.
However, we did not observe much use of these tools in (open-source) practice, including in the commercially maintained zoos.
It is unclear whether the gap is in adopting existing tools or in meeting the real needs of practitioners.
In addition,
creating large DL datasets is expensive due to high labor and time cost. Data augmentation methods are widely used in recent DL models to solve the related problems (\eg overfitting, lower accuracy)~\citep{Shorten2019SurveyonImageDataAug}.
Our results (\cref{RQ4 results}) show that engineers often have data augmentation and data quality defects. Therefore, we recommend future studies on evaluating the quality of data augmentation. 
It would be of great help if engineers can test in advance whether the data augmentation is sufficient for training.

Researchers have been working on automated DL
  testing~\citep{Tian2018DeepTest, Pei2017DeepXplore}
  and fuzzing~\citep{Zhang2020DifferentialFuzzing4DLOps, Guo2018DLFuzz} tools for lowering the cost of the testing in the training stage.
However, there are not many specific testing techniques for DL reengineering~\citep{Braiek2020onTestingMLPrograms}.
In our reengineering team (\cref{RQ5 results}), we found performance debugging challenging, especially for reproducibility defects. 
We recommend software researchers explore end-to-end testing tools for reengineering tasks which can take advantage of existing model implementations.
For example, new fuzzing technology can be developed to use adversarial inputs to test the correctness of training in the early stage by comparing to the original model which can eventually save lots of computational resources and time.
To lower end-to-end costs, improved unit testing methods will also help.
We recommend future directions on similar techniques of effective unit/differential testing for each DL stage to catch training bugs earlier, grounded in the stage-related characteristics of DL defects.

\subsubsection{Enabling Model Reuse}
DL Model reuse may mainly happen in the modeling stage.
Most modeling defects are due to the operations of pre-trained weights and models (\cref{fig:Root causes}). 
Though there have been tools for the conversion of models between different DL frameworks, notably ONNX~\citep{ONNX}, converting models 
remains costly due to
the rapidly increasing number of operators~\citep{mmdnn} and data types~\citep{ONNXBlog}.
Moreover, we found that engineers also struggle with data management and training configuration.
Thus we recommend further investment in \emph{end-to-end model conversion technology}. 

When models cannot be reused automatically, DL reengineers must manually replicate the logic, resulting in a range of defects (\cref{fig:Root causes}).
Both the open-source defects and team leaders mentioned mathematical defects, \eg sign mismatch or numerical instability. 
We suggest a domain-specific language for loss function mathematics, similar to how regular expressions support string parsing~\citep{michael_regexes_2019}.

\subsubsection{Standardized Practices}
As observed in the open-source defects, few of these reengineering efforts followed a standard process, causing novices to get confused.
We recommend step-by-step guidelines for DL reengineering, starting with~\cref{fig:process workflow}.
Though major companies provide some practices to standardize the DL model documentation, such as model card proposed by Google~\citep{Mitchell2019ModelCardGoogle} and metadata extraction from IBM~\citep{Tsay2022AIMetadataExtractionIBM}, our analysis (\cref{fig:Root causes}) and recent work~\citep{Jiang2022PTMReuse} indicate that existing implementations lack standardized documentation which results in the challenge of model analysis and reuse.
Therefore, we recommend engineers develop and adhere to a \emph{standard template}, \eg stating environment configuration and hyper-parameter tuning. We also envision further studies on automated tools to extract the training scripts, hyperparameters, and evaluation results from open-source projects to generate more standardized documentations.

\section{Threats to Validity} \label{sec:Validity}

\myparagraph{Construct Threats}
In this paper we introduced the concept of \emph{Deep Learning Reengineering}.
This concept tailors the idea of software reengineering~\citep{SEReengineering,Byrne1992ConceptualFoundation4SWReengineering} to the reengineering tasks pertinent to DL.
We used this concept as a criterion by which to select repositories and defects for analysis.
Although this concept resembles traditional problems in reuse and porting, we believe reengineering is a useful conceptualization for the activities of software engineers working with DL technologies.
\new{Through our study, we provide empirical evidence on the prevalence and nature of these concepts, demonstrating the appropriateness of this conceptual framework for understanding and improving the practice of deep learning reengineering process (\cref{sec: ReengModel}).}

\myparagraph{Internal Threats}
In measuring aspects of this concept, we relied on manual classification of GitHub issue data. 
Our methodology might bring potential bias in our results.
\new{First, our data collection filters might bias the collection of defects. 
\new{We followed previous studies to study closed issues, which allow us to understand the flow of a full discussion and can help to reveal all possible information types~\citep{arya2019analysisofInfoTypeofOSSIssueDiscussion, Sun2017RealBugsforML}. 
However, we acknowledge that issues without associated fixes can also be important.}
}
Second, the use of only one experienced rater in failure analysis and one analyzer in the interview study also increases the likelihood of subjective biases influencing the results.
To mitigate subjectivity in the GitHub issue analysis, we adapted and applied existing taxonomies. 
Ambiguity was resolved through instrument calibration and discussion.
Acceptable levels of interrater agreement were measured on a sample of the data.

We also agree that there could be some bias in the framework analysis of in the interview study. We considered the bias in our study design.
To mitigate the bias in our interview data, we build a draft of reengineering workflow based on the knowledge of ML development workflow~\citep{Amershi2019SE4MLCaseStudy}. Our observations from the failure analysis also let us tease out similarities and differences in the reengineering context. During the interview, we asked if the subjects had anything to add to our workflow.
\new{We also conducted member checking by sharing our findings with 6 team leaders from \cref{Method: CVReengineeringTeam},
who confirmed their agreement with our analysis~\citep{ritchie2013qualitative}.}


\myparagraph{External Threats}
Our case study considered DL reengineering in the context of computer vision.
Our findings may not generalize to reengineering in other DL applications. 
Within our case study, our failure analysis examined open-source CV models.
Our data may not generalize to commercial practices; to mitigate this, we focused on popular CV models.
Lack of theoretical generalization is a drawback of our case study method; the trade-off is for a deeper look at a single context. We believe findings for CV are worth having, even if they do not generalize to all DL reengineering efforts. CV is a crucial technology with many applications~\citep{Wang2021UAVBug, Garcia2020AVBugs}. 
As a point of comparison, we think that research on web applications contextualized to important frameworks like Rails~\citep{yang2019view} and Node.js~\citep{chang2019detecting} are worth having, even if not all software is a web application or if some web applications use other technologies.

Within our CV reengineering team study, our team had only two years of (undergraduate) experience in the domain. 
However, the corporate sponsor provided weekly expert coaching, and the results are now used internally by the sponsor. 
This provides some evidence that the team produced acceptable engineering results, implying that their process is worth describing in this work.


\section{Conclusions} \label{sec:Conclusion}
Software engineers are engaged in DL reengineering --- getting DL research ideas to work well in their own contexts.
Prior work mainly focus on the product view of DL systems, while our case study explored the process view of characteristics, pitfalls, and practices of DL model reengineering activities.
We analyzed \BugNum CV reengineering defects and reported on
a two-year reengineering effort.
Our analysis shows that the characteristics of reengineering defects vary by DL stage, and that the environment and training configuration are the most challenging parts.
Additionally, we identified \new{four} challenges of DL reengineering: model operationalization, performance debugging,  portability of DL operations, \new{and customized data pipeline}.

We integrated our findings through a DL reengineering process workflow, annotated with practices, challenges, and frequency of defects.
\new{
We compare our work with prior studies on DL development, traditional software reengineering process, and pre-trained model reuse.
Our work shares similar findings with prior defect studies from DL product perspectives, such as a large proportion of API defects. However, we found a higher proportion of defects caused by hyper-parameter tuning and training data quality. 
Moreover, we shows similarities between model reengineering and pre-trained DL model reuse. 
There is some overlap between these two topics and the challenges and practices can also be shareable in both domains.
Our results inform future research directions on further exploration of model reengineering, development of DL software testing tools, facilitate easier model reuse, and develop standardized practices in the field.
}




\section{Data Availability} \label{sec:Reproducibility}

Our artifact is at 
\url{https://github.com/Wenxin-Jiang/EMSE-CVReengineering-Artifact}.

\section{Compliance with Ethical Standards}

\subsection{Disclosure of potential conflicts of interest}

This work was supported by Google and Cisco and by NSF awards \#2107230, \#2229703, \#2107020, and \#2104319.
We acknowledge that Google has an interest in promoting the use of TensorFlow, hence their investment in the TensorFlow Model Garden.
Although this interest enabled the collection of some of the data (the student team), Google did not otherwise engage in the project.
No Google employees participated in the project, neither as subjects nor as researchers.

\subsection{Research involving Human Participants}

Our human subjects work (interviews) followed a protocol that was approved by Purdue University's Institutional Review Board (IRB).
The IRB protocol number is \#\emph{IRB-2021-366}).

%
%

\bibliographystyle{refs/spbasic}      
\bibliography{main}

\begin{thebibliography}{173}
\providecommand{\natexlab}[1]{#1}
\providecommand{\url}[1]{{#1}}
\providecommand{\urlprefix}{URL }
\expandafter\ifx\csname urlstyle\endcsname\relax
  \providecommand{\doi}[1]{DOI~\discretionary{}{}{}#1}\else
  \providecommand{\doi}{DOI~\discretionary{}{}{}\begingroup
  \urlstyle{rm}\Url}\fi
\providecommand{\eprint}[2][]{\url{#2}}

\bibitem[{ONN(2019{\natexlab{a}})}]{ONNX}
 (2019{\natexlab{a}}) {ONNX} {\textbar} {Home}.
  \urlprefix\url{https://onnx.ai/}

\bibitem[{ONN(2019{\natexlab{b}})}]{ONNXBlog}
 (2019{\natexlab{b}}) Portability between deep learning frameworks – with
  {ONNX}.
  \urlprefix\url{https://blog.codecentric.de/en/2019/08/portability-deep-learning-frameworks-onnx/}

\bibitem[{Git(2020)}]{GithubLabels}
 (2020) Managing labels.
  \urlprefix\url{https://docs.github.com/en/issues/using-labels-and-milestones-to-track-work/managing-labels}

\bibitem[{MLR(2020)}]{MLReproducibilityChallengeSpring2021}
 (2020) Papers with {Code} - {ML} {Reproducibility} {Challenge} 2021 {Edition}.
  \urlprefix\url{https://paperswithcode.com/rc2021}

\bibitem[{CVE(2021)}]{CVEngineer2021}
 (2021) Being a {Computer} {Vision} {Engineer} in 2021.
  \urlprefix\url{https://viso.ai/computer-vision/computer-vision-engineer/}

\bibitem[{MLO(2021)}]{MLOpsWorkflow}
 (2021) Machine {Learning} {Operations}. \urlprefix\url{https://ml-ops.org/}

\bibitem[{Abdullah et~al.(2022)Abdullah, Madain, and
  Jararweh}]{abdullah2022chatgpt}
Abdullah M, Madain A, Jararweh Y (2022) Chatgpt: Fundamentals, applications and
  social impacts. In: 2022 Ninth International Conference on Social Networks
  Analysis, Management and Security (SNAMS), IEEE, pp 1--8

\bibitem[{Alahmari et~al.(2020)Alahmari, Goldgof, Mouton, and
  Hall}]{Alahmari2020RepeatabilityofDLModels}
Alahmari SS, Goldgof DB, Mouton PR, Hall LO (2020) Challenges for the
  {Repeatability} of {Deep} {Learning} {Models}. IEEE Access

\bibitem[{AlDanial(2022)}]{ClocTool}
AlDanial (2022) cloc. \urlprefix\url{https://github.com/AlDanial/cloc}

\bibitem[{Alzubaidi et~al.(2021)Alzubaidi, Zhang, Humaidi, Al-Dujaili, Duan,
  Al-Shamma, Santamar{\'\i}a, Fadhel, Al-Amidie, and
  Farhan}]{alzubaidi2021DLreview}
Alzubaidi L, Zhang J, Humaidi AJ, Al-Dujaili A, Duan Y, Al-Shamma O,
  Santamar{\'\i}a J, Fadhel MA, Al-Amidie M, Farhan L (2021) Review of deep
  learning: Concepts, cnn architectures, challenges, applications, future
  directions. Journal of big Data 8:1--74

\bibitem[{Amershi et~al.(2019)Amershi, Begel, Bird, DeLine, and
  Gall}]{Amershi2019SE4MLCaseStudy}
Amershi S, Begel A, Bird C, DeLine R, Gall H (2019) Software {Engineering} for
  {Machine} {Learning}: {A} {Case} {Study}. In: International {Conference} on
  {Software} {Engineering}: {Software} {Engineering} in {Practice}
  ({ICSE}-{SEIP})

\bibitem[{Amusuo et~al.(2022)Amusuo, Sharma, Rao, Vincent, and
  Davis}]{amusuo2022SoftwareFailureAnalysis}
Amusuo P, Sharma A, Rao SR, Vincent A, Davis JC (2022) Reflections on software
  failure analysis. In: {ACM} {Joint} {European} {Software} {Engineering}
  {Conference} and {Symposium} on the {Foundations} of {Software} {Engineering}
  — {Ideas}, {Visions}, and {Reflections} track ({ESEC}/{FSE}-{IVR})

\bibitem[{Anandayuvaraj and
  Davis(2022)}]{Anandayuvaraj2022ReflectingonRecurringIoTDevelopmentFailures}
Anandayuvaraj D, Davis JC (2022) Reflecting on recurring failures in iot
  development. In: Proceedings of the 37th IEEE/ACM International Conference on
  Automated Software Engineering, pp 1--5

\bibitem[{Aranda and Venolia(2009)}]{Aranda2009SecretLifeofBugs}
Aranda J, Venolia G (2009) The secret life of bugs: {Going} past the errors and
  omissions in software repositories. In: International {Conference} on
  {Software} {Engineering} ({ICSE})

\bibitem[{Arya et~al.(2019)Arya, Wang, Guo, and
  Cheng}]{arya2019analysisofInfoTypeofOSSIssueDiscussion}
Arya D, Wang W, Guo JL, Cheng J (2019) Analysis and detection of information
  types of open source software issue discussions. In: 2019 IEEE/ACM 41st
  International Conference on Software Engineering (ICSE), IEEE, pp 454--464

\bibitem[{Bahdanau et~al.(2015)Bahdanau, Cho, and Bengio}]{Bahdanau2015}
Bahdanau D, Cho KH, Bengio Y (2015) Neural machine translation by jointly
  learning to align and translate. In: International {Conference} on {Learning}
  {Representations} ({ICLR})

\bibitem[{Banna et~al.(2021)Banna, Chinnakotla, Yan, Vegesana, Vivek,
  Krishnappa, Jiang, Lu, Thiruvathukal, and Davis}]{banna2021experience}
Banna V, Chinnakotla A, Yan Z, Vegesana A, Vivek N, Krishnappa K, Jiang W, Lu
  YH, Thiruvathukal GK, Davis JC (2021) An experience report on machine
  learning reproducibility: {Guidance} for practitioners and {TensorFlow} model
  garden contributors. \urlprefix\url{https://arxiv.org/abs/2107.00821}

\bibitem[{Bennett and
  Rajlich(2000)}]{bennett2000softwareMaintenanceandEvolution}
Bennett KH, Rajlich VT (2000) Software maintenance and evolution: a roadmap.
  In: Proceedings of the Conference on the Future of Software Engineering, pp
  73--87

\bibitem[{Berner et~al.(2019)Berner, Brockman, Chan, Cheung, Dębiak, Dennison,
  Farhi, Fischer, Hashme, Hesse, Józefowicz, Gray, Olsson, Pachocki, Petrov,
  Pinto, Raiman, Salimans, Schlatter, Schneider, Sidor, Sutskever, Tang,
  Wolski, and Zhang}]{Dota2019}
Berner C, Brockman G, Chan B, Cheung V, Dębiak P, Dennison C, Farhi D, Fischer
  Q, Hashme S, Hesse C, Józefowicz R, Gray S, Olsson C, Pachocki J, Petrov M,
  Pinto HPdO, Raiman J, Salimans T, Schlatter J, Schneider J, Sidor S,
  Sutskever I, Tang J, Wolski F, Zhang S (2019) Dota 2 with {Large} {Scale}
  {Deep} {Reinforcement} {Learning}. arXiv preprint arXiv:191206680
  \urlprefix\url{https://arxiv.org/abs/1912.06680}

\bibitem[{Bhatia et~al.(2023)Bhatia, Eghan, Grichi, Cavanagh, Jiang, and
  Adams}]{bhatia2023towardsMLpipelines}
Bhatia A, Eghan EE, Grichi M, Cavanagh WG, Jiang ZM, Adams B (2023) Towards a
  change taxonomy for machine learning pipelines: Empirical study of ml
  pipelines and forks related to academic publications. Empirical Software
  Engineering 28(3):60

\bibitem[{Bhavsar et~al.(2020)Bhavsar, Shah, and
  Gopalan}]{bhavsar2020MLreengineering}
Bhavsar K, Shah V, Gopalan S (2020) Machine learning: a software process
  reengineering in software development organization. International Journal of
  Engineering and Advanced Technology 9(2):4492--4500

\bibitem[{Bibal and
  Frénay(2016)}]{Bibal2016InterpretabilityofMLModelsandRepresentations}
Bibal A, Frénay B (2016) Interpretability of {Machine} {Learning} {Models} and
  {Representations}: an {Introduction}. In: European {Symposium} on
  {Artificial} {Neural} {Networks}

\bibitem[{Birt et~al.(2016)Birt, Scott, Cavers, Campbell, and
  Walter}]{birt2016MemberChecking}
Birt L, Scott S, Cavers D, Campbell C, Walter F (2016) Member checking: a tool
  to enhance trustworthiness or merely a nod to validation? Qualitative health
  research 26(13):1802--1811

\bibitem[{Boehm and Beck(2010)}]{Boehm2010ChangingNatureofSWEvolution}
Boehm B, Beck K (2010) The changing nature of software evolution; {The}
  inevitability of evolution. In: {IEEE} {Software}

\bibitem[{Borges and Valente(2018)}]{WhatsinaGithubStar}
Borges H, Valente MT (2018) What's in a {GitHub} {Star}? {Understanding}
  {Repository} {Starring} {Practices} in a {Social} {Coding} {Platform}. In:
  Journal of {Systems} and {Software} ({JSS}), \doi{10.1016/j.jss.2018.09.016}

\bibitem[{Braiek and Khomh(2020)}]{Braiek2020onTestingMLPrograms}
Braiek HB, Khomh F (2020) On testing machine learning programs. Journal of
  Systems and Software (JSS) 164:110542

\bibitem[{Breck et~al.(2017)Breck, Cai, Nielsen, Salib, and
  Sculley}]{Breck2017MLtestscore}
Breck E, Cai S, Nielsen E, Salib M, Sculley D (2017) The {ML} test score: {A}
  rubric for {ML} production readiness and technical debt reduction. In: 2017
  {IEEE} {International} {Conference} on {Big} {Data} ({Big} {Data}), pp
  1123--1132, \doi{10.1109/BigData.2017.8258038}

\bibitem[{Breck et~al.(2019)Breck, Polyzotis, Roy, Whang, and
  Zinkevich}]{Breck2019DataValidation4ML}
Breck E, Polyzotis N, Roy S, Whang S, Zinkevich M (2019) Data {Validation} for
  {Machine} {Learning}. In: the {Conference} on {Machine} {Learning} and
  {Systems} ({MLSys})

\bibitem[{Brown et~al.(2020)Brown, Mann, Ryder, Subbiah, Kaplan, Dhariwal,
  Neelakantan, Shyam, Sastry, Askell, Agarwal, Herbert-Voss, Krueger, Henighan,
  Child, Ramesh, Ziegler, Wu, Winter, Hesse, Chen, Sigler, Litwin, Gray, Chess,
  Clark, Berner, McCandlish, Radford, Sutskever, and Amodei}]{Brown2020gpt}
Brown TB, Mann B, Ryder N, Subbiah M, Kaplan J, Dhariwal P, Neelakantan A,
  Shyam P, Sastry G, Askell A, Agarwal S, Herbert-Voss A, Krueger G, Henighan
  T, Child R, Ramesh A, Ziegler DM, Wu J, Winter C, Hesse C, Chen M, Sigler E,
  Litwin M, Gray S, Chess B, Clark J, Berner C, McCandlish S, Radford A,
  Sutskever I, Amodei D (2020) Language {Models} are {Few}-{Shot} {Learners}.
  Tech. Rep. arXiv:2005.14165, arXiv,
  \urlprefix\url{http://arxiv.org/abs/2005.14165}

\bibitem[{Bubeck et~al.(2023)Bubeck, Chandrasekaran, Eldan, Gehrke, Horvitz,
  Kamar, Lee, Lee, Li, Lundberg et~al.}]{bubeck2023sparksofAGI}
Bubeck S, Chandrasekaran V, Eldan R, Gehrke J, Horvitz E, Kamar E, Lee P, Lee
  YT, Li Y, Lundberg S, et~al. (2023) Sparks of artificial general
  intelligence: Early experiments with gpt-4. arXiv preprint arXiv:230312712

\bibitem[{Byrne(1992)}]{Byrne1992ConceptualFoundation4SWReengineering}
Byrne E (1992) A conceptual foundation for software re-engineering. In:
  Conference on {Software} {Maintenance}

\bibitem[{Chang et~al.(2019)Chang, Dou, Gao, Wang, Wei, and
  Huang}]{chang2019detecting}
Chang X, Dou W, Gao Y, Wang J, Wei J, Huang T (2019) Detecting atomicity
  violations for event-driven node. js applications. In: 2019 IEEE/ACM 41st
  International Conference on Software Engineering (ICSE), IEEE, pp 631--642

\bibitem[{Chen et~al.(2022{\natexlab{a}})Chen, Wen, Shi, Lin, Rajbahadur, Ming,
  and Jiang}]{Chen2022TowardsTrainingReproducibleDLModels}
Chen B, Wen M, Shi Y, Lin D, Rajbahadur GK, Ming Z, Jiang (2022{\natexlab{a}})
  Towards {Training} {Reproducible} {Deep} {Learning} {Models}. In:
  International {Conference} on {Software} {Engineering} ({ICSE}), pp
  2202--2214, \doi{10.1145/3510003.3510163}

\bibitem[{Chen et~al.(2022{\natexlab{b}})Chen, Liang, Shen, and
  Jiang}]{Chen2022DLFrameworkBug}
Chen J, Liang Y, Shen Q, Jiang J (2022{\natexlab{b}}) Toward {Understanding}
  {Deep} {Learning} {Framework} {Bugs}.
  \urlprefix\url{http://arxiv.org/abs/2203.04026}

\bibitem[{Cheng et~al.(2022)Cheng, Misra, Schwing, Kirillov, and
  Girdhar}]{Cheng2022Mask2Former}
Cheng B, Misra I, Schwing AG, Kirillov A, Girdhar R (2022) Masked-attention
  {Mask} {Transformer} for {Universal} {Image} {Segmentation}.
  \urlprefix\url{http://arxiv.org/abs/2112.01527}

\bibitem[{Cohen et~al.(2004)Cohen, Lindvall, and
  Costa}]{cohen2004introduction2AgileMethod}
Cohen D, Lindvall M, Costa P (2004) An introduction to agile methods. Advanced
  Computing 62(03):1--66

\bibitem[{Cohen(1960)}]{cohen1960coefficient}
Cohen J (1960) A coefficient of agreement for nominal scales. Educational and
  psychological measurement 20(1):37--46

\bibitem[{Davis et~al.(2023)Davis, Jajal, Jiang, Schorlemmer, Synovic, and
  Thiruvathukal}]{davis2023PTMReuseChallengesandDirections}
Davis JC, Jajal P, Jiang W, Schorlemmer TR, Synovic N, Thiruvathukal GK (2023)
  Reusing deep learning models: Challenges and directions in software
  engineering. In: Proceedings of the IEEE John Vincent Atanasoff Symposium on
  Modern Computing (JVA’23)

\bibitem[{Devanbu et~al.(2020)Devanbu, Dwyer, Elbaum, Lowry, Moran, Poshyvanyk,
  Ray, Singh, and Zhang}]{Devanbu2020SE4DLVision}
Devanbu P, Dwyer M, Elbaum S, Lowry M, Moran K, Poshyvanyk D, Ray B, Singh R,
  Zhang X (2020) Deep {Learning} \& {Software} {Engineering}: {State} of
  {Research} and {Future} {Directions}. arXiv
  \urlprefix\url{https://arxiv.org/abs/2009.08525}

\bibitem[{Dhanya et~al.(2022)Dhanya, Subeesh, Kushwaha, Vishwakarma, Kumar,
  Ritika, and Singh}]{dhanya2022DL4SmartAgriculturalApps}
Dhanya V, Subeesh A, Kushwaha N, Vishwakarma DK, Kumar TN, Ritika G, Singh A
  (2022) Deep learning based computer vision approaches for smart agricultural
  applications. Artificial Intelligence in Agriculture

\bibitem[{Ding et~al.(2021)Ding, Reddy, and Joshi}]{CMUMLReproducibility}
Ding Z, Reddy A, Joshi A (2021) Reproducibility.
  \urlprefix\url{https://blog.ml.cmu.edu/2020/08/31/5-reproducibility/}

\bibitem[{Doshi-Velez and Kim(2017)}]{DoshiVelez2017RigoriousInterpretableML}
Doshi-Velez F, Kim B (2017) Towards {A} {Rigorous} {Science} of {Interpretable}
  {Machine} {Learning}. \urlprefix\url{https://arxiv.org/abs/1702.08608}

\bibitem[{Easterbrook et~al.(2008)Easterbrook, Singer, Storey, and
  Damian}]{easterbrook2008selectingEMMethodsforSEResearch}
Easterbrook S, Singer J, Storey MA, Damian D (2008) Selecting empirical methods
  for software engineering research. Guide to advanced empirical software
  engineering pp 285--311

\bibitem[{Eghbali and Pradel(2020)}]{Eghbali2020String-relatedBugs}
Eghbali A, Pradel M (2020) No strings attached: an empirical study of
  string-related software bugs. In: International {Conference} on {Automated}
  {Software} {Engineering} ({ASE})

\bibitem[{Fitzgerald(2006)}]{fitzgerald2006OSStransformation}
Fitzgerald B (2006) The transformation of open source software. MIS quarterly
  pp 587--598

\bibitem[{Forsyth and Ponce(2002)}]{forsyth2002ComputerVision}
Forsyth DA, Ponce J (2002) Computer vision: a modern approach. prentice hall
  professional technical reference

\bibitem[{Garcia et~al.(2020)Garcia, Feng, Shen, Almanee, Xia, and
  Chen}]{Garcia2020AVBugs}
Garcia J, Feng Y, Shen J, Almanee S, Xia Y, Chen QA (2020) A comprehensive
  study of autonomous vehicle bugs. In: International {Conference} on
  {Software} {Engineering} ({ICSE}),
  \urlprefix\url{https://dl.acm.org/doi/10.1145/3377811.3380397}

\bibitem[{Goel et~al.(2020)Goel, Tung, Lu, and Thiruvathukal}]{Goel2020LPDL}
Goel A, Tung C, Lu YH, Thiruvathukal GK (2020) A {Survey} of {Methods} for
  {Low}-{Power} {Deep} {Learning} and {Computer} {Vision}. In: {IEEE} {World}
  {Forum} on {Internet} of {Things} ({WF}-{IoT})

\bibitem[{Google(2022)}]{TFMGGithub}
Google (2022) Tensorflow model garden.
  \urlprefix\url{https://github.com/tensorflow/models}

\bibitem[{Gopalakrishna et~al.(2022)Gopalakrishna, Anandayuvaraj, Detti, Bland,
  Rahaman, and Davis}]{Gopalakrishna2022IoTPractices}
Gopalakrishna NK, Anandayuvaraj D, Detti A, Bland FL, Rahaman S, Davis JC
  (2022) ``{If} security is required'': {Engineering} and {Security}
  {Practices} for {Machine} {Learning}-based {IoT} {Devices}. In: International
  {Workshop} on {Software} {Engineering} {Research} \& {Practices} for the
  {Internet} of {Things} ({SERP4IoT})

\bibitem[{Goyal et~al.(2018)Goyal, Dollár, Girshick, Noordhuis, Wesolowski,
  Kyrola, Tulloch, Jia, and He}]{Goyal2018ImageNetin1Hour}
Goyal P, Dollár P, Girshick R, Noordhuis P, Wesolowski L, Kyrola A, Tulloch A,
  Jia Y, He K (2018) Accurate, {Large} {Minibatch} {SGD}: {Training} {ImageNet}
  in 1 {Hour}. \urlprefix\url{https://arxiv.org/abs/1706.02677}

\bibitem[{Grima-Farrell(2017)}]{grima2017RTPModel}
Grima-Farrell C (2017) The rtp model: An interactive research to practice
  framework. What Matters in a Research to Practice Cycle? Teachers as
  Researchers pp 237--250

\bibitem[{Guan et~al.(2023)Guan, Xiao, Li, Liu, and
  Bai}]{guan2023RealWorldBugsinMLmodel}
Guan H, Xiao Y, Li J, Liu Y, Bai G (2023) A comprehensive study of real-world
  bugs in machine learning model optimization. In: Proceedings of the
  International Conference on Software Engineering

\bibitem[{Gundersen and Kjensmo(2018)}]{Gundersen2018ReproducibilityinAI}
Gundersen OE, Kjensmo S (2018) State of the art: {Reproducibility} in
  artificial intelligence. AAAI Conference on Artificial Intelligence (AAAI)

\bibitem[{Gundersen et~al.(2018)Gundersen, Gil, and
  Aha}]{Gundersen2018onReproducibleAI}
Gundersen OE, Gil Y, Aha DW (2018) On reproducible {AI}: {Towards} reproducible
  research, open science, and digital scholarship in {AI} publications. AI
  Magazine

\bibitem[{Guo et~al.(2018)Guo, Jiang, Zhao, Chen, and Sun}]{Guo2018DLFuzz}
Guo J, Jiang Y, Zhao Y, Chen Q, Sun J (2018) {DLFuzz}: {Differential} {Fuzzing}
  {Testing} of {Deep} {Learning} {Systems}. In: European {Software}
  {Engineering} {Conference} and {Symposium} on the {Foundations} of {Software}
  {Engineering} ({ESEC}/{FSE})

\bibitem[{Humbatova et~al.(2020)Humbatova, Jahangirova, Bavota, Riccio, Stocco,
  and Tonella}]{Humbatova2020TaxonomyofRealFaultsinDLSystems}
Humbatova N, Jahangirova G, Bavota G, Riccio V, Stocco A, Tonella P (2020)
  Taxonomy of real faults in deep learning systems. In: International
  {Conference} on {Software} {Engineering} ({ICSE})

\bibitem[{Hutson(2018)}]{Huston2018AIfacesReproducibilityCrisis}
Hutson M (2018) Artificial intelligence faces reproducibility crisis. American
  Association for the Advancement of Science 359(6377):725--726,
  \doi{10.1126/science.359.6377.725}

\bibitem[{Islam et~al.(2016)Islam, Mondal, and
  Roy}]{Islam2016BugRelicationinCodeClones}
Islam JF, Mondal M, Roy CK (2016) Bug replication in code clones: An empirical
  study. In: International Conference on Software Analysis, Evolution, and
  Reengineering (SANER), IEEE, vol~1, pp 68--78

\bibitem[{Islam et~al.(2019)Islam, Nguyen, Pan, and
  Rajan}]{Islam2019DLBugCharacteristics}
Islam MJ, Nguyen G, Pan R, Rajan H (2019) A comprehensive study on deep
  learning bug characteristics. In: European {Software} {Engineering}
  {Conference} and {Symposium} on the {Foundations} of {Software} {Engineering}
  ({ESEC}/{FSE})

\bibitem[{Islam et~al.(2020)Islam, Pan, Nguyen, and
  Rajan}]{Islam2020RepairingDNN:FixpatternsChallenges}
Islam MJ, Pan R, Nguyen G, Rajan H (2020) Repairing deep neural networks: fix
  patterns and challenges. In: International {Conference} on {Software}
  {Engineering} ({ICSE})

\bibitem[{Jarzabek(1993)}]{Jarzabek1993SWReengineering4Reusability}
Jarzabek S (1993) Software reengineering for reusability. In: International
  {Computer} {Software} and {Applications} {Conference} ({COMPSAC})

\bibitem[{Jiang et~al.(2022{\natexlab{a}})Jiang, Ergu, Liu, Cai, and
  Ma}]{jiang2022YOLOreview}
Jiang P, Ergu D, Liu F, Cai Y, Ma B (2022{\natexlab{a}}) A review of yolo
  algorithm developments. Procedia Computer Science 199:1066--1073

\bibitem[{Jiang et~al.(2022{\natexlab{b}})Jiang, Synovic, and
  Sethi}]{Jiang2022PTMSupplyChain}
Jiang W, Synovic N, Sethi R (2022{\natexlab{b}}) An {Empirical} {Study} of
  {Artifacts} and {Security} {Risks} in the {Pre}-trained {Model} {Supply}
  {Chain}. Los Angeles p~10

\bibitem[{Jiang et~al.(2022{\natexlab{c}})Jiang, Synovic, Sethi, Indarapu,
  Hyatt, Schorlemmer, Thiruvathukal, and Davis}]{2022JiangEmpirical}
Jiang W, Synovic N, Sethi R, Indarapu A, Hyatt M, Schorlemmer TR, Thiruvathukal
  GK, Davis JC (2022{\natexlab{c}}) An empirical study of artifacts and
  security risks in the pre-trained model supply chain. In: ACM Workshop on
  Software Supply Chain Offensive Research and Ecosystem Defenses (SCORED'22),
  p 105–114, \doi{10.1145/3560835.3564547},
  \urlprefix\url{https://doi.org/10.1145/3560835.3564547}

\bibitem[{Jiang et~al.(2023{\natexlab{a}})Jiang, Synovic, Hyatt, Schorlemmer,
  Sethi, Lu, Thiruvathukal, and Davis}]{Jiang2022PTMReuse}
Jiang W, Synovic N, Hyatt M, Schorlemmer TR, Sethi R, Lu YH, Thiruvathukal GK,
  Davis JC (2023{\natexlab{a}}) An empirical study of pre-trained model reuse
  in the hugging face deep learning model registry. In: {IEEE}/{ACM} 45th
  {International} {Conference} on {Software} {Engineering} (ICSE'23)

\bibitem[{Jiang et~al.(2023{\natexlab{b}})Jiang, Synovic, Jajal, Schorlemmer,
  Tewari, Pareek, Thiruvathukal, and Davis}]{jiang2023ptmtorrent}
Jiang W, Synovic N, Jajal P, Schorlemmer TR, Tewari A, Pareek B, Thiruvathukal
  GK, Davis JC (2023{\natexlab{b}}) Ptmtorrent: A dataset for mining
  open-source pre-trained model packages. Proceedings of the 20th International
  Conference on Mining Software Repositories (MSR'23)

\bibitem[{Jing(2021)}]{ModelZooWeb}
Jing YK (2021) Model {Zoo} - {Deep} learning code and pretrained models.
  \urlprefix\url{https://modelzoo.co/}

\bibitem[{Johnson and Onwuegbuzie(2004)}]{MixedMethodsResearch}
Johnson RB, Onwuegbuzie AJ (2004) Mixed {Methods} {Research}: {A} {Research}
  {Paradigm} {Whose} {Time} {Has} {Come}. Educational Researcher

\bibitem[{Kathikar et~al.(2023)Kathikar, Nair, Lazarine, Sachdeva, and
  Samtani}]{Kathikar2023VulnerabilitiesofHFModels}
Kathikar A, Nair A, Lazarine B, Sachdeva A, Samtani S (2023) Assessing the
  vulnerabilities of the open-source artificial intelligence (ai) landscape: A
  large-scale analysis of the hugging face platform

\bibitem[{Keras(2022)}]{KerasApplication}
Keras (2022) Keras applications.
  \urlprefix\url{https://keras.io/api/applications/}

\bibitem[{Kim and Li(2020)}]{Intro2TFMG}
Kim J, Li J (2020) Introducing the model garden for tensorflow 2.
  \urlprefix\url{https://blog.tensorflow.org/2020/03/introducing-model-garden-for-tensorflow-2.html}

\bibitem[{Kitchenham et~al.(2002)Kitchenham, Pfleeger, Pickard, Jones, Hoaglin,
  El~Emam, and Rosenberg}]{kitchenham2002preliminaryGuidlinesforEMSE}
Kitchenham BA, Pfleeger SL, Pickard LM, Jones PW, Hoaglin DC, El~Emam K,
  Rosenberg J (2002) Preliminary guidelines for empirical research in software
  engineering. IEEE Transactions on software engineering 28(8):721--734

\bibitem[{Krizhevsky et~al.(2012)Krizhevsky, Sutskever, and Hinton}]{ImageNet}
Krizhevsky A, Sutskever I, Hinton GE (2012) {ImageNet} {Classification} with
  {Deep} {Convolutional} {Neural} {Networks}. In: Advances in {Neural}
  {Information} {Processing} {Systems} ({NeurIPS}), vol~6, pp 84--90

\bibitem[{Kumar et~al.(2017)Kumar, Boehm, and
  Yang}]{Kumar2017DataManagementinML}
Kumar A, Boehm M, Yang J (2017) Data {Management} in {Machine} {Learning}:
  {Challenges}, {Techniques}, and {Systems}. In: International {Conference} on
  {Management} of {Data}

\bibitem[{Leveson(1995)}]{leveson1995safeware}
Leveson NG (1995) Safeware: {System} safety and computers. ACM, New York, NY,
  USA

\bibitem[{Leveson(2016)}]{leveson2016EngineeringASaferWorld}
Leveson NG (2016) Engineering a safer world: Systems thinking applied to
  safety. The MIT Press

\bibitem[{Li et~al.(2020)Li, Jiao, Cao, Wong, and Wu}]{Li2020ModelAdaptation}
Li R, Jiao Q, Cao W, Wong HS, Wu S (2020) Model {Adaptation}: {Unsupervised}
  {Domain} {Adaptation} without {Source} {Data}. Proceedings of the IEEE
  Computer Society Conference on Computer Vision and Pattern Recognition pp
  9638--9647, \doi{10.1109/CVPR42600.2020.00966}

\bibitem[{Li et~al.(2021)Li, Liu, Yang, Peng, and Zhou}]{li2021CNNsurvey}
Li Z, Liu F, Yang W, Peng S, Zhou J (2021) A survey of convolutional neural
  networks: analysis, applications, and prospects. IEEE transactions on neural
  networks and learning systems

\bibitem[{Lin et~al.(2014)Lin, Maire, Belongie et~al.}]{COCO}
Lin TY, Maire M, Belongie S, et~al. (2014) Microsoft {COCO}: {Common} {Objects}
  in {Context}. In: European {Conference} on {Computer} {Vision} ({ECCV})

\bibitem[{Linda et~al.(1996)Linda, Rosenberg, and Hyatt}]{SEReengineering}
Linda D, Rosenberg H, Hyatt LE (1996) Software {Re}-engineering. Software
  Assurance Technology Center

\bibitem[{Liu et~al.(2021)Liu, Gao, Xia, Lo, Grundy, and
  Yang}]{Liu2020ReplicabilityandReproducibilityofDLinSE}
Liu C, Gao C, Xia X, Lo D, Grundy J, Yang X (2021) On the {Replicability} and
  {Reproducibility} of {Deep} {Learning} in {Software} {Engineering}. ACM
  Transactions on Software Engineering and Methodology 31(1):1--46

\bibitem[{Liu et~al.(2022)Liu, Ji, Fu, Tam, Du, Yang, and
  Tang}]{liu2022Ptuning}
Liu X, Ji K, Fu Y, Tam W, Du Z, Yang Z, Tang J (2022) P-tuning: Prompt tuning
  can be comparable to fine-tuning across scales and tasks. In: Proceedings of
  the 60th Annual Meeting of the Association for Computational Linguistics
  (Volume 2: Short Papers), pp 61--68

\bibitem[{Liu et~al.(2014)Liu, Xu, and
  Cheung}]{Liu2014PerformanceBugs4SmartPhoneApps}
Liu Y, Xu C, Cheung SC (2014) Characterizing and detecting performance bugs for
  smartphone applications. In: Proceedings of the 36th {International}
  {Conference} on {Software} {Engineering}, ACM, Hyderabad India, pp
  1013--1024, \doi{10.1145/2568225.2568229},
  \urlprefix\url{https://dl.acm.org/doi/10.1145/2568225.2568229}

\bibitem[{Liu et~al.(2020)Liu, Chen, Zhang, Qin, Ji, Lin, and Yang}]{mmdnn}
Liu Y, Chen C, Zhang R, Qin T, Ji X, Lin H, Yang M (2020) Enhancing the
  interoperability between deep learning frameworks by model conversion. In:
  European {Software} {Engineering} {Conference}/{Foundations} of {Software}
  {Engineering} ({ESEC}/{FSE})

\bibitem[{Lorenzoni et~al.(2021)Lorenzoni, Alencar, Nascimento, and
  Cowan}]{Lorenzoni2021MLModelDevelopmentfromSEPerspective}
Lorenzoni G, Alencar P, Nascimento N, Cowan D (2021) Machine {Learning} {Model}
  {Development} from a {Software} {Engineering} {Perspective}: {A} {Systematic}
  {Literature} {Review}. arXiv \urlprefix\url{https://arxiv.org/abs/2102.07574}

\bibitem[{Lu et~al.(2019)Lu, Yang, Chen, and
  Ren}]{Lu2019AutoDNNSelection4EdgeInference}
Lu B, Yang J, Chen LY, Ren S (2019) Automating {Deep} {Neural} {Network}
  {Model} {Selection} for {Edge} {Inference}. In: 2019 {IEEE} {First}
  {International} {Conference} on {Cognitive} {Machine} {Intelligence}
  ({CogMI}), pp 184--193, \doi{10.1109/CogMI48466.2019.00035}

\bibitem[{Ma et~al.(2018{\natexlab{a}})Ma, Juefei-Xu, Zhang, Sun, Xue, Li,
  Chen, Su, Li, Liu et~al.}]{ma2018deepgauge}
Ma L, Juefei-Xu F, Zhang F, Sun J, Xue M, Li B, Chen C, Su T, Li L, Liu Y,
  et~al. (2018{\natexlab{a}}) Deepgauge: Multi-granularity testing criteria for
  deep learning systems. In: Proceedings of the 33rd ACM/IEEE international
  conference on automated software engineering, pp 120--131

\bibitem[{Ma et~al.(2018{\natexlab{b}})Ma, Liu, Lee, Zhang, and
  Grama}]{ma2018mode}
Ma S, Liu Y, Lee WC, Zhang X, Grama A (2018{\natexlab{b}}) Mode: automated
  neural network model debugging via state differential analysis and input
  selection. In: Proceedings of the 2018 26th ACM Joint Meeting on European
  Software Engineering Conference and Symposium on the Foundations of Software
  Engineering, pp 175--186

\bibitem[{Majthoub et~al.(2018)Majthoub, Qutqut, and
  Odeh}]{majthoub2018softwarereengineering}
Majthoub M, Qutqut MH, Odeh Y (2018) Software re-engineering: An overview. In:
  2018 8th International Conference on Computer Science and Information
  Technology (CSIT), IEEE, pp 266--270

\bibitem[{McHugh(2012)}]{mchugh2012interrater}
McHugh ML (2012) Interrater reliability: the kappa statistic. Biochemia medica
  22(3):276--282

\bibitem[{Mckeeman(1998)}]{Mckeeman1998DifferentialTesting}
Mckeeman WM (1998) Differential {Testing} for {Software}. Digital Technical
  Journal

\bibitem[{Meta(2022)}]{TorchVisionGithub}
Meta (2022) Torchvision. \urlprefix\url{https://github.com/pytorch/vision}

\bibitem[{Michael et~al.(2019)Michael, Donohue, Davis, Lee, and
  Servant}]{michael_regexes_2019}
Michael LG, Donohue J, Davis JC, Lee D, Servant F (2019) Regexes are {Hard}:
  {Decision}-{Making}, {Difficulties}, and {Risks} in {Programming} {Regular}
  {Expressions}. In: 2019 34th {IEEE}/{ACM} {International} {Conference} on
  {Automated} {Software} {Engineering} ({ASE}), pp 415--426,
  \doi{10.1109/ASE.2019.00047}, iSSN: 2643-1572

\bibitem[{Mitchell et~al.(2019)Mitchell, Wu, Zaldivar, Barnes, Vasserman,
  Hutchinson, Spitzer, Raji, and Gebru}]{Mitchell2019ModelCardGoogle}
Mitchell M, Wu S, Zaldivar A, Barnes P, Vasserman L, Hutchinson B, Spitzer E,
  Raji ID, Gebru T (2019) Model {Cards} for {Model} {Reporting}. In:
  Proceedings of the {Conference} on {Fairness}, {Accountability}, and
  {Transparency}, ACM, Atlanta GA USA, pp 220--229,
  \doi{10.1145/3287560.3287596},
  \urlprefix\url{https://dl.acm.org/doi/10.1145/3287560.3287596}

\bibitem[{Montes et~al.(2022)Montes, Peerapatanapokin, Schultz, Guo, Jiang, and
  Davis}]{Montes2022DiscrepanciesAmongPTNN}
Montes D, Peerapatanapokin P, Schultz J, Guo C, Jiang W, Davis JC (2022)
  Discrepancies among pre-trained deep neural networks: a new threat to model
  zoo reliability. In: European Software Engineering Conference and Symposium
  on the Foundations of Software Engineering (ESEC/FSE-IVR track).

\bibitem[{Nahar et~al.(2022)Nahar, Zhou, Lewis, and
  Kästner}]{Nahar2022CollaborationChallengesinBuildingMLSystems}
Nahar N, Zhou S, Lewis G, Kästner C (2022) Collaboration {Challenges} in
  {Building} {ML}-{Enabled} {Systems}: {Communication}, {Documentation},
  {Engineering}, and {Process}. In: International {Conference} on {Software}
  {Engineering} ({ICSE})

\bibitem[{Nikanjam and Khomh(2021)}]{Nikanjam2021DesignSmellinDL}
Nikanjam A, Khomh F (2021) Design {Smells} in {Deep} {Learning} {Programs}:
  {An} {Empirical} {Study}. In: {IEEE} {International} {Conference} on
  {Software} {Maintenance} and {Evolution} ({ICSME})

\bibitem[{Nikitin et~al.(2022)Nikitin, Vychuzhanin, Sarafanov, Polonskaia,
  Revin, Barabanova, Maximov, Kalyuzhnaya, and
  Boukhanovsky}]{Nikitin2022AutomatedEvolutionaryApproach4theDesignofCompositeMLPipelines}
Nikitin NO, Vychuzhanin P, Sarafanov M, Polonskaia IS, Revin I, Barabanova IV,
  Maximov G, Kalyuzhnaya AV, Boukhanovsky A (2022) Automated evolutionary
  approach for the design of composite machine learning pipelines. Future
  Generation Computer Systems

\bibitem[{O'Connor(2023)}]{Ryan2023PytorchvsTF}
O'Connor R (2023) Pytorch vs tensorflow in 2023.
  \urlprefix\url{https://www.assemblyai.com/blog/pytorch-vs-tensorflow-in-2023/}

\bibitem[{O'Shea and Nash(2015)}]{oShea2015introductiontoCNN}
O'Shea K, Nash R (2015) An introduction to convolutional neural networks. arXiv
  preprint arXiv:151108458

\bibitem[{Pan et~al.(2023)Pan, Ibrahimzada, Krishna, Sankar, Wassi, Merler,
  Sobolev, Pavuluri, Sinha, and
  Jabbarvand}]{pan2023understandingLLMinCodeTranslation}
Pan R, Ibrahimzada AR, Krishna R, Sankar D, Wassi LP, Merler M, Sobolev B,
  Pavuluri R, Sinha S, Jabbarvand R (2023) Understanding the effectiveness of
  large language models in code translation. arXiv preprint arXiv:230803109

\bibitem[{Pei et~al.(2017)Pei, Cao, Yang, and Jana}]{Pei2017DeepXplore}
Pei K, Cao Y, Yang J, Jana S (2017) {DeepXplore}: {Automated} {Whitebox}
  {Testing} of {Deep} {Learning} {Systems}. In: Symposium on {Operating}
  {Systems} {Principles} ({SOSP})

\bibitem[{Perry et~al.(2004)Perry, Sim, and
  Easterbrook}]{Perry2004CaseStudy4SE}
Perry D, Sim S, Easterbrook S (2004) Case studies for software engineers. In:
  International {Conference} on {Software} {Engineering} ({ICSE})

\bibitem[{Pham et~al.(2020)Pham, Qian, Wang, Lutellier, Rosenthal, Tan, Yu, and
  Nagappan}]{Pham2020AnalysisofVarianceinDLSWSystems}
Pham HV, Qian S, Wang J, Lutellier T, Rosenthal J, Tan L, Yu Y, Nagappan N
  (2020) Problems and {Opportunities} in {Training} {Deep} {Learning}
  {Software} {Systems}: {An} {Analysis} of {Variance}. In: International
  {Conference} on {Automated} {Software} {Engineering} ({ASE}),
  \doi{10.1145/3324884.3416545}

\bibitem[{Pineau(2022)}]{FacebookAIReproducibility}
Pineau J (2022) How the {AI} community can get serious about reproducibility.
  \urlprefix\url{https://ai.facebook.com/blog/how-the-ai-community-can-get-serious-about-reproducibility/}

\bibitem[{Pineau et~al.(2020)Pineau, Vincent-Lamarre, Sinha, Lariviere, and
  Beygelzimer}]{Pineau2020}
Pineau J, Vincent-Lamarre P, Sinha K, Lariviere V, Beygelzimer A (2020)
  Improving {Reproducibility} in {Machine} {Learning} {Research}. Journal of
  Machine Learning Research

\bibitem[{Plastiras et~al.(2018)Plastiras, Kyrkou, and
  Theocharides}]{Redmon2018YOLO}
Plastiras G, Kyrkou C, Theocharides T (2018) {You Only Look Once: Unified,
  Real-Time Object Detection}. ACM International Conference Proceeding Series
  \doi{10.1145/3243394.3243692}

\bibitem[{Pressman(2005)}]{pressman2005SEBOOK}
Pressman RS (2005) Software engineering: a practitioner's approach. Palgrave
  macmillan

\bibitem[{Qi et~al.(2023)Qi, Sun, Gao, Zhang, Li, and
  Liu}]{Qi2023DNNReengineering}
Qi B, Sun H, Gao X, Zhang H, Li Z, Liu X (2023) Reusing deep neural network
  models through model re-engineering. In: International Conference on Software
  Engineering, IEEE Press, p 983–994, \doi{10.1109/ICSE48619.2023.00090},
  \urlprefix\url{https://doi.org/10.1109/ICSE48619.2023.00090}

\bibitem[{Rahman et~al.(2019)Rahman, River, Khomh, Guhneuc, and
  Lehnert}]{Rahman2019MLSEinPractice}
Rahman S, River E, Khomh F, Guhneuc YG, Lehnert B (2019) Machine learning
  software engineering in practice: {An} industrial case study. arXiv preprint
  \doi{10.48550/arXiv.1906.07154}

\bibitem[{Ralph et~al.(2021)Ralph, Ali, Baltes, Bianculli, Diaz, Dittrich,
  Ernst, Felderer, Feldt, Filieri, de~França, Furia, Gay, Gold, Graziotin, He,
  Hoda, Juristo, Kitchenham, Lenarduzzi, Martínez, Melegati, Mendez, Menzies,
  Molleri, Pfahl, Robbes, Russo, Saarimäki, Sarro, Taibi, Siegmund, Spinellis,
  Staron, Stol, Storey, Taibi, Tamburri, Torchiano, Treude, Turhan, Wang, and
  Vegas}]{SIGSOFT2020EmpiricalStandards4SEResearch}
Ralph P, Ali Nb, Baltes S, Bianculli D, Diaz J, Dittrich Y, Ernst N, Felderer
  M, Feldt R, Filieri A, de~França BBN, Furia CA, Gay G, Gold N, Graziotin D,
  He P, Hoda R, Juristo N, Kitchenham B, Lenarduzzi V, Martínez J, Melegati J,
  Mendez D, Menzies T, Molleri J, Pfahl D, Robbes R, Russo D, Saarimäki N,
  Sarro F, Taibi D, Siegmund J, Spinellis D, Staron M, Stol K, Storey MA, Taibi
  D, Tamburri D, Torchiano M, Treude C, Turhan B, Wang X, Vegas S (2021)
  Empirical {Standards} for {Software} {Engineering} {Research}. arXiv
  \urlprefix\url{https://arxiv.org/abs/2010.03525}

\bibitem[{Ren et~al.(2017)Ren, He, Girshick, and Sun}]{Ren2017FasterRCNN}
Ren S, He K, Girshick R, Sun J (2017) Faster {R}-{CNN}: {Towards} {Real}-{Time}
  {Object} {Detection} with {Region} {Proposal} {Networks}. IEEE Transactions
  on Pattern Analysis and Machine Intelligence (TPAMI)

\bibitem[{Ritchie and Spencer(2002)}]{ritchie2002qualitative}
Ritchie J, Spencer L (2002) Qualitative data analysis for applied policy
  research. In: Analyzing qualitative data, Routledge, pp 187--208

\bibitem[{Ritchie et~al.(2013)Ritchie, Lewis, Nicholls, Ormston
  et~al.}]{ritchie2013qualitative}
Ritchie J, Lewis J, Nicholls CM, Ormston R, et~al. (2013) Qualitative research
  practice: A guide for social science students and researchers. sage

\bibitem[{Rosenberg and Hyatt(1996)}]{rosenberg1996softwareReengineering}
Rosenberg LH, Hyatt LE (1996) Software re-engineering. Software Assurance
  Technology Center pp 2--3

\bibitem[{Runeson and Höst(2009)}]{Runeson2009Guidelines4CaseStudyinSE}
Runeson P, Höst M (2009) Guidelines for conducting and reporting case study
  research in software engineering. Empirical Software Engineering (EMSE)

\bibitem[{Schelter et~al.(2017)Schelter, Boese, Kirschnick, Klein, and
  Seufert}]{schelter2017automatically}
Schelter S, Boese JH, Kirschnick J, Klein T, Seufert S (2017) Automatically
  tracking metadata and provenance of machine learning experiments. In: Machine
  Learning Systems Workshop at NIPS

\bibitem[{Schelter et~al.(2018)Schelter, Biessmann, Januschowski, Salinas,
  Seufert, and Szarvas}]{Amazon2018MLModelManagementChallenge}
Schelter S, Biessmann F, Januschowski T, Salinas D, Seufert S, Szarvas G (2018)
  On {Challenges} in {Machine} {Learning} {Model} {Management}. Bulletin of the
  IEEE Computer Society Technical Committee on Data Engineering

\bibitem[{Schmidhuber(2015)}]{Schmidhuber2015DLinNN}
Schmidhuber J (2015) Deep learning in neural networks: {An} overview. Neural
  Networks

\bibitem[{Sculley et~al.(2014)Sculley, Holt, Golovin, Davydov, Phillips, Ebner,
  Chaudhary, and Young}]{Sculley2014MLTechDebt}
Sculley D, Holt G, Golovin D, Davydov E, Phillips T, Ebner D, Chaudhary V,
  Young M (2014) Machine {Learning} : {The} {High}-{Interest} {Credit} {Card}
  of {Technical} {Debt}. In: {NIPS} {Workshop} on {Software} {Engineering} for
  {Machine} {Learning} ({SE4ML})

\bibitem[{Sculley et~al.(2015)Sculley, Holt, Golovin, Davydov, Phillips, Ebner,
  Chaudhary, Young, Crespo, and
  Dennison}]{Sculley2015HiddenTechnicalDebtinMLSystems}
Sculley D, Holt G, Golovin D, Davydov E, Phillips T, Ebner D, Chaudhary V,
  Young M, Crespo JF, Dennison D (2015) Hidden {Technical} {Debt} in {Machine}
  {Learning} {Systems}. In: Advances in {Neural} {Information} {Processing}
  {Systems}, Curran Associates, Inc., vol~28,
  \urlprefix\url{https://proceedings.neurips.cc/paper/2015/hash/86df7dcfd896fcaf2674f757a2463eba-Abstract.html}

\bibitem[{Seaman et~al.(2008)Seaman, Shull, Regardie, Elbert, Feldmann, Guo,
  and Godfrey}]{Seaman2008DefectCategorization}
Seaman CB, Shull F, Regardie M, Elbert D, Feldmann RL, Guo Y, Godfrey S (2008)
  Defect categorization: making use of a decade of widely varying historical
  data. In: Empirical {Software} {Engineering} and {Measurement} ({ESEM})

\bibitem[{Serban et~al.(2020)Serban, Van Der~Blom, Hoos, and
  Visser}]{Serban2020SEBPinMLAdoptionEffects}
Serban A, Van Der~Blom K, Hoos H, Visser J (2020) {Adoption and effects of
  software engineering best practices in machine learning}. International
  Symposium on Empirical Software Engineering and Measurement
  \doi{10.1145/3382494.3410681}

\bibitem[{Shen et~al.(2021)Shen, Ma, Chen, Tian, Cheung, and
  Chen}]{Shen2021DLCompilerBugs}
Shen Q, Ma H, Chen J, Tian Y, Cheung SC, Chen X (2021) A comprehensive study of
  deep learning compiler bugs. In: European {Software} {Engineering}
  {Conference} and {Symposium} on the {Foundations} of {Software} {Engineering}
  ({ESEC}/{FSE})

\bibitem[{Shorten and Khoshgoftaar(2019)}]{Shorten2019SurveyonImageDataAug}
Shorten C, Khoshgoftaar TM (2019) A survey on {Image} {Data} {Augmentation} for
  {Deep} {Learning}. Journal of Big Data

\bibitem[{Shu et~al.(2022)Shu, Nie, Huang, Yu, Goldstein, Anandkumar, and
  Xiao}]{shu2022testtimePromptTuningforZeroShotGeneralizationinVLModels}
Shu M, Nie W, Huang DA, Yu Z, Goldstein T, Anandkumar A, Xiao C (2022)
  Test-time prompt tuning for zero-shot generalization in vision-language
  models. Advances in Neural Information Processing Systems 35:14274--14289

\bibitem[{Singh et~al.(2019)Singh, Singh, and
  Singh}]{singh2019OSSreengineering}
Singh J, Singh K, Singh J (2019) Reengineering framework for open source
  software using decision tree approach. International Journal of electrical
  and computer engineering (IJECE) 9(3):2041--2048

\bibitem[{Srivastava and Thomson(2009)}]{Srivastava2008FrameworkAnalysis}
Srivastava A, Thomson S (2009) Framework analysis: A qualitative methodology
  for applied policy research

\bibitem[{Sun et~al.(2017)Sun, Zhou, Li, Hu, Yang, and
  Li}]{Sun2017RealBugsforML}
Sun X, Zhou T, Li G, Hu J, Yang H, Li B (2017) An {Empirical} {Study} on {Real}
  {Bugs} for {Machine} {Learning} {Programs}. In: Asia-{Pacific} {Software}
  {Engineering} {Conference} ({APSEC})

\bibitem[{Szeliski(2022)}]{szeliski2022ComputerVision}
Szeliski R (2022) Computer vision: algorithms and applications. Springer Nature

\bibitem[{Tan et~al.(2014)Tan, Liu, Li, Wang, Zhou, and
  Zhai}]{Tan2014OSSBugCharacteristics}
Tan L, Liu C, Li Z, Wang X, Zhou Y, Zhai C (2014) Bug characteristics in open
  source software. Empirical Software Engineering (EMSE)

\bibitem[{Tatman et~al.(2018)Tatman, Vanderplas, and
  Dane}]{Tatman2018TaxonomyofReproducibility4MLResearch}
Tatman R, Vanderplas J, Dane S (2018) A {Practical} {Taxonomy} of
  {Reproducibility} for {Machine} {Learning} {Research}. In: Reproducibility in
  {Machine} {Learning} {Workshop} at {ICML}

\bibitem[{Thiruvathukal et~al.(2022)Thiruvathukal, Lu, Kim, Chen, and
  Chen}]{thiruvathukal2022lpcv}
Thiruvathukal GK, Lu YH, Kim J, Chen Y, Chen B (2022) Low-power Computer
  Vision: Improve the Efficiency of Artificial Intelligence

\bibitem[{Thung et~al.(2012)Thung, Wang, Lo, and
  Jiang}]{Thung2012BugsinMLSystems}
Thung F, Wang S, Lo D, Jiang L (2012) An empirical study of bugs in machine
  learning systems. In: International {Symposium} on {Software} {Reliability}
  {Engineering} ({ISSRE})

\bibitem[{Tian et~al.(2018)Tian, Pei, Jana, and Ray}]{Tian2018DeepTest}
Tian Y, Pei K, Jana S, Ray B (2018) {DeepTest}: automated testing of
  deep-neural-network-driven autonomous cars. In: International {Conference} on
  {Software} {Engineering} ({ICSE})

\bibitem[{Touvron et~al.(2023{\natexlab{a}})Touvron, Lavril, Izacard, Martinet,
  Lachaux, Lacroix, Rozi{\`e}re, Goyal, Hambro, Azhar
  et~al.}]{touvron2023llama}
Touvron H, Lavril T, Izacard G, Martinet X, Lachaux MA, Lacroix T, Rozi{\`e}re
  B, Goyal N, Hambro E, Azhar F, et~al. (2023{\natexlab{a}}) Llama: Open and
  efficient foundation language models. arXiv preprint arXiv:230213971

\bibitem[{Touvron et~al.(2023{\natexlab{b}})Touvron, Martin, Stone, Albert,
  Almahairi, Babaei, Bashlykov, Batra, Bhargava, Bhosale
  et~al.}]{touvron2023llama2}
Touvron H, Martin L, Stone K, Albert P, Almahairi A, Babaei Y, Bashlykov N,
  Batra S, Bhargava P, Bhosale S, et~al. (2023{\natexlab{b}}) Llama 2: Open
  foundation and fine-tuned chat models. arXiv preprint arXiv:230709288

\bibitem[{Tsay et~al.(2022)Tsay, Braz, Hirzel, Shinnar, and
  Mummert}]{Tsay2022AIMetadataExtractionIBM}
Tsay J, Braz A, Hirzel M, Shinnar A, Mummert T (2022) Extracting enhanced
  artificial intelligence model metadata from software repositories. Empirical
  Software Engineering 27(7):176, \doi{10.1007/s10664-022-10206-6},
  \urlprefix\url{https://link.springer.com/10.1007/s10664-022-10206-6}

\bibitem[{Tucker and Devon(2010)}]{Tucker2010CaseStudyinSWREengineering}
Tucker DC, Devon MS (2010) A {Case} {Study} in {Software} {Reengineering}. In:
  International {Conference} on {Informatio} ({ITNG})n {Technology}: {New}
  {Generations}

\bibitem[{Unceta et~al.(2020)Unceta, Nin, and
  Pujol}]{Unceta2020EnvironmentalAdaptationDifferentialReplicationinML}
Unceta I, Nin J, Pujol O (2020) Environmental adaptation and differential
  replication in machine learning. Entropy

\bibitem[{Valett and McGarry(1989)}]{Valett1989SWMeasurementExperiencesinSELab}
Valett JD, McGarry FE (1989) A {Summary} of {Software} {Measurement}
  {Experiences} in the {Software} {Engineering} {Laboratory}. Journal of
  Systems and Software 9:137 -- 148

\bibitem[{Vartak et~al.(2016)Vartak, Subramanyam, Lee, Viswanathan, Husnoo,
  Madden, and Zaharia}]{vartak2016modeldb}
Vartak M, Subramanyam H, Lee WE, Viswanathan S, Husnoo S, Madden S, Zaharia M
  (2016) Modeldb: a system for machine learning model management. In: the
  Workshop on Human-In-the-Loop Data Analytics

\bibitem[{Villa and Zimmerman(2018)}]{DeterminedAIReproducibility}
Villa J, Zimmerman Y (2018) Reproducibility in {ML}: why it matters and how to
  achieve it. \urlprefix\url{https://determined.ai/blog/reproducibility-in-ml}

\bibitem[{Vinyals et~al.(2019)Vinyals, Babuschkin, Czarnecki, Mathieu, Dudzik,
  Chung, Choi, Powell, Ewalds, Georgiev et~al.}]{vinyals2019grandmaster}
Vinyals O, Babuschkin I, Czarnecki WM, Mathieu M, Dudzik A, Chung J, Choi DH,
  Powell R, Ewalds T, Georgiev P, et~al. (2019) Grandmaster level in starcraft
  ii using multi-agent reinforcement learning. Nature

\bibitem[{Voulodimos et~al.(2018)Voulodimos, Doulamis, Doulamis, and
  Protopapadakis}]{DL4CV}
Voulodimos A, Doulamis N, Doulamis A, Protopapadakis E (2018) Deep {Learning}
  for {Computer} {Vision}: {A} {Brief} {Review}. Computational Intelligence and
  Neuroscience

\bibitem[{Wang et~al.(2021)Wang, Li, Xiao, Liu, and Sui}]{Wang2021UAVBug}
Wang D, Li S, Xiao G, Liu Y, Sui Y (2021) An exploratory study of autopilot
  software bugs in unmanned aerial vehicles. In: {ACM} {Joint} {European}
  {Software} {Engineering} {Conference} and {Symposium} on the {Foundations} of
  {Software} {Engineering} ({ESEC}/{FES}), \doi{10.1145/3468264.3468559}

\bibitem[{Wang et~al.(2017)Wang, Dou, Gao, Gao, Qin, Yin, and
  Wei}]{Wang2017ConcurrencyBugsinNodejs}
Wang J, Dou W, Gao Y, Gao C, Qin F, Yin K, Wei J (2017) A comprehensive study
  on real world concurrency bugs in {Node}.js. In: 2017 32nd {IEEE}/{ACM}
  {International} {Conference} on {Automated} {Software} {Engineering} ({ASE}),
  pp 520--531, \doi{10.1109/ASE.2017.8115663}

\bibitem[{Wang et~al.(2023)Wang, Lu, Yuan, Chen, Liang, De~Sa, Re, and
  Zhang}]{wang2023cocktailsgd}
Wang J, Lu Y, Yuan B, Chen B, Liang P, De~Sa C, Re C, Zhang C (2023)
  Cocktailsgd: Fine-tuning foundation models over 500mbps networks. In:
  International Conference on Machine Learning, PMLR, pp 36058--36076

\bibitem[{Wang et~al.(2020{\natexlab{a}})Wang, Brown, Jennings, and
  Stolee}]{Wang2020RegexBugs}
Wang P, Brown C, Jennings JA, Stolee KT (2020{\natexlab{a}}) An {Empirical}
  {Study} on {Regular} {Expression} {Bugs}. International Conference on Mining
  Software Repositories (MSR)

\bibitem[{Wang et~al.(2020{\natexlab{b}})Wang, Huang, Ge, Zhang, Feng, Li,
  Zhang, and Ng}]{wang2020synergybetweenDLandSE}
Wang S, Huang L, Ge J, Zhang T, Feng H, Li M, Zhang H, Ng V
  (2020{\natexlab{b}}) Synergy between machine/deep learning and software
  engineering: How far are we? arXiv preprint arXiv:200805515

\bibitem[{Wardat et~al.(2021)Wardat, Le, and Rajan}]{Wardat2021DeepLocalize}
Wardat M, Le W, Rajan H (2021) {DeepLocalize}: {Fault} {Localization} for
  {Deep} {Neural} {Networks}. In: 2021 {IEEE}/{ACM} 43rd {International}
  {Conference} on {Software} {Engineering} ({ICSE}), pp 251--262,
  \doi{10.1109/ICSE43902.2021.00034}

\bibitem[{Wei et~al.(2021)Wei, Bosma, Zhao, Guu, Yu, Lester, Du, Dai, and
  Le}]{wei2021finetuned}
Wei J, Bosma M, Zhao VY, Guu K, Yu AW, Lester B, Du N, Dai AM, Le QV (2021)
  Finetuned language models are zero-shot learners. arXiv preprint
  arXiv:210901652

\bibitem[{Wei et~al.(2022)Wei, Wang, Yang, and Chan}]{wei2022sebox4dl}
Wei Z, Wang H, Yang Z, Chan W (2022) Sebox4dl: a modular software engineering
  toolbox for deep learning models. In: Proceedings of the ACM/IEEE 44th
  International Conference on Software Engineering: Companion Proceedings, pp
  193--196

\bibitem[{Willemink et~al.(2020)Willemink, Koszek, Hardell, Wu, Fleischmann,
  Harvey, Folio, Summers, Rubin, and
  Lungren}]{Willemink2020MedicalImagingData4ML}
Willemink MJ, Koszek WA, Hardell C, Wu J, Fleischmann D, Harvey H, Folio LR,
  Summers RM, Rubin DL, Lungren MP (2020) Preparing {Medical} {Imaging} {Data}
  for {Machine} {Learning}. Radiological Society of North America

\bibitem[{Wu et~al.(2023)Wu, Yin, Qi, Wang, Tang, and
  Duan}]{wu2023visualChatGPT}
Wu C, Yin S, Qi W, Wang X, Tang Z, Duan N (2023) Visual chatgpt: Talking,
  drawing and editing with visual foundation models. arXiv preprint
  arXiv:230304671

\bibitem[{Wu et~al.(2016)Wu, Schuster, Chen, Le, Norouzi, Macherey, Krikun,
  Cao, Gao, Macherey et~al.}]{wu2016google}
Wu Y, Schuster M, Chen Z, Le QV, Norouzi M, Macherey W, Krikun M, Cao Y, Gao Q,
  Macherey K, et~al. (2016) Google's neural machine translation system:
  Bridging the gap between human and machine translation. arXiv

\bibitem[{Xin et~al.(2021)Xin, Miao, Parameswaran, and
  Polyzotis}]{Xin2021ProductionMLPipelines}
Xin D, Miao H, Parameswaran A, Polyzotis N (2021) Production machine learning
  pipelines: Empirical analysis and optimization opportunities. In: Proceedings
  of the 2021 International Conference on Management of Data, pp 2639--2652

\bibitem[{Xu et~al.(2021)Xu, Wang, Shou, Ngo, Sadick, and
  Wang}]{Xu2021CVinConstructionACriticalReview}
Xu S, Wang J, Shou W, Ngo T, Sadick AM, Wang X (2021) Computer {Vision}
  {Techniques} in {Construction}: {A} {Critical} {Review}. Archives of
  Computational Methods in Engineering

\bibitem[{Yang et~al.(2019)Yang, Yan, Wan, Lu, and Cheung}]{yang2019view}
Yang J, Yan C, Wan C, Lu S, Cheung A (2019) View-centric performance
  optimization for database-backed web applications. In: 2019 IEEE/ACM 41st
  International Conference on Software Engineering (ICSE), IEEE, pp 994--1004

\bibitem[{You et~al.(2021{\natexlab{a}})You, Liu, Wang, Jordan, and
  Long}]{You2021RankingandTuningPTMs}
You K, Liu Y, Wang J, Jordan MI, Long M (2021{\natexlab{a}}) Ranking and
  {Tuning} {Pre}-trained {Models}: {A} {New} {Paradigm} of {Exploiting} {Model}
  {Hubs}. The Journal of Machine Learning Research (JMLR) 23(1):9400--9446,
  \urlprefix\url{http://arxiv.org/abs/2110.10545}

\bibitem[{You et~al.(2021{\natexlab{b}})You, Liu, Wang, and
  Long}]{You2021LogME}
You K, Liu Y, Wang J, Long M (2021{\natexlab{b}}) {LogME}: {Practical}
  {Assessment} of {Pre}-trained {Models} for {Transfer} {Learning}. In:
  International {Conference} on {Machine} {Learning} ({ICML}), PMLR, pp
  12133--12143, \urlprefix\url{https://proceedings.mlr.press/v139/you21b.html}

\bibitem[{Yuan et~al.(2021)Yuan, Chen, Chen, Codella, Dai, Gao, Hu, Huang, Li,
  Li et~al.}]{yuan2021florence}
Yuan L, Chen D, Chen YL, Codella N, Dai X, Gao J, Hu H, Huang X, Li B, Li C,
  et~al. (2021) Florence: A new foundation model for computer vision. arXiv
  preprint arXiv:211111432

\bibitem[{Yuan(2023)}]{yuan2023powerofFM}
Yuan Y (2023) On the power of foundation models. In: International Conference
  on Machine Learning, PMLR, pp 40519--40530

\bibitem[{Zhang et~al.(2020{\natexlab{a}})Zhang, Harman, Ma, and
  Liu}]{Zhang2019MLTesting}
Zhang JM, Harman M, Ma L, Liu Y (2020{\natexlab{a}}) Machine learning testing:
  Survey, landscapes and horizons. IEEE Transactions on Software Engineering
  48(1):1--36

\bibitem[{Zhang et~al.(2020{\natexlab{b}})Zhang, Xiao, Zhang, Liu, Lin, and
  Yang}]{Zhang2020ProgramFailuresofDLjobs}
Zhang R, Xiao W, Zhang H, Liu Y, Lin H, Yang M (2020{\natexlab{b}}) An
  empirical study on program failures of deep learning jobs. In: International
  {Conference} on {Software} {Engineering} ({ICSE})

\bibitem[{Zhang et~al.(2019)Zhang, Gao, Ma, Lyu, and
  Kim}]{Zhang2019CommonChallengesinDevelopingDLApplications}
Zhang T, Gao C, Ma L, Lyu M, Kim M (2019) An {Empirical} {Study} of {Common}
  {Challenges} in {Developing} {Deep} {Learning} {Applications}. In:
  International {Symposium} on {Software} {Reliability} {Engineering} ({ISSRE})

\bibitem[{Zhang et~al.(2021)Zhang, Liu, Sun, Fang, Liu, Wang, Chai, and
  Chen}]{Zhang2020DifferentialFuzzing4DLOps}
Zhang X, Liu J, Sun N, Fang C, Liu J, Wang J, Chai D, Chen Z (2021) Duo:
  {Differential} {Fuzzing} for {Deep} {Learning} {Operators}. IEEE Transactions
  on Reliability

\bibitem[{Zhang et~al.(2018)Zhang, Chen, Cheung, Xiong, and
  Zhang}]{Zhang2018TFBugs}
Zhang Y, Chen Y, Cheung SC, Xiong Y, Zhang L (2018) An empirical study on
  {TensorFlow} program bugs. International Symposium on Software Testing and
  Analysis (ISSTA)

\bibitem[{Zhang et~al.(2020{\natexlab{c}})Zhang, Ren, Chen, Xiong, Cheung, and
  Xie}]{Zhang2020NumericalBugsinNN}
Zhang Y, Ren L, Chen L, Xiong Y, Cheung SC, Xie T (2020{\natexlab{c}})
  Detecting numerical bugs in neural network architectures. European Software
  Engineering Conference and Symposium on the Foundations of Software
  Engineering (ESEC/FSE)

\bibitem[{Zhou et~al.(2023)Zhou, Li, Li, Yu, Liu, Wang, Zhang, Ji, Yan, He
  et~al.}]{zhou2023comprehensive}
Zhou C, Li Q, Li C, Yu J, Liu Y, Wang G, Zhang K, Ji C, Yan Q, He L, et~al.
  (2023) A comprehensive survey on pretrained foundation models: A history from
  bert to chatgpt. arXiv preprint arXiv:230209419

\bibitem[{Zhou et~al.(2022)Zhou, Zhang, and Konz}]{zhou2022CVinManufacturing}
Zhou L, Zhang L, Konz N (2022) Computer vision techniques in manufacturing.
  IEEE Transactions on Systems, Man, and Cybernetics: Systems 53(1):105--117

\bibitem[{Zou et~al.(2023)Zou, Yang, Zhang, Li, Li, Gao, and
  Lee}]{Zou2023MetaSegmentEverything}
Zou X, Yang J, Zhang H, Li F, Li L, Gao J, Lee YJ (2023) Segment {Everything}
  {Everywhere} {All} at {Once}. \urlprefix\url{http://arxiv.org/abs/2304.06718}

\end{thebibliography}

%
%


\end{document}